\documentclass[pra,singlecolumn,amssymb,amsmath]{revtex4-1}

\usepackage{graphicx}
\usepackage{epsfig,color}
\usepackage{amsmath,amssymb,amsfonts,bbold}
\usepackage{amsthm}
\usepackage{hyperref}
\usepackage{enumerate}

\newtheorem{sttmnt}{Statement}
\newcommand{\be}{\begin{equation}}
\newcommand{\ee}{\end{equation}}

\newcommand{\Id}{\mathbb{1}}
\newcommand{\ket}[1]{|{#1}\rangle}
\newcommand{\bra}[1]{\langle{#1}|}

\newcommand{\bracket}[2]{\langle#1|#2\rangle}
\newcommand{\mr}{\mathrm}
\newcommand{\dg}{\dagger}
\newcommand{\mb}{\mathbf}
\newcommand{\mc}{\mathcal}

\begin{document}

\title{Fermionic Projected Entangled Pair States and Local $U(1)$ Gauge Theories}

\date{\today}

\author{Erez Zohar}

\author{Michele Burrello}

\author{Thorsten B. Wahl}

\author{J. Ignacio Cirac}
\address{Max-Planck-Institut f\"ur Quantenoptik, Hans-Kopfermann-Stra\ss e 1, 85748 Garching, Germany.}

\begin{abstract}
Tensor networks, and in particular Projected Entangled Pair States (PEPS), are a powerful tool for the study of quantum many body physics, thanks to both their built-in ability of classifying and studying symmetries, and the efficient numerical calculations they allow. In this work, we introduce a way to extend the set of symmetric PEPS  in order to include local gauge invariance and investigate lattice gauge theories with fermionic matter. To this purpose, we provide as a case study and first example, the construction of a fermionic PEPS, based on Gaussian schemes, invariant under both global and local $U(1)$ gauge transformations. The obtained states correspond to a truncated $U(1)$ lattice gauge theory in $2+1$ dimensions, involving both the gauge field and fermionic matter. For the global symmetry (pure fermionic) case, these PEPS can be studied in terms of spinless fermions subject to a p-wave superconducting pairing. For the local symmetry (fermions and gauge fields) case, we find confined and deconfined phases in the pure gauge limit, and we discuss the screening properties of the phases arising in the presence of dynamical matter.
\end{abstract}

\maketitle

\tableofcontents

\section{Introduction}

Within the framework of the standard model of particle physics, the three fundamental forces are described by \emph{gauge bosons}, which are the excitations of \emph{gauge fields}. Gauge fields are vector fields, which manifest a very special \emph{local} continuous symmetry, called \emph{local gauge invariance}. This symmetry gives the matter fields gauge charges, and its local nature induces local interactions of the gauge currents with the charged matter. The conservation of local charges, manifested by local constraints which are extensions of the well-known \emph{Gauss law} from electrodynamics, implies a very rich, complicated structure of the Hilbert space of quantum gauge theories, dictated by superselection rules governed by these local charges. This makes such theories, in general, very challenging and difficult to solve - just as much as they are interesting and important for the description of nature.

Described within the framework of quantum field theory, gauge theories come along with a very important computational tool: perturbation theory and its Feynman diagrams. However, despite the great success and accuracy achieved for Quantum Electrodynamics (QED) with perturbative methods, they apply only partially to QCD. Unlike QED, which is the Abelian gauge theory associated with the group $U(1)$, QCD is a non-Abelian, $SU(3)$ gauge theory, which makes it behave in  a completely different manner: due to an important property of such non-Abelian theories, \emph{Asymptotic Freedom} \cite{Gross1973}, the strong coupling constant flows to zero for high energies (or short distances) - allowing, therefore, for perturbative calculations in these scales, such as within the nuclei (e.g., the parton model, or Bjorken's scaling \cite{Bjorken1969b,Bjorken1969}). On the other hand, at low energies or large distances, the coupling constant is strong and perturbative physics is impossible; this may be seen as both the cause and the effect of \emph{Quark Confinement} \cite{Wilson,Polyakov}, the phenomenon responsible for holding quarks bound together into hadrons, and for the absence of free quarks in the spectrum of the theory.

This has significant implications on the study of the theory, and, indeed, over the years many non-pertubative techniques have been developed and applied for the study of QCD, and non-Abelian gauge theories in general. One of them, perhaps the most fruitful, is lattice gauge theory (LGT) \cite{Wilson,KogutSusskind,KogutLattice,Kogut1983}, in which either spacetime, or space, is discretized, allowing either for a regularization of the theory for analytical purposes, or very efficient and fruitful numerical (Monte Carlo) calculations. While having a great success with many different types of calculations and predictions (e.g., low-energy hadronic spectrum \cite{FLAG2013}, among others), Monte Carlo calculations are problematic in some cases: first, with fermions with a finite chemical potential (required, for example, for the phases of color superconductivity and quark-gluon plasma \cite{McLerran1986,Fukushima2011}), due to the computationally hard sign problem \cite{Troyer2005}, and second, as the calculations are carried out in Euclidean spacetime, real-time dynamics in Minkowski spacetime cannot be achieved (see, on the other hand, the recent works \cite{Hebenstreit2013,Hebenstreit2013a}).

 A complementary way of overcoming the computational difficulties may be the use of tensor network techniques, or
 tensor network states (TNs), and in particular Matrix Product States (MPS) \cite{Perez2007} and Projected Entangled Pair States (PEPS) \cite{Verstraete2004,Verstraete2008,Cirac2009}.
 One may, for example, use TN variational techniques, in which a TN state with variational parameters is used as an ansatz for the ground state of a given Hamiltonian, as well as calculate dynamics of such states in very efficient ways, exploiting methods like DMRG (Density Matrix Renormalization Group) \cite{Schollwock2005}.
 This approach has been recently applied with MPS for $1+1$ dimensional lattice gauge theories, either Abelian or non-Abelian, and used for the study of their spectrum, dynamics (including string-breaking) and finite temperature effects   \cite{Banuls2013,Banuls2013a,Rico2013,Kuhn2014,Silvi2014,Buyens2014,Saito2014,Banuls2015,Kuhn2015,Pichler2015}. Furthermore also tensor renormalization group techniques have been recently applied to the study of such models \cite{Shimizu2014}.
 The MPS studies have shown many of the static and dynamic properties of some well known theories (such as the Schwinger model \cite{Schwinger1962I,Schwinger1962}, for example), and allowed to reach better precision than analogous Monte Carlo calculations and to perform dynamical simulations, holding the promise for even better accuracy and computational possibilities.

 However, the great computational power is not the only reason for TNs to be candidates for the study of gauge theories.  In a somewhat change of paradigm, one may describe a physical system from the point of view of its most representative states, instead of starting from its Lagrangian or Hamiltonian formulation. Tensor networks are, indeed, well suited to define families of states, as functions of a set of variational parameters, which fulfill precise symmetry constraints. Therefore they provide a natural way to encode all the symmetries of a system \cite{Perez2008} and to describe its possible thermodynamical phases in terms of representative states, allowing to investigate the main physical properties within the universality class of the problem under scrutiny \cite{Schuch2010}.  Starting from the tensor network construction, it is also possible to show that such states constitute the ground states of local parent Hamiltonians. These Hamiltonians may be explicitly derived in the simplest cases, and offer, as a function of the variational parameters, suitable examples to study the properties of the thermodynamical phases in a certain universality class. Previous works in this direction include two-dimensional PEPS schemes for lattice pure-gauge theories (without dynamical matter) \cite{Tagliacozzo2014}, as well as a general framework useful to the study of lattice gauge theories with bosonic matter \cite{Haegeman2014}.

 The next reasonable step is to consider lattice gauge theories with \emph{fermionic dynamical matter}, as in the case of high energy physics theories, for example. This paper addresses precisely this problem; more specifically, we analyze how  one could utilize PEPS for the study of lattice gauge theories with dynamical matter. Could one classify locally gauge invariant states using PEPS, in a way that allows, eventually, to study the theories described by the Hilbert spaces to which they belong?

For that purpose, we systematically construct, in this paper, fermionic  PEPS (fPEPS)  \cite{Kraus2010}  in $2+1$ dimensions, which have both local gauge symmetry and fundamental physical symmetries - rotation, translation and charge conjugation. To demonstrate the strength of fPEPS for studying such theories, we consider, as a case study, a truncated compact $U(1)$ gauge theory. Although simple,  such states encode all the crucial ingredients for our demonstration: fermionic matter with bosonic gauge fields; nontrivial manifestation of the spatial symmetries, due to the fermions; and a rich, interesting phase diagram.

 We shall hereby show that using PEPS, one may capture the symmetry properties of a gauge theory, which are essentially the ones which define it, as can be deduced from its name; that both global and local symmetries may be manifested by PEPS - i.e., that one may use this method to treat both matter and gauge fields; and that PEPS allow to study the phase diagram of a gauge theory, and in some cases, using standard techniques, also to derive parent Hamiltonians. We shall emphasize, on the other hand, that the goal is not to study a compact $U(1)$ lattice gauge theory in $2+1$ dimensions, but rather to show that PEPS may be used for the study of gauge theories, once the formalism we introduce is combined with efficient numerical methods. And, eventually, since PEPS allow us to find local parent Hamiltonians, most likely within the universality class of the model in question \cite{Schuch2010}, one may deduce that even if the parent Hamiltonian of a state in question, using the methods presented below, is not the one of the desired lattice gauge theory, it is highly probable that this parent Hamiltonian will be in the same universality class, which means, that the two Hamiltonians would share many features. As the states in study are exact, i.e. exact ground statee of parent Hamiltonians, and one can perfoem numerical calculations, this could be used as a "lab" for other theoretical methods used in high energy physics.

Another possible avenue of exploring lattice gauge theories, which suggests a way of overcoming these difficulties, is quantum simulation. In recent years, many proposals have been made, for the mapping of lattice gauge theories, Abelian and non-Abelian, to atomic and optical systems (such as ultracold atoms in optical lattices, for example) \cite{Wiese2014,Zohar2015a}.
  These systems, called \emph{quantum simulators}, may be built in the laboratory and serve as quantum computers especially tailored for the purpose of lattice gauge theory calculations. Due to experimental requirements, one mostly has to approximate the simulated model by another one, with truncated local Hilbert spaces for the gauge degrees of freedom \cite{Horn1981,Orland1990,Brower1999,Wiese2014,Zohar2015a,Tagliacozzo2014,Zohar2015}. A very important issue is the evaluation of the truncated approximation, which may also be done by the use of tensor network states \cite{Kuhn2014,Kuhn2015}.

\section{Outline}
The work presented in this paper is organized as follows. First, in section \ref{sec:global}, we introduce a class of states for staggered fermions \cite{Susskind1977} on a two dimensional spatial lattice, constructed as fermionic Gaussian PEPS \cite{Kraus2010}. These states will depend on a set of three parameters we shall introduce, in a way that guarantees both the spatial symmetries of translation and rotation invariance, and a global $U(1)$ symmetry. As PEPS, these states are the ground states of local \emph{parent Hamiltonians}. By constructing them as Gaussian PEPS, we will be able to easily derive these quadratic Hamiltonians, which will be, in our case, BdG Hamiltonians of p-wave superconductors. We will study the phase diagram of these states as a function of their parameters, and see that they exhibit gapped phases with a strong p-wave pairing, separated by gapless lines with either strong or weak pairing.

Then, in section \ref{sec:local} we will introduce new degrees of freedom, corresponding to a lattice Abelian gauge field, as in compact QED \cite{KogutSusskind,KogutLattice}, but with finite (truncated) local Hilbert spaces \cite{Zohar2013,Rishon2012,AngMom}. This will allow us to \emph{gauge} the $U(1)$ symmetry of the states and make it local. Although these states, as these of an interacting theory, will no longer be Gaussian, they can still be parameterized exactly as the fermionic states with the global symmetry, manifesting the rotational invariance again and extending the translation invariance to a charge conjugation symmetry. From the general theory of PEPS \cite{Schuch2010} we know that, also in this case, the state $\left|\psi_b\right\rangle$ can be described as the ground state of a local parent Hamiltonian. There is a standard method for constructing the parent Hamiltonian, but deriving its explicit form is a lengthy procedure, out of the scope of our work which aims at describing the states.

  After the construction of the set of locally gauge-invariant states, we will be able to determine a phase diagram, as a function of the parameters of the states, by monitoring the transfer matrix of the PEPS as shall be explained. Furthermore, the tensor network approach allows us to calculate the expectation value of several important observables for these states, as, for example, the Wilson loops. We will find that the obtained states exhibit a rich variety of different behaviors, consistently with the expectations for the thermodynamical phases of a gauge theory - such as phases which confine and do not confine static charges in the case of pure-gauge states, and phases with different screening properties for states which involve, besides the gauge field, dynamical fermions.

We do not assume familiarity of the reader with PEPS or tensor network states, therefore the paper is written and structured such that the results and their significance are clear even for people who are not experts in the field. The appendices include detailed derivations and proofs in PEPS language, supporting the main paper, and are written in a self-contained way, aimed both at the expert and the non-expert readers. Throughout the paper, the Einstein summation convention (summation on double indices) is assumed.

\section{Gaussian fPEPS: Global Symmetry} \label{sec:global}

\subsection{PEPS construction of globally invariant Gaussian states}
Hereby we shall describe the construction of globally invariant Gaussian states which shall fulfill several symmetries, yet to be classified. We assume no acquaintance of the reader with PEPS, and will thus describe in detail the process of constructing the state.

\subsubsection{The PEPS construction}
Consider a two dimensional lattice (corresponding to a $2+1$ dimensional
spacetime), whose vertices (sites) are denoted by $\mathbf{x} \in \mathbb{Z}^2$, with unit vectors $\mathbf{\hat e}_{1,2}$. On each vertex $\mathbf{x}$ one defines a fermionic Fock space
$\mathcal{H}_{\mathbf{x}}$ \footnote{ Note that fermionic Hilbert spaces should be defined, rigorously, in a way which considers the fermionic statistics, and thus, for example, does not allow for a tensor product structure. We hereby use simpler notations, but one should be aware of this formality.}, with a single \emph{physical} mode annihilated by $\psi_{\mathbf{x}}$, such that its vacuum state $\left|\Omega_p \left(\mathbf{x}\right)\right\rangle$ satisfies
\begin{equation}
\psi_{\mathbf{x}} \left|\Omega_p\left(\mathbf{x}\right)\right\rangle = 0.
\end{equation}
We decompose this lattice into two sublattices. The even one (were  the indices $x_1$ and $x_2$ have the same parity) may be occupied by \emph{particles}, and the odd by \emph{antiparticles}. We define their charges accordingly,
\begin{equation}
Q_{\mathbf{x}} = s_{\mathbf{x}} \psi^{\dagger}_{\mathbf{x}}\psi_{\mathbf{x}}\,,
\end{equation}
where $s_{\mathbf{x}} \equiv \left(-1\right)^{x_1+x_2}$.
This is a definition of a staggered charge for \emph{staggered fermions}. A particle-hole transformation, which we shall not carry out, maps the model into the Fock space of the Kogut-Susskind staggered fermions \cite{Susskind1977}. The staggering procedure allows us to analyze the system as a discretization of a theory with two-component spinors, staggered between the two sublattices. Such theory can be considered as embedded in the $3+1$ dimensional theory by Susskind, for example, by taking the restriction $x_3=0$. In such a restriction only two spinorial components are required, which are, in the convention we use, the first and the fourth in the $3+1$ dimensional construction of \cite{Susskind1977} (see also \cite{Burden1987,Cox2000} for the Hamiltonian formulation in $2+1$ dimensions).

Before considering the state of the entire lattice, we shall focus on the state of a single vertex $\mathbf{x}$, called the \emph{fiducial state} $\left|F\left(\mathbf{x}\right)\right\rangle$,  involving the already introduced physical mode $\psi^{\dagger}_{\mathbf{x}}$, and other modes, called the
\emph{virtual} ones, which shall now be introduced and later will be traced out, while tailoring the vertices to each other, constructing the desired state $\left|\psi\right\rangle$.

\begin{figure}
  \centering
  \includegraphics[scale=0.3]{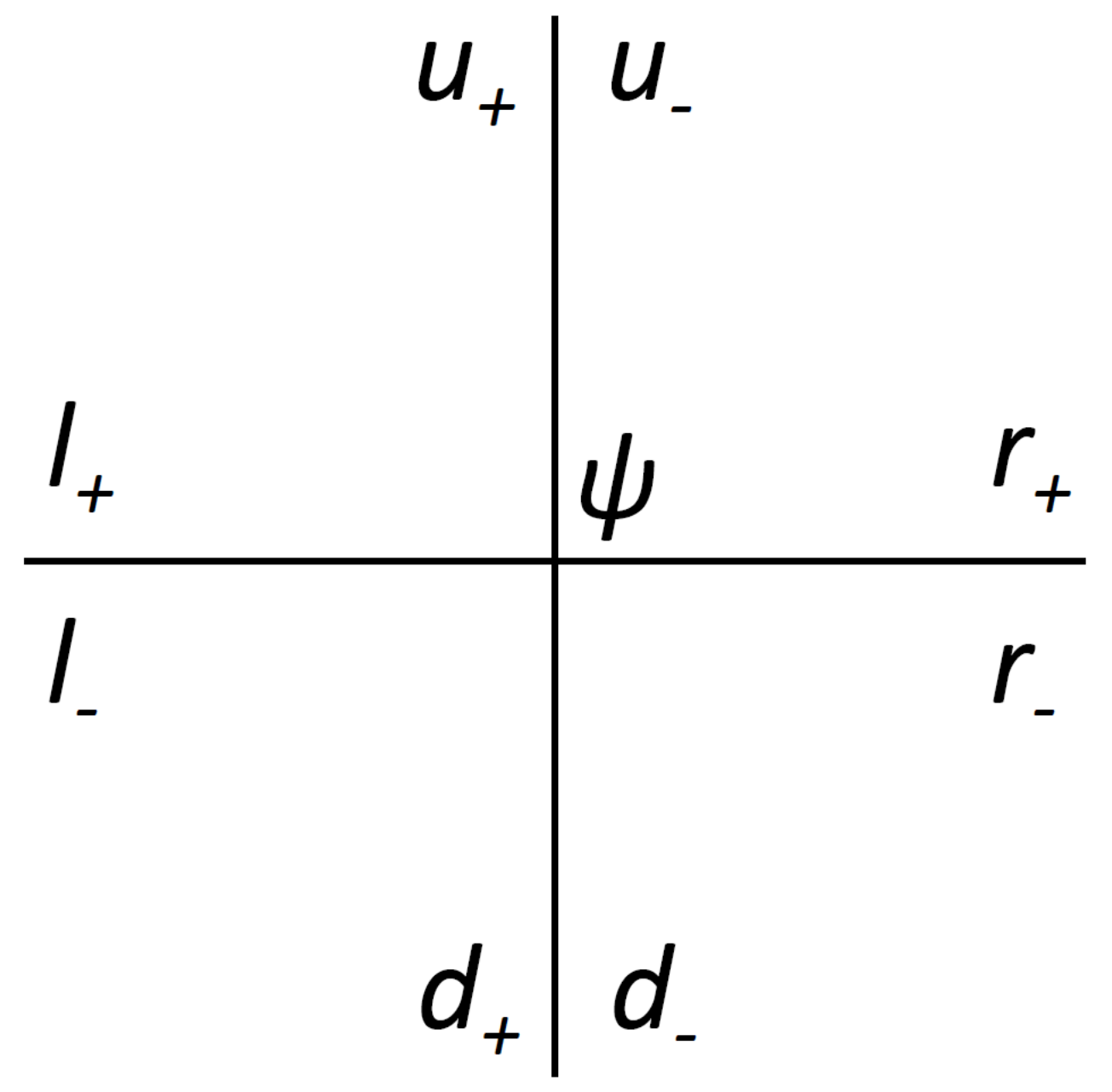}
	\caption{The Hilbert space on a vertex (composing the fiducial state): a single physical fermion $\psi$, with eight virtual fermions surrounding it, two on each of the edges intersecting at the vertex.}
  \label{fig1}
\end{figure}

On each vertex we define eight  virtual modes, located on the edges of the links intersecting at the vertex. They are all fermionic, denoted by the annihilation operators
$l_{\pm},r_{\pm},u_{\pm},d_{\pm}$, which stand for ``left'', ``right'', ``up'' and ``down'' (see figure \ref{fig1}). Their vacuum state is denoted by $\left|\Omega_v\right\rangle$,
and the vacuum state of all the local fermions (both physical and virtual) by $\left|\Omega\right\rangle$.

If one concatenates all the physical and virtual creation operators on a given vertex to a vector of operators $\alpha_i^\dagger$, the most general fiducial state, which is Gaussian, takes the form \cite{Kraus2010}
\begin{equation}
\left|F\left(\mathbf{x}\right)\right\rangle = A\left(\mathbf{x}\right)\left|\Omega\left(\mathbf{x}\right)\right\rangle
\label{fst}
\end{equation}
with
\begin{equation}
A = \exp\left(\underset{ij}{\sum} \hat T_{ij}\alpha_{i}^{\dagger}\alpha_{j}^{\dagger}\right).
\end{equation}
Here and in the following, while working at a given vertex, we will neglect the position index $\mathbf{x}$.

\begin{figure}[t]
  \centering
  \includegraphics[width=0.44\textwidth]{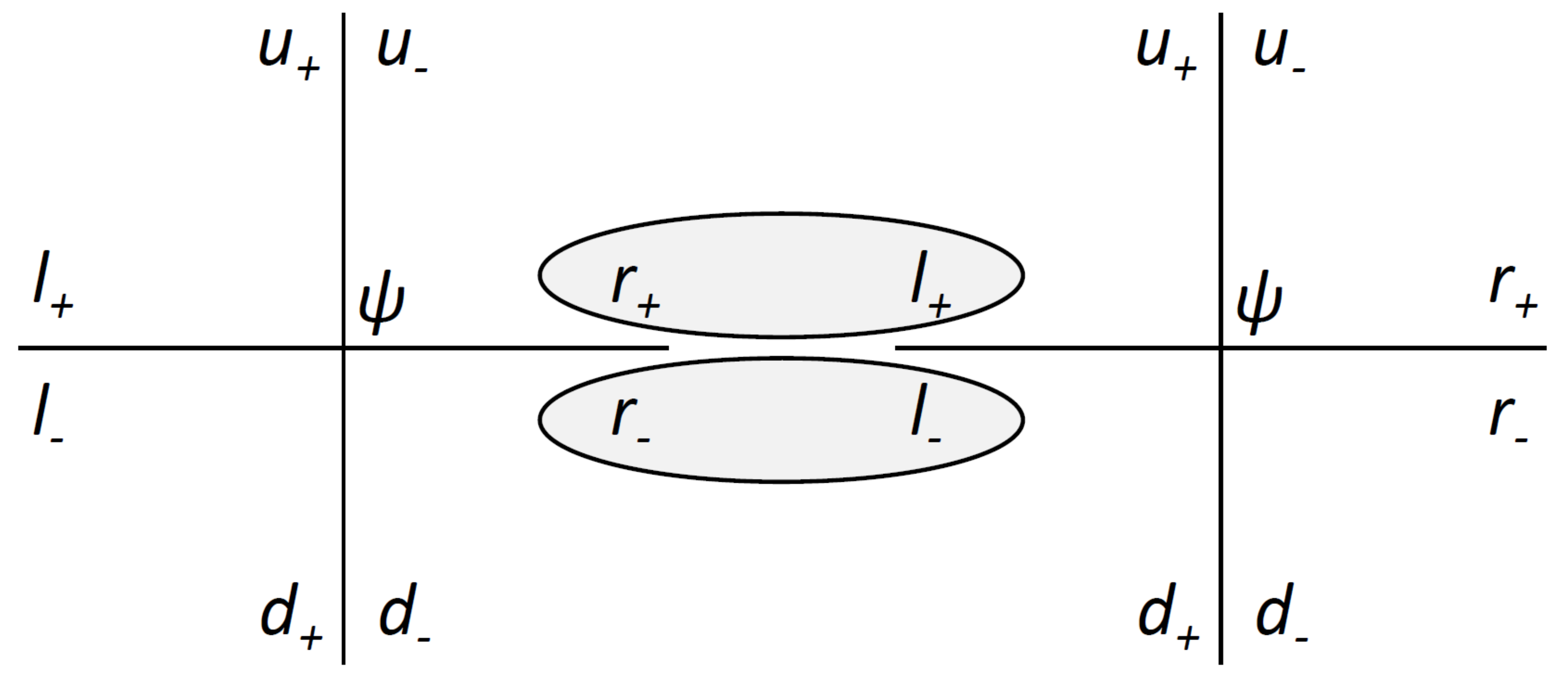}
	\caption{The ellipses denote the entangled state $\left|H\right\rangle$, onto which the virtual fermions are projected, for the contraction of the fiducial states into the PEPS.}
  \label{fig1a}
\end{figure}

The next step is to connect the fiducial states lying on the lattice vertices among themselves, and project out the virtual degrees of freedom, for the creation of a physical state
$\left|\psi\right\rangle$ for the entire lattice. To this purpose, we project the virtual states on both sides of a bond into joint entangled states (see figure \ref{fig1a}), which allow to connect the whole network of fiducial states, hence the name \emph{Projected Entangled Pair States} (PEPS). In particular, for each horizontal and vertical bond, we project both the positive and negative virtual modes on the following entangled bond states:
\begin{equation}
\begin{aligned}
\left|H_{\mathbf{x}}\right\rangle
=\frac{1}{2}
\exp
\left(l_{+,\mathbf{x+\hat e}_1}^{\dagger}r_{+,\mathbf{x}}^{\dagger}\right)
\exp\left(l_{-,\mathbf{x+\hat e}_1}^{\dagger}r_{-,\mathbf{x}}^{\dagger}\right)
\left|\Omega_{H,\mathbf{x}}\right\rangle \\
\left|V_{\mathbf{x}}\right\rangle =\frac{1}{2}\exp\left(u_{+,\mathbf{x}}^{\dagger}d_{+,\mathbf{x+\hat e}_2}^{\dagger}\right)\exp\left(u_{-,\mathbf{x}}^{\dagger}d_{-,\mathbf{x+\hat e}_2}^{\dagger}\right)\left|\Omega_{V,\mathbf{x}}\right\rangle
\end{aligned}
\end{equation}
where $\left|\Omega_{H,V}\right\rangle$ are the respective vacua on the links. From these states, one constructs the projection operators
\begin{equation}
\omega \left(\mathbf{x}\right) = \left|H_{\mathbf{x}}\right\rangle \left\langle H_{\mathbf{x}}\right| \,,\quad
\eta \left(\mathbf{x}\right) = \left|V_{\mathbf{x}}\right\rangle \left\langle V_{\mathbf{x}}\right|\,.
\label{projdef}
\end{equation}
Finally, the physical PEPS is given by:
\begin{equation}
\left|\psi\left(T\right)\right\rangle =
\left\langle \Omega_v \right| \underset{\mathbf{x}}{\prod}\omega\left(\mathbf{x}\right)\eta\left(\mathbf{x}\right) A\left(\mathbf{x}\right)\left|\Omega\right\rangle.
\label{globPEPS}
\end{equation}
where $\left|\Omega\right\rangle$ is the global vacuum, while $\left|\Omega_v\right\rangle$ is the one of the virtual fermions only.

\subsubsection{The required symmetries}

Our goal is to construct a physical state $\left|\psi \left(\left\{t_i\right\}\right)\right\rangle$ of the physical modes,
parameterized by a set of variational parameters $\left\{t_i\right\}$, which satisfies the following symmetries:
\begin{enumerate}
  \item \emph{Translational invariance}. Define the unitary operators $U_T$, such that
  \begin{equation}
  \begin{aligned}
    U_T\left(\mathbf{\hat e}_1\right) \psi_{\mathbf{x}}^{\dagger} U^{\dagger}_T\left(\mathbf{\hat e}_1\right) & = \psi^{\dagger}_{\mathbf{x+\hat e}_1} \\
    U_T\left(\mathbf{\hat e}_2\right) \psi_{\mathbf{x}}^{\dagger} U^{\dagger}_T\left(\mathbf{\hat e}_2\right) & = \psi^{\dagger}_{\mathbf{x+\hat e}_2}
    \end{aligned}
    \label{Transop}
    \end{equation}
    The state $\ket{\psi \left(\left\{t_i\right\}\right)}$ is translationally invariant if and only if
    \begin{equation}
    U_T\left(\mathbf{\hat e}_n\right)\ket{\psi \left(\left\{t_i\right\}\right)}  = \left|\psi \left(\left\{t_i\right\}\right)\right\rangle\,, \quad n = 1,2\\
    \end{equation}
    (in general, invariance might also involve a change of the state by a global phase, but we shall neglect this option here and in the following symmetries; this makes sense in this case, since one naturally uses PEPS for zero momentum states; otherwise, different tensors are required, e.g. \cite{Vanderstraeten2015}).

    \item \emph{Rotational invariance}. As we are dealing with a model defined on a square lattice, only $C_4$ rotations are relevant. We denote a counter-clockwise $\pi/2$ rotation by $\varLambda$, i.e.
    \begin{equation}
\mathbf{x}=\left(x_1,x_2\right)\longrightarrow\varLambda\mathbf{x}=\left(-x_2,x_1\right).
\label{latrot}
\end{equation}

Denote by $U_p$ the quantum unitary operator which rotates the physical modes:
\begin{equation}
U_p \psi_{\mathbf{x}} U^{\dagger}_p = \eta_p \psi_{\varLambda\mathbf{x}}
\label{physrot}
\end{equation}
where, in general, $\left|\eta_p\right|=1$. We will make, however, the choice of
\begin{equation}
\eta_p = e^{i \pi /4}
\label{phasedef}
\end{equation}
which satisfies the set of physical requirements imposed by the staggered fermion discretization \cite{Susskind1977} of the Dirac theory
\footnote{The Dirac theory, for a continuous Dirac field $\Psi$, is described by the Dirac equation,
$\left(i \gamma^{\mu}\partial_{\mu} - m\right)\Psi = 0$,
where $\mu$ are Lorentz indices, $\Psi$ is a Dirac spinor with an even number of components,
and $\gamma^{\mu}$ is the vector of Dirac matrices, satisfying the Clifford algebra
$\left\{\gamma^{\mu},\gamma^{\nu}\right\} = 2g^{\mu \nu}$
($g^{\mu \nu}$ is the spacetime metric, e.g. the Minkowski metric in the conventional formulation).
The Dirac spinor $\Psi$ may undergo Lorentz tranformations, including rotations, generated by
$\frac{i}{4}\left[\gamma^{\mu},\gamma^{\nu}\right]$.
We work on the lattice, and in particular staggered fermions, and thus we follow the conventions set in the original
formulation of staggered fermions given in \cite{Susskind1977}, for $3+1$ dimensions, with 4 component spinors.
In the process of discretization, the four components of each spinor are distributed along four lattice sites.
In our case, we are only interested in $2+1$ dimensions, and thus may with two component spinors, spread along two lattice sites.
We choose to do so by restricting the convention of \cite{KogutSusskind} to the plane $x_3 = 0$, and thus our 2 component spinors are composed of the first and fourth
components of the original ones, corresponding to a particle and an anti particle. Following the conventions given there,
\begin{equation}
\gamma_0 = \sigma_z ;\; \gamma_1 = i\sigma_y ;\; \gamma_2 = i\sigma_x.
\end{equation}
and thus we  get that our rotation generator (in the $1-2$ plane) is given by
\begin{equation}
\frac{i}{4}\left[\gamma_1,\gamma_2\right] = -\frac{1}{2}\sigma_z
\end{equation}
Therefore a particle annihilation operator gets a phase of $e^{i \pi /4}$ when rotated by $\frac{\pi}{2}$,
and an anti-particle operator gets the conjugate phase; hence, the ``even'' annihilation operators in our theory will be rotated with $e^{i \pi /4}$,
as they correspond to particles. The antiparticles of our theory differ from those of \cite{Susskind1977} by a particle-hole transformation; thus, in our case they should be rotated with the same phase, $e^{i \pi /4}$.
}.
 A rotationally-invariant state has to satisfy
\begin{equation}
U_p\left(\varLambda\right) \left|\psi \left(\left\{t_i\right\}\right)\right\rangle = \left|\psi \left(\left\{t_i\right\}\right)\right\rangle
\label{transrot}
\end{equation}
(again, we are not allowing for a global phase).

\item \emph{Global $U(1)$ Invariance}. We wish our state to be invariant under a \emph{global} $U(1)$ gauge transformation. Due to the staggering, this is the transformation generated by
\begin{equation}
\mathcal{G}_0 = \underset{\mathbf{x}}{\sum}{Q_{\mathbf{x}}} = \underset{\mathbf{x}}{\sum} s_{\mathbf{x}} \psi^{\dagger}_{\mathbf{x}}\psi_{\mathbf{x}},
\label{Gglobgen}
\end{equation}
i.e., under a gauge transformation with the parameter $\phi$,
\begin{equation}
\psi^{\dagger}_{\mathbf{x}} \rightarrow e^{i s_{\mathbf{x}} \phi}\psi^{\dagger}_{\mathbf{x}}.
\label{globtrans}
\end{equation}

A \emph{globally} gauge invariant state satisfies:
\begin{equation}
e^{i \phi \mathcal{G}_0} \left|\psi \left(\left\{t_i\right\}\right)\right\rangle = \left|\psi \left(\left\{t_i\right\}\right)\right\rangle.
\end{equation}
\end{enumerate}

While these symmetries can be met by states which are either Gaussian or not, we shall use Gaussian PEPS as described above which shall satisfy these symmetries by construction. We will see how to obtain these symmetries first on the virtual level of the fiducial states, and then, using the PEPS projectors, to make them a real physical symmetry.

\subsubsection{Imposing global invariance}
On each of the tensor network links we define a \emph{virtual electric field},
\begin{equation}
E_{\alpha} = \alpha^{\dagger}_+\alpha_+ - \alpha^{\dagger}_-\alpha_-
\end{equation}
(where $\alpha = l,r,u,d$), with eigenvalues $-1,0,1$. We further define a \emph{local virtual Gauss operator},
\begin{equation}
\begin{aligned}
&G_0 = \text{div} E - Q  = E_r + E_u - E_l - E_d - Q = \\
&r^{\dagger}_+ r_+ - r^{\dagger}_- r_- +
u^{\dagger}_+ u_+ - u^{\dagger}_- u_- -
l^{\dagger}_+ l_+ + l^{\dagger}_- l_- -
d^{\dagger}_+ d_+ + d^{\dagger}_- d_- - s_{\mathbf{x}}\psi^{\dagger}\psi
\label{fiducialG}
\end{aligned}
\end{equation}

We wish to define a fiducial state of the physical and virtual fermions, $\left|F\right\rangle$. This will be an eigenstate of the Gauss operator,
\begin{equation} \label{globalvirtualg}
G_0\left|F\right\rangle = q\left|F\right\rangle
\end{equation}
- a virtual Gauss law, with a static charge $q$. As we shall see, this will ensure, once we propely construct the PEPS, that we will eventually have a globally invariant state.

We can decompose the physical and virtual modes into two sets with respect to this Gauss law: modes whose number operators appear in \eqref{fiducialG} with a negative sign will be called \emph{negative modes}, and
modes with a positive sign - \emph{positive modes}. We shall denote the virtual negative modes by $\left\{a_i\right\}_{i=1}^4 = \left\{l_+,r_-,u_-,d_+\right\}$,
and the virtual positive modes by $\left\{b_i\right\}_{i=1}^4 = \left\{l_-,r_+,u_+,d_-\right\}$. The physical mode is negative on the even sublattice and positive on the odd one,
and shall be denoted, in this notation, by either $a_0$ or $b_0$ for even or odd vertices, respectively.

Note that, in general, one could introduce further virtual fermions, and update the Gauss law accordingly. In the general case, one may consider $N_n$ negative modes and $N_p$ positive ones; we will focus on the case $\left|N_n-N_p\right|=1$, with two fermions per bond.

We wish to find the most general Gaussian state $\left|F\right\rangle$ which is an eigenstate of $G_0$, and of the form \eqref{fst}, with two fermions per bond. Gaussian states are fully characterized by their covariance matrix $\Gamma$, and thus one may exploit the covariance matrix for their parametrization, as done in \ref{app:covariance}. Below we shall describe an  equivalent way for their parametrization, which we shall utilize next.

\begin{sttmnt}\label{th:gauss}:
The most general gauge invariant Gaussian fiducial state $\left|F\right\rangle$ with no static charges takes the form
\begin{equation} \label{fiducial1}
\left|F\left(\mathbf{x}\right)\right\rangle = A\left(\mathbf{x}\right)\left|\Omega_p \left(\mathbf{x}\right)\right\rangle \left|\Omega_v \left(\mathbf{x}\right)\right\rangle
\equiv A\left(\mathbf{x}\right)\left|\Omega \left(\mathbf{x}\right)\right\rangle
\end{equation}
using the Gaussian operator
\begin{equation} \label{eqA}
A\left(\mathbf{x}\right) =
\left\{
  \begin{array}{ll}
    \exp\left(\underset{ij}{\sum}T_{ij}a_{i}^{\dagger}b_{j}^{\dagger}\right), & \hbox{$\mathbf{x}$ even;} \\
    \exp\left(\underset{ij}{\sum}T_{ij}b_{i}^{\dagger}a_{j}^{\dagger}\right), & \hbox{$\mathbf{x}$ odd.}
  \end{array}
\right.
\end{equation}
\end{sttmnt}
$i=0,...,4,j=1,...,4$. For even vertices, $N_n=5,N_p=4$, and the opposite for odd vertices. Thus, in general, before demanding any other symmetries, we obtain that the state depends on $20$ complex parameters - the elements of the $T$ matrix.

\begin{proof} The most general Gaussian fiducial state (even or odd) is constructed out of the Gaussian operator \cite{Bravyi05,Kraus2010}
\begin{equation}
A = \exp\left(\underset{ij}{\sum} \hat T_{ij}\alpha_{i}^{\dagger}\alpha_{j}^{\dagger}\right).
\end{equation}
with $i,j=1,...,9$.
The fiducial state is an eigenstate of $G_0$, if and only if
\begin{equation}
e^{i \phi G_0} A e^{-i \phi G_0} = e^{i \phi q} A\,.
\end{equation}

Let us consider an infinitesimal transformation, with $\phi = \epsilon \ll 1$. The transformation reads
\begin{equation}
A \rightarrow A' \approx \exp\left(\underset{ij}{\sum} \hat T_{ij}
\left(\alpha_{i}^{\dagger}\alpha_{j}^{\dagger}+i\epsilon\left[G_0,\alpha_{i}^{\dagger}\alpha_{j}^{\dagger}\right]\right)\right)
\end{equation}
The desired result is obtained if  the commutators $\left[G_0,\alpha_{i}^{\dagger}\alpha_{j}^{\dagger}\right]$ are c-numbers for
any $i,j$. $G_0$ consists of number operators with different sign. If $\alpha_i^{\dagger}=a_i^{\dagger}$ and $\alpha_j^{\dagger}=b_j^{\dagger}$, the commutator vanishes. If, instead, both the creation operators are either positive or negative, the commutator results in a term which is quadratic in the creation operators, and these terms cannot cancel each other. Thus we conclude that $\hat T_{ij}$ is nonzero only if $\alpha_i^{\dagger}$ and $\alpha_j^{\dagger}$ create opposite charges, therefore $\alpha_i^{\dagger}=a_i^{\dagger},\,\alpha_j^{\dagger}=b_j^{\dagger}$ (or the other way around). We also obtain that $q=0$.
\end{proof}

One can easily verify that the state \eqref{globPEPS} is gauge-invariant: the global transformation we wish to apply on the state is
\begin{equation}
\mathcal{U} = \text{exp}\left(i\phi\underset{\mathbf{x}}{\sum}s_{\mathbf{x}}\psi^{\dagger}_{\mathbf{x}}\psi_{\mathbf{x}}\right);
\end{equation}
and thanks to statement \ref{th:gauss}, we know that
\begin{equation}
 U_{\psi}\left|F\left(\mathbf{x}\right)\right\rangle \equiv e^{i \phi s_{\mathbf{x}}\psi^{\dagger}_{\mathbf{x}}\psi_{\mathbf{x}}}\left|F\left(\mathbf{x}\right)\right\rangle =
e^{i \phi \text{div}E_{\mathbf{x}}}\left|F\left(\mathbf{x}\right)\right\rangle \equiv
 U_V^{r}\left(\mathbf{x}\right)U_V^{u}\left(\mathbf{x}\right)U_V^{l\dagger}\left(\mathbf{x}\right) U_V^{d\dagger}\left(\mathbf{x}\right) \left|F\left(\mathbf{x}\right)\right\rangle
\end{equation}
(where $U_V^i = e^{i \phi E_i}$), and
\begin{equation}
e^{i \phi s_{\mathbf{x}}\psi^{\dagger}_{\mathbf{x}}\psi_{\mathbf{x}}} A\left(\mathbf{x}\right) e^{-i \phi s_{\mathbf{x}}\psi^{\dagger}_{\mathbf{x}}\psi_{\mathbf{x}}}=
U_V^{r}\left(\mathbf{x}\right)U_V^{u}\left(\mathbf{x}\right)U_V^{l\dagger}\left(\mathbf{x}\right) U_V^{d\dagger}\left(\mathbf{x}\right)
A\left(\mathbf{x}\right)
U_V^{r\dagger}\left(\mathbf{x}\right)U_V^{u\dagger}\left(\mathbf{x}\right)U_V^{l}\left(\mathbf{x}\right) U_V^{d}\left(\mathbf{x}\right).
\end{equation}
On the other hand, the projectors are invariant under these virtual operations, since
\begin{equation}
U_{V,\mathbf{x}}^{r}U_{V,\mathbf{x + \hat e}_1}^{l \dagger}\left|H_{\mathbf{x}}\right\rangle = \left|H_{\mathbf{x}}\right\rangle \,,\quad
U_{V,\mathbf{x}}^{u}U_{V,\mathbf{x + \hat e}_2}^{d \dagger}\left|V_{\mathbf{x}}\right\rangle = \left|V_{\mathbf{x}}\right\rangle
\label{prjinv}
\end{equation}
 and thus one obtains that the state is, indeed, globally invariant:
\begin{equation}
\mathcal{U}\left|\psi\left(T\right)\right\rangle = \left|\psi\left(T\right)\right\rangle.
\end{equation}

\subsubsection{Virtual phase symmetries}
The construction of the PEPS is not unique, and includes a virtual symmetry, which is due to a redundancy in the definition of the virtual modes of such tensor networks
\cite{Verstraete2008}.
In our construction, we have virtual symmetries which are defined by the following phase transformations on the virtual modes:
\begin{equation}
\begin{aligned}
U_S\left(\alpha,\beta,\gamma,\delta\right) a^{\dagger}_i U^{\dagger}_S\left(\alpha,\beta,\gamma,\delta\right)=S_{Aij}\left(\alpha,\beta,\gamma,\delta\right) a^{\dagger}_j \\
U_S\left(\alpha,\beta,\gamma,\delta\right) b^{\dagger}_i U^{\dagger}_S\left(\alpha,\beta,\gamma,\delta\right)=S_{Bij}\left(\alpha,\beta,\gamma,\delta\right) b^{\dagger}_j
\label{eqUS}
\end{aligned}
\end{equation}
where
\begin{equation}
S^e_A  = \left(
        \begin{array}{ccccc}
          1 & 0 & 0 & 0 & 0 \\
          0 & e^{i \alpha} & 0 & 0 & 0 \\
          0 & 0 & e^{i \beta} & 0 & 0 \\
          0 & 0 & 0 & e^{i \gamma} & 0 \\
          0 & 0 & 0 & 0 & e^{i \delta} \\
        \end{array}
      \right)
      \,,\quad
S^e_B  = \left(
        \begin{array}{cccc}
          e^{-i \tilde\beta} & 0 & 0 & 0 \\
          0 & e^{-i \tilde\alpha} & 0 & 0 \\
          0 & 0 & e^{-i \tilde\delta} & 0 \\
          0 & 0 & 0 & e^{-i \tilde\gamma} \\
        \end{array}
      \right)
\end{equation}
for an even vertex, and
\begin{equation}
S^o_A  = \left(
        \begin{array}{cccc}
          e^{i \tilde\alpha} & 0 & 0 & 0 \\
          0 & e^{i \tilde\beta} & 0 & 0 \\
          0 & 0 & e^{i \tilde\gamma} & 0 \\
          0 & 0 & 0 & e^{i \tilde\delta} \\
        \end{array}
      \right)
      \,,\quad
S^o_B  = \left(
        \begin{array}{ccccc}
          1 & 0 & 0 & 0 & 0 \\
          0 & e^{-i \beta} & 0 & 0 & 0 \\
          0 & 0 & e^{-i \alpha} & 0 & 0 \\
          0 & 0 & 0 & e^{-i \delta} & 0 \\
          0 & 0 & 0 & 0 & e^{-i \gamma} \\
        \end{array}
      \right)
\end{equation}
for an odd one.
Note that the projectors are invariant under this transformation:
\begin{equation}
\begin{aligned}
U_S\left(\alpha,\beta,\gamma,\delta\right) \omega U^{\dagger}_S\left(\alpha,\beta,\gamma,\delta\right) = \omega \\
U_S\left(\alpha,\beta,\gamma,\delta\right) \eta U^{\dagger}_S\left(\alpha,\beta,\gamma,\delta\right) = \eta
\end{aligned}
\end{equation}
The application of this transformation on the virtual modes does not affect the physical state.
Let us define $|\psi(\widetilde T)\rangle$ as the state created using the operators $\widetilde A = \exp\left( i \tilde T^{\left(e\right)}_{ij} a^{\dagger}_i b^{\dagger}_j\right) $ (and similarly for odd vertices) where
\begin{equation}
\begin{aligned}
 \widetilde T^{e} &= S^{eT}_A\left(\alpha,\beta,\gamma,\delta\right)TS^{e}_B\left(\alpha,\beta,\gamma,\delta\right) \,,\\
 \widetilde T^{o} &= S^{oT}_B\left(\alpha,\beta,\gamma,\delta\right)TS^{o}_A\left(\alpha,\beta,\gamma,\delta\right)
\end{aligned}
\end{equation}
for an arbitrary choice of the phases $\alpha,\beta,\gamma,\delta$.
Since $\widetilde T^{e,o}$ result from $\tilde A\left(\mathbf{x}\right) = U_S\left(\mathbf{x}\right)
A\left(\mathbf{x}\right) U^{\dagger}_S\left(\mathbf{x}\right)$, we obtain
\begin{equation}
\begin{aligned}
|\psi(\widetilde T)\rangle =
& \left\langle \Omega_v \right|\underset{\mathbf{x}}{\prod}\omega\left(\mathbf{x}\right)\eta\left(\mathbf{x}\right)\tilde A\left(\mathbf{x}\right)\left|\Omega_v\right\rangle
\left|\Omega_p\right\rangle =
\left\langle \Omega_v \right|\underset{\mathbf{x}}{\prod}\omega\left(\mathbf{x}\right)\eta\left(\mathbf{x}\right) U_S\left(\mathbf{x}\right)
A\left(\mathbf{x}\right) U^{\dagger}_S \left(\mathbf{x}\right)  \left|\Omega_v\right\rangle
\left|\Omega_p\right\rangle = \\
& \left\langle \Omega_v \right| \underset{\mathbf{x}}{\prod}U^{\dagger}_S\left(\mathbf{x}\right)\omega\left(\mathbf{x}\right)
\eta\left(\mathbf{x}\right) U_S\left(\mathbf{x}\right) A\left(\mathbf{x}\right)  \left|\Omega_v\right\rangle
\left|\Omega_p\right\rangle
= \left| \psi\left( T\right)\right\rangle
\end{aligned}
\end{equation}
where we used the invariance of the virtual vacuum under the transformation.
Therefore, we conclude that the transformations $S$ are just virtual symmetries of the state $\ket{\psi(T)}$, or consequences of a redundant description, which will allow us to reduce, eventually, the number of relevant parameters (especially the phases) of the parametrization matrix $T$.

\subsubsection{Imposing the Physical Symmetries}
So far, $\left|\psi\left(T\right)\right\rangle$ fulfills the global symmetry requirement, and depends on $20$ complex parameters - the matrix elements of $T$. Demanding the other two required physical symmetries discussed in the beginning of this section, and using the virtual symmetry presented above, we shall now reduce the number of degrees of freedom further.

We begin with rotational invariance. It follows from the definition of $A\left(\mathbf{x}\right)$ in the form of a fermionic Gaussian operator, that such invariance cannot induce a global phase, consistently with \eqref{transrot}. We now proceed to the construction of the rotation transformation presented in (\ref{latrot}),(\ref{physrot}),(\ref{phasedef}).

\begin{figure}
  \centering
  \includegraphics[width=0.44\textwidth]{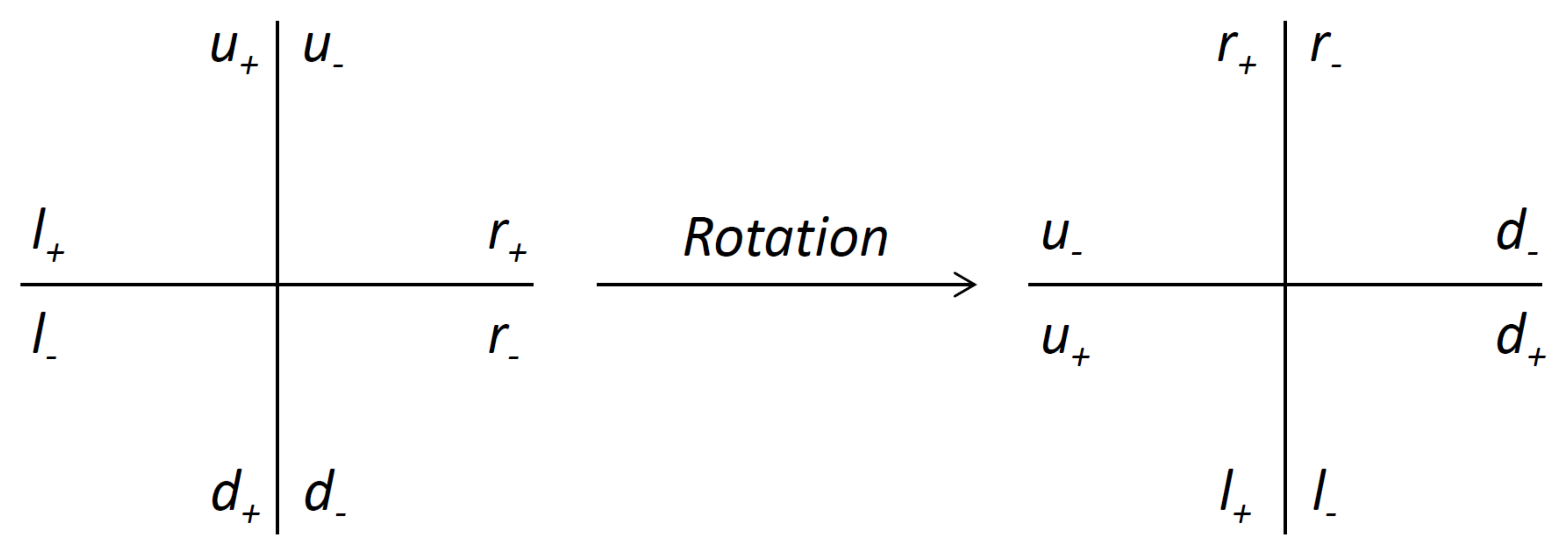}
	\caption{Rotation of a single fiducial state.}
  \label{fig2}
\end{figure}

In order to parameterize $T$ such that rotational invariance is fulfilled, we have to define a rotation transformation for the virtual modes,
in terms of a quantum unitary operator $U_{R}\left(\varLambda\right)$, as follows:
\begin{equation}
\begin{array}{c}
l_{\pm}^{\dagger}\left(\mathbf{x}\right)\longrightarrow
U_{R}^{\dagger}\left(\varLambda\right)l_{\pm}^{\dagger}\left(\mathbf{x}\right)U_{R}\left(\varLambda\right)=\eta_{u}u_{\mp}^{\dagger}\left(\varLambda^{-1}\mathbf{x}\right)\\
r_{\mp}^{\dagger}\left(\mathbf{x}\right)\longrightarrow
U_{R}^{\dagger}\left(\varLambda\right)r_{\mp}^{\dagger}\left(\mathbf{x}\right)U_{R}\left(\varLambda\right)=\eta_{d}d_{\pm}^{\dagger}\left(\varLambda^{-1}\mathbf{x}\right)\\
u_{\mp}^{\dagger}\left(\mathbf{x}\right)\longrightarrow
U_{R}^{\dagger}\left(\varLambda\right)u_{\mp}^{\dagger}\left(\mathbf{x}\right)U_{R}\left(\varLambda\right)=\eta_{r}r_{\mp}^{\dagger}\left(\varLambda^{-1}\mathbf{x}\right)\\
d_{\pm}^{\dagger}\left(\mathbf{x}\right)\longrightarrow
U_{R}^{\dagger}\left(\varLambda\right)d_{\pm}^{\dagger}\left(\mathbf{x}\right)U_{R}\left(\varLambda\right)=\eta_{l}l_{\pm}^{\dagger}\left(\varLambda^{-1}\mathbf{x}\right)
\end{array}
\label{virtrot}
\end{equation}
where $\eta_{l},\eta_{r},\eta_{u},\eta_{d}$ are phases we will specify in the following, and we described a clockwise rotation for convenience.
This rotation preserves the Gauss law orientation (see Fig. \ref{fig2}). In particular, we impose $\eta_{u}\eta_{d}=1$ and $\eta_{r}\eta_{l}=-1$, in order to obtain the expected rotation transformation for the projectors:
\begin{equation}
\begin{aligned}
& U_{R}\left(\varLambda\right)\omega\left(\mathbf{x}\right)U_{R}^{\dagger}\left(\varLambda\right)=\eta\left(\varLambda\mathbf{x}\right) \\
& U_{R}\left(\varLambda\right)\eta\left(\mathbf{x}\right)U_{R}^{\dagger}\left(\varLambda\right)=\omega\left(\varLambda\mathbf{x}\right)
\end{aligned}
\label{projrot}
\end{equation}

Finally, to fulfill the rotational invariance, we adopt the parametrization of $T$ specified by the following statement:

\begin{sttmnt} \label{lemma:rot} The PEPS $\left|\psi\left(T\right)\right\rangle$ is invariant under the rotation transformation $U_p$, if
\begin{equation}
T  =
\left(\begin{array}{cccc}
t & -\eta_p^{-2} t & - \eta_p^{-3} t & - \eta_p^{-1}  t\\
 x & y &  z &  w\\
-y &  x & - w  &  z\\
- w &  z &   x & y\\
- z & -  w & -y &   x
\end{array}\right)
\end{equation}
\textnormal{- reducing the number of free parameters to five complex ones in total.}
\end{sttmnt}

\emph{Proof:}
If we parametrize $T$ (and therefore $A$) such that the entire fiducial state is rotated as a whole piece, with no change to its internal structure,
\begin{equation}
U_{p}\left(\varLambda\right)U_{R}\left(\varLambda\right) A\left(\mathbf{x}\right) U^{\dagger}_{p}\left(\varLambda\right)U_{R}^{\dagger}\left(\varLambda\right) = A\left(\varLambda\mathbf{x}\right)\,,
\end{equation}
then we get the required symmetry by using Eq. (\ref{projrot}):
\begin{equation}
\begin{aligned}
U_p\left(\varLambda\right) \left|\psi\right\rangle & =
\left\langle \Omega_v \right| \underset{\mathbf{x}}{\prod}\omega\left(\mathbf{x}\right)\eta\left(\mathbf{x}\right)
U_p\left(\varLambda\right) A\left(\mathbf{x}\right) U^{\dagger}_p\left(\varLambda\right)  \left|\Omega_v\right\rangle
 \left|\Omega_p\right\rangle \\
&= \left\langle \Omega_v \right| \underset{\mathbf{x}}{\prod}\omega\left(\mathbf{x}\right)\eta\left(\mathbf{x}\right) U^{\dagger}_R\left(\varLambda\right)  A\left(\varLambda\mathbf{x}\right)   U_R\left(\varLambda\right) \left|\Omega_v\right\rangle
 \left|\Omega_p\right\rangle \\
&= \left\langle \Omega_v \right| \prod_{\mathbf{x}} U_R\left(\varLambda\right) \omega\left(\mathbf{x}\right) \eta\left(\mathbf{x}\right)
U^{\dagger}_R\left(\varLambda\right) A\left(\varLambda\mathbf{x}\right) \left|\Omega_v\right\rangle
 \left|\Omega_p\right\rangle \\
 &=\left\langle \Omega_v \right|\underset{\mathbf{x}}{\prod} \eta\left(\varLambda\mathbf{x}\right)
 \omega\left(\varLambda \mathbf{x}\right)
  A\left(\varLambda\mathbf{x}\right) \left|\Omega_v\right\rangle
 \left|\Omega_p\right\rangle = \left|\psi\right\rangle.
\end{aligned}
\end{equation}
Therefore, the only remaining task is to parameterize $T$ accordingly.
The transformation $U_{p}\left(\varLambda\right)U_{R}\left(\varLambda\right)$ acts on the creation operators as
\begin{equation}
\begin{aligned}
U_{p}^{\dagger}\left(\varLambda\right)U^{\dagger}_{R}\left(\varLambda\right) a^{\dagger}_i U_{R}\left(\varLambda\right)U_{p}\left(\varLambda\right)
=R_{A,ij} a^{\dagger}_j \\
U_{p}^{\dagger}\left(\varLambda\right)U^{\dagger}_{R}\left(\varLambda\right) b^{\dagger}_i U_{R}\left(\varLambda\right)U_{p}\left(\varLambda\right)
=R_{B,ij} b^{\dagger}_j
\end{aligned}
\end{equation}
with the matrices
\begin{equation}
R_{A}=\left(\begin{array}{ccccc}
\eta_{p} & 0 & 0 & 0 & 0\\
0 & 0 & 0 & \eta_{u} & 0\\
0 & 0 & 0 & 0 & \eta_{d}\\
0 & 0 & \eta_{r} & 0 & 0\\
0 & \eta_{l} & 0 & 0 & 0
\end{array}\right)
\end{equation}
(recall that $\eta_d=\overline{\eta}_u$ and $\eta_l=-\overline{\eta}_r$) and
\begin{equation}
R_{B}=\left(\begin{array}{cccc}
0 & 0 & \eta_{u} & 0\\
0 & 0 & 0 & \eta_{d}\\
0 & \eta_{r} & 0 & 0\\
\eta_{l} & 0 & 0 & 0
\end{array}\right)
\end{equation}
for an even vertex; for odd vertices, the two matrices have to be exchanged.
Thus, the rotational symmetry is equivalent to imposing
\begin{equation}
T=R^{\intercal}_ATR_{B}
\label{rotcond}
\end{equation}
(where $\intercal$ labels the transposed matrix) for even vertices, and an analogous condition for odd vertices is obtained upon exchanging $A$ and $B$.
This is satisfied if and only if $\eta_{p}^{4}=-1$, consistently with our physical demand $\eta_p = e^{i \pi /4}$.

The most general parametrization satisfying (\ref{rotcond}) takes the form:
\begin{equation} \label{rotinvt}
T_\eta=\left(\begin{array}{cccc}
t & -\eta_r \eta_u \eta_p^{-2}t & - \eta_u \eta_p^{-3}t & -\eta_r \eta_p^{-1}t\\
x & y & z & w\\
-y & \eta_r^2 \eta_u^2 x & -\eta_u^2 w & \eta_r^2 z\\
-\eta_r^{-1} \eta_u w & \eta_r \eta_u z & \eta_u^2 x & y\\
-\eta_u^{-1} \eta_r z & -\eta_r \eta_u w & -y & \eta_r^2 x
\end{array}\right).
\end{equation}

Some phases of $T_\eta$, though, can be eliminated using the fact that the state is left invariant under the transformation (\ref{eqUS}). By properly choosing the eight phases
$\alpha,\beta,\gamma,\delta,\tilde\alpha,\tilde\beta,\tilde\gamma,\tilde\delta$.
 One may find out that the parametrization (\ref{rotinvt}) is left invariant only if some requirements are fulfilled by the eight phases $\alpha,\beta,\gamma,\delta,\tilde\alpha,\tilde\beta,\tilde\gamma,\tilde\delta$, namely, one is only left with a single phase, out of which all the others may be expressed. Then, using the transformation matrices $S_A^e = S_B^o = 1 \oplus S, S_A^o = S_B^e = S$, with
\begin{equation}
S = \left(
  \begin{array}{cccc}
    \sqrt{\eta_r \eta_u} & 0 & 0 & 0 \\
    0 & \frac{1}{\sqrt{\eta_r \eta_u}} & 0 & 0 \\
    0 & 0 & \sqrt{\frac{\eta_r}{\eta_u}} & 0 \\
    0 & 0 & 0 & \sqrt{\frac{\eta_u}{\eta_r}} \\
  \end{array}
\right)
\end{equation}
we find the equivalent parametrization
\begin{equation}
T  =
\left(\begin{array}{cccc}
\sqrt{\eta_r \eta_u} t & -\eta_p^{-2} \sqrt{\eta_r \eta_u} t & - \eta_p^{-3} \sqrt{\eta_r \eta_u} t & - \eta_p^{-1} \sqrt{\eta_r \eta_u} t\\
\eta_r \eta_u x & y & \eta_r z & \eta_u w\\
-y & \eta_r \eta_u x & -\eta_u w  & \eta_r z\\
-\eta_u w & \eta_r z & \eta_r \eta_u  x & y\\
-\eta_r z & - \eta_u w & -y & \eta_r \eta_u  x
\end{array}\right)
\label{sqrtphase}
\end{equation}
for both even and odd vertices. From this form, it is clearly seen that one may absorb the phases $\eta_r,\eta_u$ into the definitions of $t,x,z,w$, and thus we can set $\eta_u = \eta_r = 1$, without any loss of physical generality, which completes the proof. Furthermore,  thanks to a suitable application of additional virtual symmetries $S$, we can choose the original phases $\eta_r,\eta_u$ such that
$\eta_r \eta_u \eta^2_t=1$, where $\eta_t=\arg{t}$ is the phase of $t$ in \eqref{sqrtphase}: in this case we find that, without any loss of
generality, we can set $t \geq 0$, and even $t>0$ (since for $t=0$ the physical and virtual modes are decoupled).
$\square$
\\

Next, we wish to incorporate translational invariance as well, on top of the other symmetries, leading to the final parametrization of the PEPS.
Now we have translational invariance with unit cell size of $2 \times 2$, but we wish to incorporate an invariance without blocking, such that translational invariance also means charge conjugation: we require that the state will be invariant to first translating the particles by a single lattice site
(which is the fermionic part of the charge conjugation) and afterwards exchanging the positive and negative virtual modes, which is an inversion of the virtual electric field (to become physical
once we introduce bosons and make the gauge symmetry local).

For that, we decompose the operator $A$ for even and odd vertices as follows:
\begin{equation}
A\left(\mathbf{x}\right) = \left\{
      \begin{array}{ll}
        e^{t_j \psi^{\dagger}\left(\mathbf{x}\right) b_j^{\dagger}\left(\mathbf{x}\right)}e^{\tau_{ij} a_i^{\dagger}\left(\mathbf{x}\right) b_j^{\dagger}\left(\mathbf{x}\right)}, & \hbox{even;} \\
        e^{t_j \psi^{\dagger}\left(\mathbf{x}\right) a_j^{\dagger}\left(\mathbf{x}\right)}e^{\tau_{ij} b_i^{\dagger}\left(\mathbf{x}\right) a_j^{\dagger}\left(\mathbf{x}\right)}, & \hbox{odd.}
      \end{array}
    \right.
\end{equation}

We now act with the operator $U_T\left(\mathbf{\hat e}_1\right)$ on the state $\left|\psi\left(T\right)\right\rangle$. The result is a somewhat peculiar state,
in which the virtual fermions at $\mathbf{x}$ are connected with physical fermions at $\mathbf{x} +  \mathbf{\hat e}_1$:
\begin{equation}
U_T\left(\mathbf{\hat e}_1\right) \left|\psi\right\rangle =
\left\langle \Omega_v \right|\underset{\mathbf{x}}{\prod}\omega\left(\mathbf{x}\right)\eta\left(\mathbf{x}\right)  A\left(\mathbf{x}_p=\mathbf{x}+\mathbf{\hat e}_1,\mathbf{x}_v=\mathbf{x}   \right)  \left|\Omega\right\rangle
\end{equation}
We also define a similar translation operator for the virtual fermions, $U_v$. If
\begin{equation}
\tau_{ij}=-\tau_{ji}
\label{eqtau}
\end{equation}
we obtain that acting with $U_T\left(\mathbf{\hat e}_1\right)$ on the physical level is equivalent to acting with $U_v^{\dagger}\left(\mathbf{\hat e}_1\right)$
on the virtual level, and similarly for the other direction $\mathbf{\hat e}_2$; then, since the projectors and the vacuum are invariant under the virtual translation, we deduce that if (\ref{eqtau}) is fulfilled, the state $\left|\psi\right\rangle$ is translationally invariant if $T$ has the form
\begin{equation}
T  =
\left(\begin{array}{cccc}
t & \eta_p^{2} t &  \eta_p t &  \eta_p^{3}  t\\
 0 & y &  z/\sqrt{2} &  z/\sqrt{2}\\
-y &  0 & - z/\sqrt{2}  &  z/\sqrt{2}\\
- z/\sqrt{2} &  z/\sqrt{2} &   0 & y\\
- z/\sqrt{2} & -  z/\sqrt{2} & -y &   0
\end{array}\right)
\label{Tmatrix}
\end{equation}
where $z,y \in \mathbb{C}$ and $t>0$, $\eta_p=e^{i \pi/4}$, and the different normalization of $z$ is chosen for later convenience.

No further phases could be reduced, since we have exploited all the virtual redundancy symmetries which respect translational and rotational invariance. Thus, we arrive at the final parametrization of our PEPS, which shall be given in the following summarizing statement.

\begin{sttmnt} \label{th:Tmatrix}
 The most general fermionic Gaussian PEPS with two virtual fermions per bond, with global $U(1)$ gauge invariance, translational invariance and rotational invariance may be parameterized by three parameters - $t>0, y,z \in \mathbb{C}$, and is given by \eqref{Tmatrix}.
\end{sttmnt}

\subsubsection{The Covariance Matrix}

After having characterized the state in terms of the matrix $T$ in \eqref{Tmatrix}, we start investigating some of the main properties of the PEPS $\ket{\psi(T)}$. In particular we will define, in the following, its covariance matrix, parent Hamiltonian and correlation functions. To this purpose,
we define a set of Majorana operators associated to the physical fermions as:
\begin{equation}
c_{\mathbf{x},1} = \psi_{\mathbf{x}} + \psi_{\mathbf{x}}^{\dagger} \,;\quad c_{\mathbf{x},2} = i\left(\psi_{\mathbf{x}} - \psi_{\mathbf{x}}^{\dagger}\right)\,,
\end{equation}
and their Fourier transformed (complex) operators:
\begin{equation}
d_{\mathbf{k},a} = \frac{1}{L}\sum_\mathbf{x} e^{-i \mathbf{k x}}c_{\mathbf{x},a} \label{FT_Majorana}
\end{equation}
where we consider a system of $L \times L$ lattice vertices with periodic boundary conditions.
Note that for $\mathbf{k}=\left(0,0\right),\left(\pi,0\right),\left(0,\pi\right),\left(\pi,\pi\right)$, the resulting operators are still Majorana operators, whereas for the rest they are complex fermions. The redundancy appears as $d^\dagger_{\mathbf{k},a}=d_{-\mathbf{k},a}$.
Since the state $\left|\psi\right\rangle$ is Gaussian \cite{Bravyi05}, it is fully described by its covariance matrix,
\begin{equation} \label{Gamma}
\Gamma^{\mathbf{xx'}}_{ab} = \frac{i}{2}\left\langle\left[c_{\mathbf{x},a},c_{\mathbf{x'},b}\right]\right\rangle
\end{equation}
which may be calculated using a Gaussian mapping (see \ref{app:gaussian}).
 This is a real, antisymmetric matrix fulfilling $\Gamma^\dagger \Gamma \le \Id$, where the equality occurs for pure states.
 As the state is translationally invariant, the covariance matrix may be decomposed into blocks in momentum space \cite{Kraus2010}, which we denote as
\begin{equation}
\left(G_{\text{out}}\left(\mathbf{k}\right)\right)_{ab} = \frac{i}{2}\left\langle\left[d_{\mathbf{k},a}^{\phantom{\dag}},d^{\dagger}_{\mathbf{k},b}\right]\right\rangle \equiv \begin{pmatrix}
        iP\left(\mathbf{k}\right) & Q\left(\mathbf{k}\right)\\
        -\overline{Q}\left(\mathbf{k}\right) & -iP\left(\mathbf{k}\right)\end{pmatrix}, \label{Gamma_k}
\end{equation}
where $P\left(\mathbf{k}\right)$ is a real function whereas $Q\left(\mathbf{k}\right)$ is a complex function which may be decomposed into its real and imaginary parts $Q\left(\mathbf{k}\right)=R\left(\mathbf{k}\right)+i I\left(\mathbf{k}\right)$.
Due to the properties of  pure fermionic Gaussian states~\cite{Kraus2009} $G_\mr{out}^2(\mb k) = -\Id$, therefore the functions $P,R$ and $I$ are normalized in such a way that $R^2(\mathbf{k})+I^2(\mathbf{k})+P^2(\mathbf{k})= 1$.

Inserting $d_{\mathbf{k},a}^\dg = d_{-\mathbf{k},a}$  into Eq.~\eqref{Gamma_k} gives $G_\mr{out}(-\mathbf{k}) = \overline{G_\mr{out}}(\mathbf{k})$, implying that
\begin{equation}
\begin{aligned}
 P\left(-\mathbf{k}\right)&= -P\left(\mathbf{k}\right), \label{Psym} \\
 R\left(-\mathbf{k}\right)&= R\left(\mathbf{k}\right),  \\
 I\left(-\mathbf{k}\right)&= -I\left(\mathbf{k}\right).
\end{aligned}
\end{equation}
As shown in \ref{app:gaussian}, the functions $P(\mb k)$, $R(\mb k)$ and $I(\mb k)$ are fractions of \emph{trigonometric} polynomials,
 $P_0(\mb k)$, $R_0(\mb k)$, $I_0(\mb k)$ and $\mathcal{D}(\mb k)$, which have maximum order 4 in $k_1$ and $k_2$ individually, of the form
\be
P(\mb k) = \frac{P_0(\mb k)}{\mathcal{D}(\mb k)}, \;\; R(\mb k) = \frac{R_0(\mb k)}{\mathcal{D}(\mb k)}, \;\; I(\mb k) = \frac{I_0(\mb k)}{\mathcal{D}(\mb k)}
\ee
where $\mathcal{D}(\mb k) = \mathcal{D}(- \mb k) \geq 0$. Hence, the relations~\eqref{Psym}  also hold for $P_0(\mb k)$, $R_0(\mb k)$ and $I_0(\mb k)$ respectively. Due to the normalization of $G_\mr{out}(\mb k)$, they fulfill $\mathcal{D}\left(\mathbf{k}\right) = \sqrt{R_0^2\left(\mathbf{k}\right)+I_0^2\left(\mathbf{k}\right)+P_0^2\left(\mathbf{k}\right)}$.

\subsubsection{The Parent Hamiltonian}

Starting from the unnormalized covariance matrix
\begin{equation}
\mathfrak{g}\left(\mathbf{k}\right) \equiv \left(\begin{array}{cc}
i P_0(\mb k)& R_0(\mb k) + I_0(\mb k) \\
-R_0(\mb k) + I_0(\mb k)& -i P_0(\mb k)
\end{array}\right)
\end{equation}
one can define a local \emph{Parent Hamiltonian} \cite{Wahl2013,Wahl2014} for the state $\ket{\psi(T)}$:
\begin{equation}
H = -\frac{i}{2} \underset{\mathbf{k},a,b}{\sum}\mathfrak{g}_{ab}\left(\mathbf{k}\right)d_{\mathbf{k},a}d^{\dagger}_{\mathbf{k},b}
\end{equation}
with the dispersion relation $\mathcal{D}\left(\mathbf{k}\right)$.
By construction, the many-body state $\left|\psi\left(T\right)\right\rangle$ is a ground state of the Hamiltonian $H$.
To obtain the Hamiltonian in real space, we define the Fourier transform of $\mathfrak{g}\left(\mathbf{k}\right)$,
 \begin{equation}
\hat{\mathfrak{g}}_{ab}\left(\mathbf{x}\right) = \frac{1}{L^2}\sum_\mathbf{k} e^{i \mathbf{k  x}}\mathfrak{g}_{ab}\left(\mathbf{k}\right)
\label{gfour}
\end{equation}
with the normalization chosen such that:
\begin{equation}
H = -\frac{i}{2} \underset{\mathbf{x,x}',a,b}{\sum}\hat{\mathfrak{g}}_{ab}\left(\mathbf{x'-x}\right)c_{\mathbf{x},a}c_{\mathbf{x}',b}
\end{equation}
or, in terms of the physical fermionic operators $\psi_\mathbf{x}$:
\begin{equation} \label{ham1}
H = \underset{\mathbf{x,x}'}{\sum}\hat R_0 \left(\mathbf{x'-x}\right) \left(\psi^{\dagger}_{\mathbf{x}}\psi_{\mathbf{x'}}+\psi^{\dagger}_{\mathbf{x'}}\psi_{\mathbf{x}} - \delta_{\mathbf{x,x'}}\right)
+ \underset{\mathbf{x,x}'}{\sum} \left(\hat \Delta_0 \left(\mathbf{x'-x}\right) \psi^{\dagger}_{\mathbf{x}}\psi^{\dagger}_{\mathbf{x'}} + {\rm H.c.}\right)
\end{equation}
where
\begin{equation}
\hat \Delta_0 \left(\mathbf{x}\right) \equiv \hat P_0 \left(\mathbf{x}\right) - i \hat I_0 \left(\mathbf{x}\right).
\end{equation}
Due to the fact that $P_0(\mb k)$, $R_0(\mb k)$ and $I_0(\mb k)$ are trigonometric polynomials of maximum order 4 in $k_1$ and $k_2$ individually, their Fourier transforms can be non-vanishing only for $|x_{1,2}' - x_{1,2}| \leq 4$, giving rise to a finite hopping range of the above Parent Hamiltonian.

We observe that, in order to fulfill the global invariance, the Hamiltonian \eqref{ham1} has to commute with the staggered charge \eqref{Gglobgen}.
This will be verified later by explicit calculation. Here we observe that the conservation of the staggered charge for a generic Hamiltonian of the kind \eqref{ham1} is verified if
 $R_0(\mathbf{x'-x})=0$ for $\mathbf{x}$ and $\mathbf{x'}$ belonging to different sublattices and $\Delta_0(\mathbf{x'-x})=0$ for $\mathbf{x}$ and $\mathbf{x'}$ in the same sublattice. Furthermore, the global U(1) gauge symmetry implies the conservation of the fermionic parity in the system that is indeed evident in the Hamiltonian \eqref{ham1}, which describes a system of spinless fermions subject to a p-wave pairing interaction, as we will discuss in Sec. \ref{sec:Ham}.

\subsubsection{The Correlation Functions}
The covariance matrix may also be used for the calculation of the correlation functions of the physical fermionic operators $\psi_{\mathbf{x}}$.
In particular, substituting $\psi_{\mathbf{x}}^{\dagger}=(c_{\mathbf{x},1}+ic_{\mathbf{x},2})/2$
we can express them in terms of the real-space covariance matrix $\Gamma$ defined in Eq. \eqref{Gamma}. We have
\begin{equation}
\left\langle \psi_{\mathbf{x}}^{\dagger}\psi_{\mathbf{x}'}\right\rangle =\frac{1}{2}\delta_{\mathbf{x}\mathbf{x'}}+\frac{1}{4}\left(-i\varGamma_{11}^{\mathbf{x}\mathbf{x'}}+\varGamma_{21}^{\mathbf{x}\mathbf{x'}}
-\varGamma_{12}^{\mathbf{x}\mathbf{x'}}-i\varGamma_{22}^{\mathbf{x}\mathbf{x'}}\right)
\end{equation}
or simply
\begin{equation}
\left\langle \psi_{\mathbf{x}}^{\dagger}\psi_{\mathbf{x}'}\right\rangle =\frac{1}{2}\left(\delta_{\mathbf{x}'\mathbf{x}}-\hat{R}\left(\mathbf{x}'-\mathbf{x}\right)\right)
\label{dagger}
\end{equation}
and similarly
\begin{equation}
\left\langle \psi_{\mathbf{x}}^{\dagger}\psi_{\mathbf{x}'}^{\dagger}\right\rangle \equiv -\frac{1}{2}\hat\Delta\left(\mathbf{x}'-\mathbf{x}\right) = -\frac{1}{2}\left(\hat{P} \left(\mathbf{x}\right) - i  \hat{I} \left(\mathbf{x}\right)\right) .
\label{daggerdagger}
\end{equation}
(where $\hat R, \hat \Delta, \hat P, \hat I$ are defined similarly to $\hat{\mathfrak{g}}$ in \eqref{gfour}, as Fourier transforms).
The first kind of correlation (\ref{dagger}) is expected to vanish
if the two vertices are of different parities (sublattices), as it would not preserve the global symmetry;
Similarly, the correlation (\ref{daggerdagger}) must vanish in the other case. Let us show this explicitly.

A gauge transformation of a Majorana operator is simply an orthogonal rotation
matrix in $SO(2)$, as it corresponds to a U(1)
(phase) transformation of the fermionic operators. As the symmetry
is global, there is a single parameter for these transformations everywhere; however,
the rotation will have two different orientations, due to the staggering.
And thus, in general, one may write the gauge transformation as
\begin{equation}
\varGamma_{ab}^{\mathbf{x}\mathbf{x'}}\longrightarrow O_{aa'}^{\mathbf{x}}O_{bb'}^{\mathbf{x'}}\varGamma_{a'b'}^{\mathbf{x}\mathbf{x'}}.
\end{equation}
(no summation on $\mathbf{x,x'}$). Denote
\begin{equation}
 O=\left(\begin{array}{cc}
\cos\phi & -\sin\phi\\
\sin\phi & \cos\phi
\end{array}\right)
\end{equation}
as the transformation matrix of an even vertex, and thus $O^{T}$
corresponds to the transformation matrix of an odd vertex. Now consider
an infinitesimal transformation $(\phi=\epsilon\ll 1)$ of a subblock of $\Gamma$ involving two
sites of the same sublattice (without loss of generality
we take them to be even, otherwise one has to invert the sign). Then,
\begin{equation}
\varGamma^{\mathbf{x}\mathbf{x'}}\longrightarrow O^{\mathbf{x}}\varGamma^{\mathbf{x}\mathbf{x'}}O^{\mathbf{x'}T}=\varGamma^{\mathbf{x}\mathbf{x'}}+
\frac{2i\epsilon}{L^2}\underset{\mathbf{k}}{\sum}\left(\begin{array}{cc}
-I\left(\mathbf{k}\right) & P\left(\mathbf{k}\right)\\
-P\left(\mathbf{k}\right) & I\left(\mathbf{k}\right)
\end{array}\right)e^{-i\mathbf{k}\left(\mathbf{x-x}'\right)}+ O\left(\epsilon^{2}\right)
\end{equation}
$\varGamma^{\mathbf{x}\mathbf{x'}}$ must be invariant, since the overall state cannot get a global phase under the transformation.
As the second term must vanish for an even $\mathbf{x-x}'$ separation, we deduce that both $P\left(\mathbf{k}\right)$ and $I\left(\mathbf{k}\right)$
(as the imaginary part of $Q\left(\mathbf{k}\right)$) must contain
only odd harmonics.

Similarly, if we consider the covariance matrix elements for two sites
on different sublattices, we obtain
\begin{equation}
\varGamma^{\mathbf{x}\mathbf{x'}}\longrightarrow O^{\mathbf{x}}\varGamma^{\mathbf{x}\mathbf{x'}}O^{\mathbf{x'}}=\varGamma^{\mathbf{x}\mathbf{x'}}+
\frac{2 \epsilon}{L^2} \sum_\mathbf{k} \left(\begin{array}{cc}
R\left(\mathbf{k}\right) & 0\\
0 & R\left(\mathbf{k}\right)
\end{array}\right)e^{-i\mathbf{k}\left(\mathbf{x-x}'\right)}+O\left(\epsilon^{2}\right)
\end{equation}
and thus $R\left(\mathbf{k}\right)$ must contain only even harmonics.

The last property we wish to discuss is the behavior of the functions $P,R,I$ - either in real or momentum space - under rotations.

\begin{sttmnt} Under a rotation $\varLambda$, for $\left|\psi\left(T\right)\right\rangle$,
\begin{equation}
\begin{aligned}
&\hat R\left(\varLambda \mathbf{x}\right) = \hat R\left(\mathbf{x}\right),\quad R\left(\varLambda \mathbf{k}\right) = R\left(\mathbf{k}\right) \label{rotR} \\
&\hat \Delta\left(\varLambda \mathbf{x}\right) = - i \hat \Delta \left(\mathbf{x}\right),\quad \Delta\left(\varLambda \mathbf{k}\right) = -i \Delta\left(\mathbf{k}\right)
\end{aligned}
\end{equation}
\end{sttmnt}
\emph{Proof}:
From Eq. \eqref{dagger} we have:
\begin{equation}
\left\langle \psi_{\varLambda\mathbf{x}}^{\dagger}\psi_{\varLambda\mathbf{x}'}\right\rangle =
\frac{1}{2}\left(\delta_{\varLambda\mathbf{x}',\varLambda\mathbf{x}}-\hat{R}\left(\varLambda \left(\mathbf{x}'- \mathbf{x}\right)\right)\right)
\end{equation}
and, on the other hand,
\begin{equation}
\left\langle \psi_{\varLambda\mathbf{x}}^{\dagger}\psi_{\varLambda\mathbf{x}'}\right\rangle =
\left\langle U_p\left(\varLambda\right) \psi_{\mathbf{x}}^{\dagger}\psi_{\mathbf{x}'} U^{\dagger}_p\left(\varLambda\right)\right\rangle.
\end{equation}
Since the state $\left|\psi\left(T\right)\right\rangle$ is invariant under rotations and satisfies Eq. \eqref{transrot} due to statement \ref{th:Tmatrix}, it follows that $\langle \psi_{\varLambda\mathbf{x}}^{\dagger}\psi_{\varLambda\mathbf{x}'}\rangle = \langle  \psi_{\mathbf{x}}^{\dagger}\psi_{\mathbf{x}'} \rangle$ which implies Eq. \eqref{rotR}.

In an analogous way, concerning the correlator $\Delta$ we obtain form Eqs. \eqref{daggerdagger} and (\ref{physrot},\ref{phasedef}):
\begin{equation}
\left\langle \psi_{\varLambda\mathbf{x}}^{\dagger}\psi_{\varLambda\mathbf{x}'}^{\dagger}\right\rangle =
-\frac{1}{2}\hat{\Delta}\left(\varLambda \left(\mathbf{x}'- \mathbf{x}\right)\right) =
- i \left\langle U_p\left(\varLambda\right) \psi_{\mathbf{x}}^{\dagger}\psi_{\mathbf{x}'}^{\dagger} U^{\dagger}_p\left(\varLambda\right)\right\rangle.
\end{equation}
and Eq. \eqref{rotR} follows again from the rotational invariance of $\ket{\psi(T)}$.

In particular, Eq. \eqref{rotR} implies that
\begin{equation} \label{PIrot}
 P\left(\varLambda\mathbf{k}\right) = - I\left(\mathbf{k}\right) , \quad I\left(\varLambda\mathbf{k}\right) = P\left(\mathbf{k}\right),
\end{equation}
and similar relations hold for the real-space functions. Furthermore, in accordance with the odd parity of these functions, we arrive at:
\begin{equation}
P\left(-\mathbf{k}\right)=P\left(\varLambda^{2}\mathbf{k}\right)=- I\left(\varLambda\mathbf{k}\right)=-P\left(\mathbf{k}\right)
\end{equation}

From the rotational invariance of $\ket{\psi(T)}$ one obtains also
\begin{equation}
d\left(\varLambda\mathbf{k}\right)=d\left(\mathbf{k}\right),
\end{equation}
as the dispersion relation should be rotationally invariant as well,
and therefore:
\begin{equation}
\begin{aligned}
R_{0}\left(\varLambda\mathbf{k}\right)&= R_{0}\left(\mathbf{k}\right), \\
I_{0}\left(\varLambda\mathbf{k}\right)&= P_{0}\left(\mathbf{k}\right),\\
P_{0}\left(\varLambda\mathbf{k}\right)&= -I_{0}\left(\mathbf{k}\right),
\label{RIProt}
\end{aligned}
\end{equation}
along with similar relations in real space, which imply that the parent Hamiltonian (\ref{ham1}) is also symmetric under rotations. $\Box$

\begin{figure}
  \centering
  \includegraphics[width=0.6\textwidth]{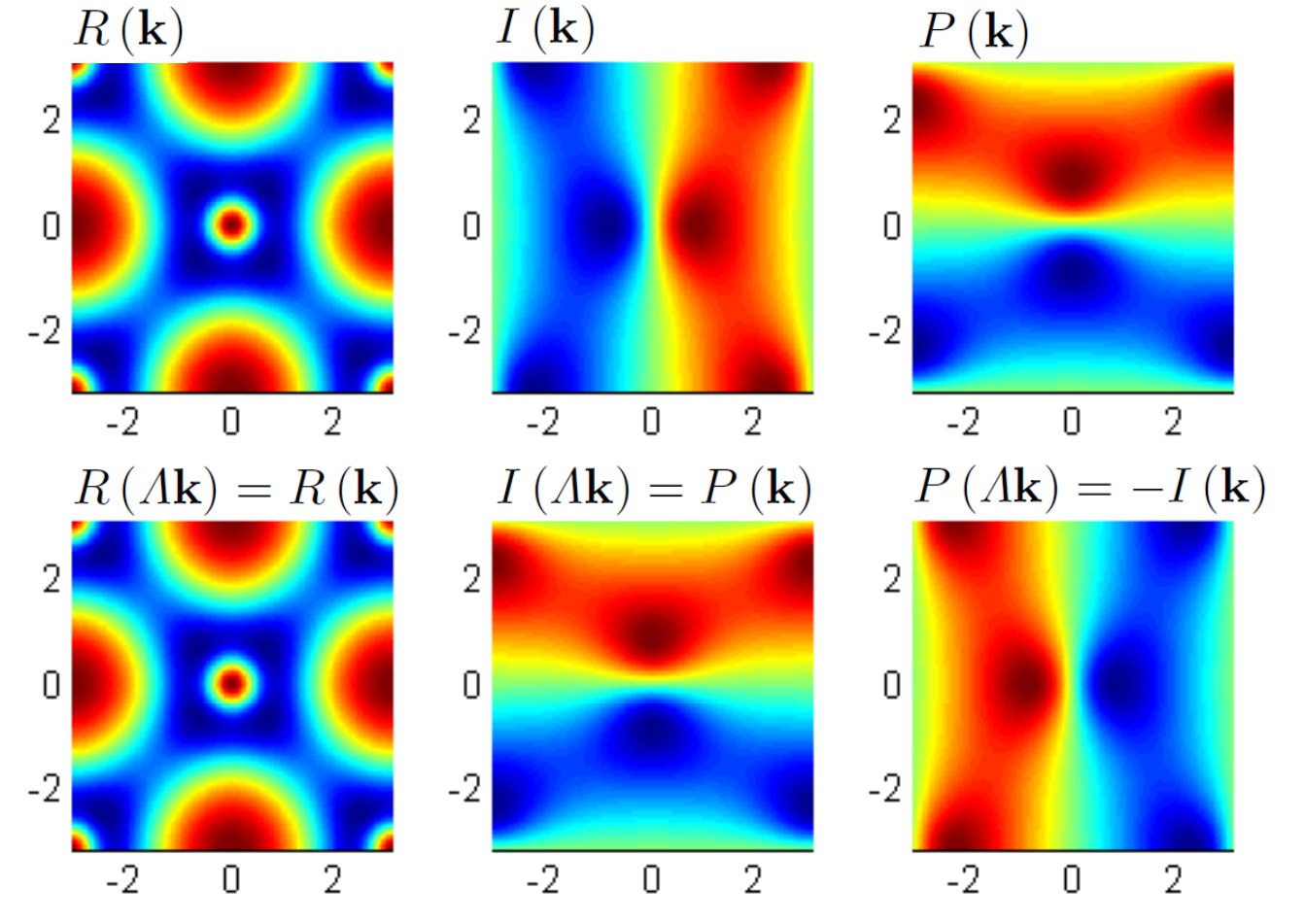}
\caption{An example, with randomly generated parameters ($t=0.2785,y=0.5469,z=0.9575$), for the momentum-space matrix elements of the covariance matrix, showing explicitly the rotational symmetry. The colors represent the amplitude of the functions and are only important for illustrative purposes. $N=200$.}
  \label{fig3}
\end{figure}

The rotation relation linking $P$ and $I$ is a consequence of the simultaneous requirements of conservation of the staggered charge \eqref{Gglobgen} and rotational invariance. In particular, as we will discuss in the next subsection, $P$ and $I$ correspond to hopping amplitudes of the fermionic matter after a staggered particle-hole transformation that maps the real space Hamiltonian \eqref{ham1} into a standard U(1) invariant Hamiltonian with a conserved number of fermions, reminiscent of the Kogut-Susskind model. The transformation rules of $P$ and $I$ under rotations are indeed consistent with the differences of the hopping amplitudes
in the Kogut-Susskind Hamiltonian for staggered fermions along different directions (resulting from the different Dirac matrices) \cite{Susskind1977}. An example of these symmetry properties can be seen in figure \ref{fig3}.

\subsection{The Globally Invariant PEPS as a p-wave Paired State: An Analytical Treatment}

\label{sec:Ham}

\subsubsection{The PEPS-BCS State in Momentum Space}

So far, we have considered some of the general properties of the correlation functions and the parent Hamiltonian by exploiting the Gaussian formalism and the physical Majorana modes. In this Section we calculate their explicit form as a function of the parameters in the $T$ matrix.

Before entering the PEPS details again, let us examine further some symmetry properties of the parent Hamiltonian in momentum space. For that, we adopt the following convention for the Fourier transform of the physical fermionic operators:
\begin{equation}
\psi_{\mathbf{x}}^{\dagger} = \frac{1}{\sqrt{L_1L_2}} \sum_\mathbf{k} e^{i \mathbf{k}  \mathbf{x}} \psi_{\mathbf{k}}^{\dagger}.
\label{FTconv}
\end{equation}
where $L_1$ and $L_2$ are the system width and length.
Note that the momentum operators $\psi_{\mathbf{k}}$ mix particles and anti-particles.

Using this convention, and the definition of Nambu spinors $\Psi_{\mathbf{k}} = (\psi_{\mathbf{k}},\psi^\dag_{-\mathbf{k}})^{\intercal}$, we obtain,
from Eq. (\ref{ham1}), the Hamiltonian:
\begin{equation}
H = {\sum}_\mathbf{k}\Psi^{\dagger}_{\mathbf{k}}\mathcal{H}\left(\mathbf{k}\right)\Psi_{\mathbf{k}} \equiv
\underset{\mathbf{k}}{\sum}\Psi^{\dagger}_{\mathbf{k}}\left(
R_0\left(\mathbf{k}\right)\sigma_z + I_0\left(\mathbf{k}\right)\sigma_y + P_0\left(\mathbf{k}\right)\sigma_x
 \right)\Psi_{\mathbf{k}}. \label{ripham}
\end{equation}
The Bogoliubov-de Gennes (BdG) Hamiltonian $\mathcal{H}\left(\mathbf{k}\right)$ describes a spinless complex p-wave superconductor with order parameter $\Delta_0(\mathbf{k})=P_0\left(\mathbf{k}\right)-iI_0\left(\mathbf{k}\right)$.
A rotation ($H \rightarrow U_p\left(\varLambda\right)H U^{\dagger}_p\left(\varLambda\right)$) corresponds to
\begin{equation}
\mathcal{H}\left(\mathbf{k}\right) \rightarrow W^{\dagger}\mathcal{H}\left(\mathbf{k}\right)W\,, \quad {\rm with} \quad W = e^{i \frac{\pi}{4} \sigma_z}\,.
\end{equation}
This leads to
\begin{equation}
W^{\dagger}\mathcal{H}\left(\mathbf{k}\right)W = \mathcal{H}\left(\varLambda\mathbf{k}\right)
\end{equation}
which results in the rotation rules (\ref{RIProt}) discussed above.

Such a p-wave pairing Hamiltonian enjoys the following particle-hole symmetry:
\begin{equation}
 \sigma_x \mathcal{H}(\mathbf{k}) \sigma_x = -\overline{\mathcal{H}}\left(-\mathbf{k}\right)
\end{equation}
whereas there is no time-reversal symmetry;
therefore, in the general case, the Hamiltonian defines a system in the topological class D within the classification of topological insulators and superconductors (see, for example, \cite{Schnyder2008,Kitaev2009,Hasan2010}).

As $H$ is a BdG Hamiltonian, its ground state, $\left|\psi\left(T\right)\right\rangle$ is a BCS state pairing modes with momenta  $\mathbf{k}$ and $-\mathbf{k}$,  having the normalized form
\begin{equation}
\prod_{\mathbf{k}\geq0}\left(u\left(\mathbf{k}\right) + v\left(\mathbf{k}\right)\psi^{\dagger}_{\mathbf{k}}\psi^{\dagger}_{-\mathbf{k}}\right)\left|\Omega_p\right\rangle
\label{BCSnormd}
\end{equation}
with
\begin{equation}
\begin{aligned}
&2\left|v\left(\mathbf{k}\right)\right|^2 = \left(1-R\left(\mathbf{k}\right)\right),\label{vu1}\\
&2 \bar u\left(\mathbf{k}\right) v\left(\mathbf{k}\right) = \Delta\left(\mathbf{k}\right).
\end{aligned}
\end{equation}
These relations can be obtained by either diagonalizing the BdG Hamiltonian or with an analytical evaluation of the correlation functions, given by the covariance matrix obtained in the Gaussian mapping. In \ref{app:gaussian} we follow the latter strategy to find explicit formulae for $P,R$ and $I$ as  functions of the parameters of the matrix $T$.

The PEPS $\ket{\psi(T)}$, however, is not normalized, and thus takes the form
\begin{equation}
\left|\psi\left(\mathbf{k}\right)\right\rangle = \left(\alpha\left(\mathbf{k}\right) + \beta\left(\mathbf{k}\right)\psi^{\dagger}_{\mathbf{k}}\psi^{\dagger}_{-\mathbf{k}}\right)\left|\Omega_{\mathbf{k}}\right\rangle
\label{BCSnnormd}
\end{equation}
where the functions $\alpha\left(\mathbf{k}\right),\beta\left(\mathbf{k}\right)$ being are unnormalized versions of the BCS functions
$u\left(\mathbf{k}\right),v\left(\mathbf{k}\right)$ and can be explicitly calculated with the Gaussian formalism.

One may obtain from the previous equations:
\begin{equation}
\begin{aligned}
&g(\mathbf{k})\equiv\frac{v\left(\mathbf{k}\right)}{u\left(\mathbf{k}\right)} = \frac{\beta\left(\mathbf{k}\right)}{\alpha\left(\mathbf{k}\right)} =
\frac{1- R\left(\mathbf{k}\right)}{P\left(\mathbf{k}\right) + i I\left(\mathbf{k}\right)} \,,\label{BCSfuncs0} \\
&R\left(\mathbf{k}\right) = \frac{\left|\alpha\left(\mathbf{k}\right)\right|^2 - \left|\beta\left(\mathbf{k}\right)\right|^2}{\left|\alpha\left(\mathbf{k}\right)\right|^2 + \left|\beta\left(\mathbf{k}\right)\right|^2} \,,\\
&\Delta\left(\mathbf{k}\right) = \frac{2\bar \alpha\left(\mathbf{k}\right)\beta\left(\mathbf{k}\right)}{\left|\alpha\left(\mathbf{k}\right)\right|^2 + \left|\beta\left(\mathbf{k}\right)\right|^2}\,.
\end{aligned}
\end{equation}
where we defined the pairing function in momentum space $g(\mathbf{k})$.

Four modes, $\mathbf{k} = \left(0,0\right) , \left(\pi,\pi\right), \left(\pi,0\right), \left(0,\pi\right)$, are left unpaired: these modes remain in the vacuum state
(e.g. $\left|\psi\left(0,0\right)\right\rangle = \tilde \alpha\left(0,0\right)\left|\Omega\right\rangle$), i.e.
$\beta = 0$ for these momenta. This means, in particular,  that $R\left(\mathbf{k}=\mathbf{0}\right) = 1$ and $\Delta\left(\mathbf{k}=\mathbf{0}\right) = 0$,
in accordance with the previous results. The value of the pairing function $g(\mathbf{k})$ in these points is instead non-trivial as we will discuss in the following.
This particular behavior of the unpaired modes is dictated by the PEPS construction: occupying these unpaired modes is indeed impossible in $\ket{\psi(T)}$, since the state was created by the operators $A$ which involve only even products of creation operators.  Despite that, we stress that, in the PEPS construction, a well-defined amplitude $\tilde \alpha$ is associated to these unpaired states (see \ref{app:C}).

Finally, the PEPS construction allows us to obtain a full analytical understanding of the globally invariant state, and to calculate explicitly all these functions. In particular the following results hold:

\begin{sttmnt} \label{th:dispersion}
The dispersion relation $\mathcal{D}\left(\mathbf{k}\right)$  is proportional to
\begin{equation}
E\left(\mathbf{k}\right) = \left|\alpha\left(\mathbf{k}\right)\right|^2+\left|\beta\left(\mathbf{k}\right)\right|^2,
\end{equation}
\end{sttmnt}
which is the dispersion relation we shall work with.

\begin{sttmnt} \label{th:BCScoeff}
The BCS coefficients are
\begin{multline}
 \alpha(k_x,k_y)= A_0 + A_1 \left[\cos\left(k_x+k_y \right) + \cos\left(k_x-k_y \right) \right]\\
 +  A_2 \left[\cos\left(2k_x \right) + \cos\left(2k_y \right) \right]
 +  A_3 \left[\cos\left(2k_x+2k_y \right) + \cos\left(2k_x-2k_y \right) \right]
\label{alpha}
\end{multline}
and
\begin{multline} \label{beta}
 \beta(k_x,k_y)= 2t^2 B_1 \left(\sin k_x-i\sin k_y\right) + \\
  2t^2 B_2 \left[\sin\left(k_x+2k_y \right)-i \sin\left(k_y-2k_x\right) \right]
 -2t^2 B^*_2 \left[\sin\left(2k_y-k_x \right)+i \sin\left(k_y+2k_x\right) \right]
\end{multline}
From these equations it is evident that:
\begin{equation} \label{rotalphabeta}
 \alpha(\varLambda\mathbf{k}) = \alpha(\mathbf{k}) \,,\quad \beta(\varLambda\mathbf{k})=-i\beta(\mathbf{k})
\end{equation}
The coefficients $A_i$ and $B_i$ depend only on $y$ and $z$:
\begin{equation}
\begin{aligned}
 A_0&=(1 + y^4)^2 - 4 y^6 z^2 + 3 (1 + 2 y^4) z^4 - 4 y^2 z^6 + z^8 \\
 A_1&=-2z^2\left(1+y^4+z^4-2y^2\left(1+z^2 \right)  \right) \\
 A_2&=4 y^4 z^2-2 y^2 \left(z^4+1\right)+z^4-2 y^6 \\
 A_3&=\frac{1}{2}\left(z^2-2y^2 \right)^2 \\
 B_1&=(1+z^2)(1+z^4-2yz^3) + y^2(1+2z^2-z^4)+ 2zy^3(2z^2-1)-y^4(z^2-1)+y^5(y-2z)\\
 B_2&=y (z-y) (1 + y^2 - z^2) + \frac{i}{2}z (-2 y + 2 y^3 + z - 3 y^2 z + z^3) \\
 B^*_2&=y (z-y) (1 + y^2 - z^2) - \frac{i}{2}z (-2 y + 2 y^3 + z - 3 y^2 z + z^3)
\end{aligned}
\end{equation}
\end{sttmnt}

\begin{sttmnt}\label{th:unpaired}
The amplitude $\tilde{\alpha}$ associated to the unpaired modes $\ket{\psi(\mathbf{k}_0)}=\tilde{\alpha}(\mathbf{k}_0)\ket{\Omega}$ of the PEPS at the momenta $\mathbf{k}_0 = \left(0,0\right) , \left(\pi,\pi\right), \left(\pi,0\right), \left(0,\pi\right)$ is related to the coefficient $\alpha$ of the paired modes by
\begin{equation}
\tilde \alpha\left(\mathbf{k}_0\right) = \sqrt{ \alpha\left(\mathbf{k}_0\right)}.
\end{equation}
\end{sttmnt}
The proofs of these three statements rely on the PEPS construction, and are given in \ref{app:C}.

We also observe, from the previous results, that $\alpha \left(\mathbf{k}\right) \ge 0$ for all the values of the parameters $y$ and $z$, and its minima lie at either $\mathbf{k}=(0,0),(\pi,\pi)$ or $\mathbf{k}=(0,\pi),(\pi,0)$. In particular, the extreme values of $\alpha$ are
\begin{equation}
\begin{aligned}
\alpha_0 \equiv \alpha(0,0) &= \alpha(\pi,\pi) = \tilde \alpha^2(0,0)= \nonumber  \left(1-(y+z)^2 \right)^2\left(1-(y-z)^2 \right)^2 ,\label{alphagpls1}\\
\alpha_\pi \equiv \alpha(\pi,0) &= \alpha(0,\pi) = \tilde \alpha^2(\pi,0)= (1 - y^2 + z^2)^4
\end{aligned}
\end{equation}

Following these results, using  eqs. (\ref{BCSfuncs0}), we can redefine the Hamiltonian coefficients (up to a constant) as
\begin{equation}
 R_0(\mathbf{k}) = \left|\alpha (\mathbf{k})\right|^2 - \left|\beta (\mathbf{k})\right|^2 \,,\quad \Delta_0(\mathbf{k}) = 2\overline{\alpha}(\mathbf{k})\beta(\mathbf{k})
\end{equation}
in full agreement with the spectrum of the BdG Hamiltonian
\begin{equation}
   E\left(\mathbf{k}\right) = \pm \sqrt{R_0^2\left(\mathbf{k}\right) + P_0^2\left(\mathbf{k}\right) + I_0^2\left(\mathbf{k}\right)} =
   \pm \left(\left|\alpha (\mathbf{k})\right|^2 + \left|\beta (\mathbf{k})\right|^2 \right).
 \end{equation}

\subsubsection{The Phase Diagram}

\begin{figure}
  \centering
  \includegraphics[width=0.5\textwidth]{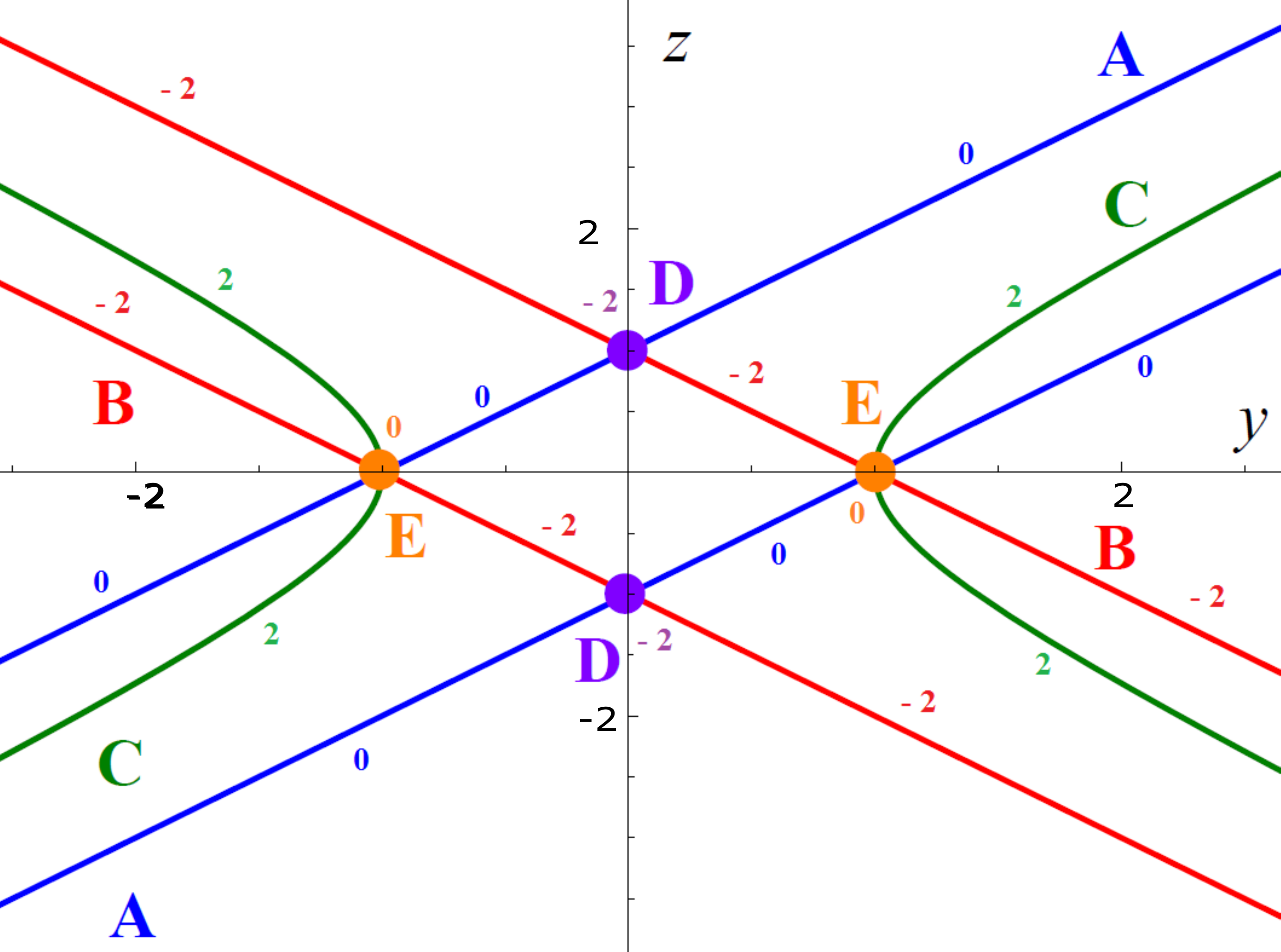}
	\caption{The phase diagram of the fermionic, globally invariant model. The lines represent the gapless phases (A)-(E) described in the text, while the other areas are gapped.
    As explained in the text, this is only the plane of $y,z \in \mathbb{R}$, but the results are valid for an analytical continuation for any $y,z \in \mathbb{C}$. The Chern numbers of the gapless lines are indicated along them. They were calculated from the vector $\left(P_0\left(\mathbf{k}\right),I_0\left(\mathbf{k}\right),R_0\left(\mathbf{k}\right)\right)$ characteruzubg the BdG Hamiltonian \eqref{ripham}.}
  \label{figphase}
\end{figure}

Exploiting the analytical expressions (\ref{alpha}) and (\ref{beta}), we present here the detailed analysis of the phase diagram associated with the globally invariant PEPS $\left|\psi\left(T\right)\right\rangle$, which is shown in Fig. \ref{figphase}. Such an analysis involves, in principle, three parameters, $t > 0$ and $y,z \in \mathbb{C}$. $t$, however, is irrelevant in the definition of the thermodynamical phases: the closing of the gap $E(\mathbf{k})$ may occur only at the four momenta defining the unpaired states, and it does not depend on $t$ since $\alpha\left(\mathbf{k}\right)$ is independent of this parameter and $\beta\left(\mathbf{k}\right)$, which is proportional to $t^2$, is always zero in these points of the Brillouin zone. Therefore only $y$ and $z$ are relevant.

For the sake of simplicity, we will mainly refer to the phase diagram in the plane defined by $y,z\in\mathbb{R}$. We wish to emphasize, though, that the previous definitions of both the functions $\alpha\left(\mathbf{k}\right)$ and $\beta\left(\mathbf{k}\right)$ and the spectrum $E\left(\mathbf{k}\right)$ are valid for any $y,z \in \mathbb{C}$. Thus, the extension of the phase diagram to complex values of $y$ and $z$ is straightforward.

As argued above, band touching points may only exist for the unpaired momenta, $\mathbf{k} = \left(0,0\right),\left(\pi,\pi\right),\left(\pi,0\right),\left(0,\pi\right)$.
The system becomes gapless ($E\left(\mathbf{k}\right) = 0$ for some $\mathbf{k}$) if and only if $\alpha_0 = 0$ or $\alpha_\pi=0$ [see Eqs. (\ref{alphagpls1})]. $\alpha_0=0$ along the four lines $z=\pm1\pm y$, whereas $\alpha_\pi=0$ along the two branches of the hyperbola $y^2-z^2=1$. In the first case band touching points will appear at $\mathbf{k}=(0,0),(\pi,\pi)$, in the second for $\mathbf{k}=(0,\pi),(\pi,0)$. These results hold also for complex values of $y$ and $z$.

To investigate further the characteristics of the system, we observe that the functions $u\left(\mathbf{k}\right)$ and $v\left(\mathbf{k}\right)$ are related to the Hamiltonian by
\begin{equation}
 |u(\mathbf{k})|^2 = \frac{1}{2}\left(1+\frac{R_0(\mathbf{k})}{E(\mathbf{k})} \right), \quad |v(\mathbf{k})|^2 = \frac{1}{2}\left(1-\frac{R_0(\mathbf{k})}{E(\mathbf{k})} \right)
\end{equation}
Consistently with the usual analysis of p-wave superconducting systems, we determine the phases appearing in the model as a function of $y$ and $z$, by considering the gap and the behavior of $u\left(\mathbf{k}\right)$ and $v\left(\mathbf{k}\right)$ in its minima. The minima of the dispersion relation are given by $\alpha_0$ or $\alpha_\pi$, therefore these parameters are related to the chemical potential in the system. In particular, since $\alpha \left(\mathbf{k}\right) \ge 0$, the chemical potential is always negative.
\\

\begin{center}
\emph{Gapped Regions of the Phase Diagram}
\end{center}

In all gapped regions, where $z \neq \pm y \pm 1$ and $y^2-z^2\neq 1$, $R=R_0/E \rightarrow 1$ for $\mathbf{k}$ in the minima of the energy $E(\mathbf{k})$. This implies that in the gapped phases, in a neighborhood of these minima, $u \rightarrow 1$ and $v \rightarrow 0$. Following Read and Green \cite{Read2000}, this corresponds to a gapped regime with a strong p-wave pairing, such that the Cooper pairs are localized. The opposite situation, with $v \rightarrow 1$ and $u \rightarrow 0$ can never be realized in the gapped phases of our model since $\alpha_0\left(\mathbf{k}\right),\alpha_\pi\left(\mathbf{k}\right) \ge 0$. This implies also that non-trivial gapped topological phases cannot appear in the system, consistently with the previous studies of chiral fermionic PEPS~\cite{Wahl2013,Dubail2013}. Indeed, the winding number of the spinor $(u(\mathbf{k}),v(\mathbf{k}))^\intercal$ in the Brillouin zone is always $0$ in the gapped phases, because $u\left(\mathbf{k}\right) \neq 0$ everywhere.

In all gapped phases, $g(\mathbf{k})$ in Eq. (\ref{BCSfuncs0}) is an analytic function of $k_1,k_2$, since $\alpha(\mathbf{k})>0$. This implies that in real space the pairing function $\hat g(\mb x)$ has to decay faster than any inverse polynomial of $|\mb x|$ (see Theorem 3.2.9. in Ref.~\cite{Grafakos2008}).

\begin{figure*}
  \centering
  \includegraphics[width=0.7\textwidth]{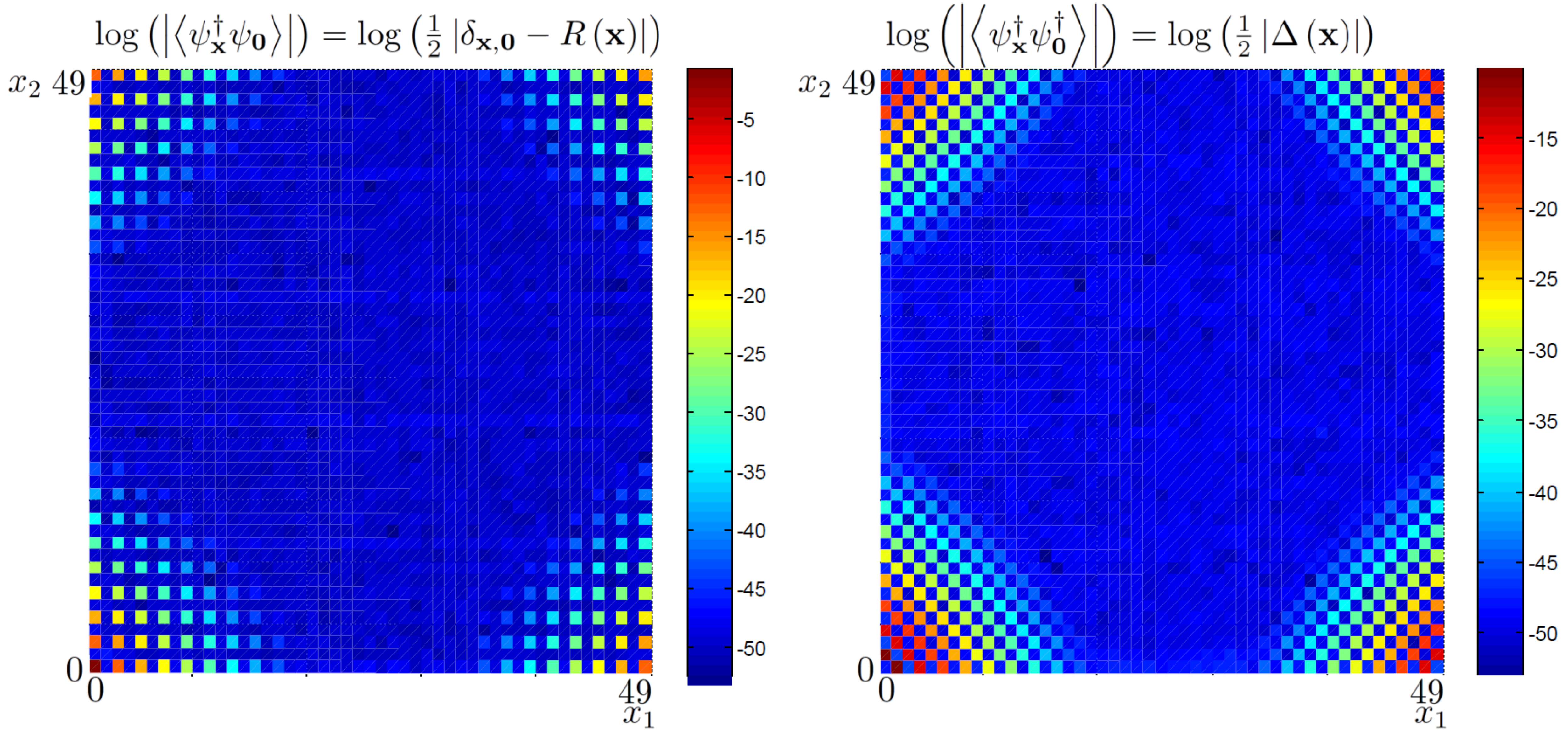}
	\caption{Correlation functions in coordinate space (logarithmic plots) for the magic point $y=1,z=\sqrt{2}$, with $N=50$ and a randomly generated $t=0.6324$.
    Local gauge invariance is manifested by the vanishing of the relevant correlation functions for odd/even separations $\mathbf{x}$; Locality is manifested by the
    exponential decay, which is apparent taking the color scale into account.}
  \label{rsp}
\end{figure*}

The general study for arbitrary $y$ and $z$ is nevertheless complicated. Therefore, we focus here on two specific cases, as paradigmatic examples of the gapped phases. The first is given by $y=1$ and $z=\sqrt{2}$, which we call the \emph{magic} point:
In this point (or in the equivalent ones obtained for $y=\pm 1$ and $z = \pm \sqrt{2}$), $\alpha \left(\mathbf{k}\right)$ becomes independent of the momentum, and one obtains:
\begin{equation}
\begin{aligned}
 \alpha\left(\mathbf{k}\right)&=16 \,,\\
 \beta\left(\mathbf{k}\right)&=16\left(2 - \sqrt{2} \right)t^2\left(\sin k_1 -i \sin k_2 \right)\,.
\end{aligned}
\end{equation}
By rescaling $\alpha\left(\mathbf{k}\right)$ and $\beta\left(\mathbf{k}\right)$ by the uninfluential factor 16, one obtains
\begin{equation}
\begin{aligned}
 R_0\left(\mathbf{k}\right)=& 1 - \left(2 - \sqrt{2} \right)^2 t^4 + \nonumber
 \frac{\left(2 - \sqrt{2} \right)^2}{2}t^4\left(\cos 2k_1 + \cos 2k_2 \right)\,,\\
 P_0\left(\mathbf{k}\right)=& 2\left(2 - \sqrt{2} \right)t^2\sin k_1\,,\\
 I_0\left(\mathbf{k}\right)=& 2\left(2 - \sqrt{2} \right)t^2\sin k_2\,.
\end{aligned}
\end{equation}
and for the pairing function
\begin{equation}
 g(\mathbf{k})\propto \left(\sin k_1 -i \sin k_2 \right),
\end{equation}
such that $\hat{g}(\mathbf{x}_1-\mathbf{x}_2)$ is nonzero only when $\mathbf{x}_1$ and $\mathbf{x}_2$ are nearest neighbors. Therefore, only maximally local Cooper pairs appear.
In Fig. \ref{rsp}, one may see plots of the correlation functions in coordinate-space, which manifest both locality and global invariance, for the case of the magic point.

When we consider different values of $y$ and $z$ in the gapped phases, we expect the superconducting pairs to be no longer maximally localized, but their space distribution becomes exponentially decaying in the distance between the two fermions, consistently with a strong-coupling.

One may also easily consider a second case, the ``trivial'' one, in which $y=z=0$, and, in fact, half of the virtual modes do not participate (since the virtual subblock $\tau$ of the $T$ matrix vanishes, only the virtual modes whose sign is the opposite of the physical modes participate). In this case, $\alpha=1$ and $\beta=2t^2(\sin k_1 -i \sin k_2 )$, therefore
\begin{equation}
\begin{aligned}
 &R_0\left(\mathbf{k}\right)=1-4t^4+2t^4\left(\cos2k_1 + \cos 2k_2 \right)  \,,\\
 &I_0\left(\mathbf{k}\right)=4t^2\sin k_2\,,\\
 &P_0\left(\mathbf{k}\right)=4t^2\sin k_1
\end{aligned}
\end{equation}
and the spectrum is
\begin{equation}
 E\left(\mathbf{k}\right)=\pm \left( 1+4t^4-2t^4\left(\cos(2k_1) + \cos(2k_2) \right) \right)
\end{equation}
which, as expected, is always gapped.

In real space the corresponding $p-ip$ pairing Hamiltonian reads
\begin{equation}
 H=\left(1-4t^4 \right)\underset{\mathbf{x}}{\sum} \psi^\dagger_{\mathbf{x}} \psi_{\mathbf{x}} +
  t^4 \underset{\mathbf{x}}{\sum} \left(\psi^\dagger_{\mathbf{x}+2\hat{e}_1} \psi_{\mathbf{x}} + \psi^\dagger_{\mathbf{x}+2\hat{e}_2} \psi_{\mathbf{x}}  + {\rm H.c.}\right) 
+2t^2\underset{\mathbf{x}}{\sum}\left(i\psi^\dagger_{\mathbf{x}} \psi^\dagger_{\mathbf{x}+\hat{e}_1} + \psi^\dagger_{\mathbf{x}} \psi^\dagger_{\mathbf{x}+\hat{e}_2} + {\rm H.c.} \right)
\end{equation}

If one applies the particle-hole transformation $\psi_{\mathbf{x}}^{\dagger} \rightarrow \psi_{\mathbf{x}}$ on the odd sites,
the previous Hamiltonian is transformed into:
\begin{multline} \label{hamdiracreal}
 H =\left(1-4t^4 \right) \sum_\mathbf{x} \left(-1\right)^{x_1+x_2}\psi^\dagger_{\mathbf{x}} \psi_{\mathbf{x}} +
  t^4 \underset{\mathbf{x}}{\sum}\left(-1\right)^{x_1+x_2} \left(\psi^\dagger_{\mathbf{x}+2\hat{e}_1} \psi_{\mathbf{x}} + \psi^\dagger_{\mathbf{x}+2\hat{e}_2} \psi_{\mathbf{x}}  + {\rm H.c.}\right) \\
 -2t^2\sum_\mathbf{x}\left(i\psi^\dagger_{\mathbf{x}} \psi_{\mathbf{x}+\hat{e}_1} + \left(-1\right)^{x_1+x_2}\psi^\dagger_{\mathbf{x}} \psi_{\mathbf{x}+\hat{e}_2} + {\rm H.c.} \right).
\end{multline}
The first and last terms can easily be recognized as the staggered mass and hopping terms of the $2+1$ dimensional Hamiltonian for free Kogut-Susskind fermions \cite{Susskind1977}. In particular, we observe that the complex p-wave pairing gives rise to hopping phases that are consistent with the rotational requirements of the Kogut-Susskind model and constitute the remnants of the Dirac matrices in the continuum limit.

The second term, instead, does not belong to this well-known model; furthermore we observe that these next-nearest-neighbor tunnelings cannot be minimally coupled to a static gauge field on the links of the lattice.

The spectrum coincides, non-surprisingly, with the one previously calculated, which means that the ground state is a state at half filling. For $t=0$, all the particles occupy the odd sites (Dirac sea), corresponding to the vacuum state in the Kogut-Susskind Hamiltonian \cite{Susskind1977}. By increasing $t$, however, the even sites become progressively more and more populated. This corresponds to the formation of maximally localized pairs. Due to the form of $T$, indeed, the pairs cannot spread on the lattice, but they are composed of pairs of neighboring particles, in the superconducting picture, or by the displacement of single fermions from an odd to a neighboring even site for the number-conserving Hamiltonian \eqref{hamdiracreal}.
\\

\begin{center}
\emph{Gapless Regions of the Phase Diagram}
\end{center}

Let us start with a Taylor expansion of $\alpha\left(\mathbf{k}\right)$ and $\beta\left(\mathbf{k}\right)$ around $\mathbf{k} \to 0$,
\begin{equation}
\begin{aligned}
 \beta\left(\mathbf{k}\right) =& \beta_1 (k_1-ik_2) + \beta_3 k_1k_2(k_1+ik_2) +\beta_3'(k_1^3-ik_2^3) + O(k^5)\\
 \alpha\left(\mathbf{k}\right) =& \alpha_0 + \alpha_2 \left(k_1^2+k_2^2 \right) + O(k^4)  \,,
\end{aligned}
\label{betaseries}
\end{equation}
with
\begin{equation}
\begin{aligned}
 \beta_1&=2t^2\left(1+(y+z)^2\right) \left(1-(y-z)^2\right)^2 \\
 \beta_3&= -2i t^2 [4 y^4 - 2 y^3 z + z^2 + z^4 + y^2 (4 - 7 z^2) +  y (-6 z + 4 z^3)] \\
 \beta_3'&=\frac{t^2}{3} \left[ 2 y (-9 + 8 y^2 + y^4) z + (7 - 28 y^2 + y^4) z^2 +  (y^2-1)(1-y^4-4y z^3) + (7 + y^2) z^4 + z^5(2y-z)\right]
\end{aligned}
\end{equation}
and
\begin{equation}
\begin{aligned}
 \alpha_0&=\left(1-(y+z)^2 \right)^2\left(1-(y-z)^2 \right)^2\\
 \alpha_2&=2\left[2 y^6 + z^2 (-1 + z^2)^2 + 2 y^2 (1 + z^2) - y^4 (4 + 3 z^2) \right].
 \end{aligned}
\end{equation}

For $\alpha_0\neq 0$ we recover that $g\left(\mathbf{k}\right) \propto (k_1-ik_2)$, therefore the system is in a strongly paired gapped phase with the Cooper pair wavefunction, proportional to $\hat{g}\left(\mathbf{x}\right)$, decaying exponentially in real space \cite{Read2000}.
We distinguish the behavior of the gapless lines according to the values of the $\alpha_i,\beta_i$ coefficients, and identify five different regions:
\begin{enumerate}
  \item \emph{Strong Pairing Gapless Lines (A)} : $z=y \pm 1$, except for the points $(y=0,z=\pm1)$ and $(y=\pm1,z=0)$.
  Here both $\alpha_0$ and $\beta_1$ disappear. In this case the leading term of $g(\mathbf{k}\to 0)$ is of order $O(k)$, corresponding, at large distances, to a real space pairing function dominated by $|\hat{g}(\mathbf{x})|=1/|\mathbf{x}|^3$. Since the average distance between paired fermions is finite, this is again consistent with a picture of localized Cooper pairs, thus with a strong-coupling regime. Furthermore, in this case too, $u\left(\mathbf{k}\right) \rightarrow 1$ for $\mathbf{k} \rightarrow 0$, and the Chern number associated to the spinor is trivial.

  The dispersion relation for $\mathbf{k} \rightarrow 0$ is expanded, in this case, as
    \begin{equation}
    E\left(\mathbf{k}\right) \approx \alpha_1^2 k^4
    \end{equation}
    - a quartic dispersion in terms of $k=\left|\mathbf{k}\right|$.

  \item \emph{Weak Pairing Gapless Lines (B)}: $z=-y \pm 1$, except for the intersection points $(y=0,z=\pm1)$ and $(y=\pm1,z=0)$.
  Here, $\alpha_0=0,\beta_1\neq0$. In this case the leading behavior of $g\left(\mathbf{k}\right)$ is proportional to $1/(k_1+ik_2)$, since the linear term $\beta_1$ does not vanish. Therefore $\hat{g}(\mathbf{x})\propto(x_1+ix_2)^{-1}$ for large distances and the system is, indeed, in a weak pairing p-wave gapless phase. In particular, since $\beta$ dominates over $\alpha$ for small momenta, we obtain $v\left(\mathbf{k}\right)\rightarrow 1$ and $u\left(\mathbf{k}\right)\rightarrow 0$, and the Chern number becomes non-trivial (-2).

    In this phase, the behavior of the band touching points is described by
    \begin{equation}
    E\left(\mathbf{k}\right) \approx \beta_1^2 k^2
    \end{equation}
    - a quadratic dispersion relation.
    Furthermore, close to $\mathbf{k}=0$, the leading behavior of the superconducting order parameter is given by
    \begin{equation}
    \Delta_0 \propto \alpha_2\beta_1 k^2 (k_1-ik_2)
    \end{equation}
    and therefore the dominant pairing is of the $p_1-ip_2$ type.

  \item \emph{The Gapless branches of the hyperbola } $y^2-z^2=1$ \emph{(C)}, except for the points $(y=\pm1,z=0)$.
   For $y^2-z^2=1$ the band touching points are $\mathbf{k}=(0,\pi),(\pi,0)$. Considering the expressions for $\beta$ in Eq. \eqref{beta}, it is easy to see that a translation of $k_2$ (or, analogously $k_1$) by $\pi$ in momentum space corresponds to a transformation from a $p_x-ip_y$ superconductor to a $p_x+ip_y$ superconductor. Such a translation simply corresponds to the mapping $\psi_\mathbf{x}\to (-1)^{x_2} \psi_\mathbf{x}$  which does not affect  $\left|\hat{g}(\mathbf{x})\right|$ in real space.

    Let us consider the band touching point at $(0,\pi)$ for $y^2-z^2=1$ and expand $\alpha$ and $\beta$ in series of the momenta around that point (we define $\tilde{k}_y=k_y-\pi$):
    \begin{equation}
    \begin{aligned}
    &\alpha\left(\mathbf{k}\right) \approx z^4(k_1^4+\tilde{k}_2^4) + 2 (8 + 8 z^2 + z^4)k_1^2\tilde{k}_2^2 \,,\\
    &\beta\left(\mathbf{k}\right) \approx -4 i t^2 (  yz-z^2-1)  (z^2+4)  k_1\tilde{k}_2(k_1-i\tilde{k_2}) + 4 t^2 z^2 (1 - y z + z^2)\left(k_1^3+i\tilde{k}_2^3 \right),
    \end{aligned}
    \end{equation}
    where the first terms of the series vanish due to $y^2-z^2=1$.
    With the exception of the points $y=\pm 1,z=0$, the pairing function $g(\mathbf{k})$ around the band-touching points has always a non-analytical behavior of the kind $1/k$, therefore these critical hyperbola branches describe weak pairing gapless phases.

    The leading behavior of the dispersion relation close to the band touching points $\mathbf{k}=(0,\pi),(\pi,0)$ is of order  $k^6$. The Chern number along the hyperbola is nontrivial again (+2).

  \item \emph{Weak Coupling Intersection Points}, $(y=0,z=\pm1)$, \emph{(D)}.
  Here, $\alpha_0=\alpha_2=\beta_1=0$, and  therefore the leading term of $g(\mathbf{k})$ is of order $O(1/k)$. This brings the system into a weak pairing gapless phase where the wavefunctions of the Cooper pairs decay as the inverse of the distance in real space. The Chern number here is calculated to be -2.

  \item \emph{Weak Coupling Intersection Points}, $(y=\pm1,z=0)$, \emph{(E)}.
  These are the intersections between the critical straight lines and hyperbola branches.  These are the only points  for which the gap closes at both $\mathbf{k}=(0,0)$ and $\mathbf{k}=(0,\pi)$ since $\alpha_0=\alpha_\pi=0$. We focus on the case $(y=1,z=0)$: The behavior of the other point, $(y=-1,z=0)$ is similar.

  The functions $\alpha\left(\mathbf{k}\right)$ and $\beta\left(\mathbf{k}\right)$ are, in this case,
  \begin{equation}
  \begin{aligned}
  \alpha\left(\mathbf{k}\right)&=16\sin^2k_1 \sin^2k_2\,,\\
  \beta\left(\mathbf{k}\right)&=16 t^2 \sin k_1 \sin k_2 (\sin k_2 - i \sin k_1) \,,\\
  \end{aligned}
  \end{equation}

  Due to their common factor $16 \sin k_1 \sin k_2$, the Hamiltonian assumes the form
  \begin{equation}
  \mathcal{H}\left(\mathbf{k}\right) = 64 \left(\sin^2 k_1 \sin^2 k_2\right) \left(\tilde{R}_0\left(\mathbf{k}\right) \sigma_x + \tilde{I}_0\left(\mathbf{k}\right) \sigma_y + \tilde{P}_0\left(\mathbf{k}\right)\sigma_x \right) \equiv \left(\sin^2 k_1 \sin^2 k_2\right) \mathcal{\tilde H}\left(\mathbf{k}\right)
  \end{equation}
  with
  \begin{equation}
  \begin{aligned}
  &\tilde{R}_0\left(\mathbf{k}\right)=1-4t^4+(2t^4-1)\left(\cos2k_1 + \cos 2k_2 \right) + \frac{1}{2}\left[\cos\left(2k_1+2k_2 \right)+\cos\left(2k_1-2k_2\right)   \right]  \,,\\
  &\tilde{I}_0\left(\mathbf{k}\right)= 4t^2\sin k_2 -2t^2\left[\sin\left(2k_1+k_2 \right)+\sin\left(k_2-2k_1 \right)   \right] \,,\\
  &\tilde{P}_0\left(\mathbf{k}\right)= 4t^2\sin k_1 -2t^2\left[\sin\left(2k_2+k_1 \right)+\sin\left(k_1-2k_2 \right)   \right]
  \end{aligned}
  \end{equation}

  The spectrum of $\mathcal{\tilde H}\left(\mathbf{k}\right)$ is
  \begin{equation}
  \tilde{E}=\pm \left[ 1+4t^4-(1+2t^4)\left(\cos(2k_1) + \cos(2k_2) \right) + \cos(2k_1)\cos(2k_2) \right]
  \end{equation}
  which is always gapless with quadratic band touching points in both $\mathbf{k}=(0,0)$ and $\mathbf{k}=(0,\pi)$ for every value of $t$. Therefore, the leading term of the spectrum of $\mathcal{H}\left(\mathbf{k}\right)$  for $\mathbf{k} \rightarrow 0$ satisfies
  \begin{equation}
  E\approx 256t^4k_1^2k_2^2(k_1^2+k_2^2)
  \end{equation}
  and an analogous behavior is found close to $(0,\pi)$.

  Let us now evaluate the pairing function:
  \begin{equation}
  g(\mathbf{k})= \frac{t^2\left(\sin k_2 -i \sin k_1 \right) }{\sin k_1 \sin k_2}=\frac{t^2}{\sin k_1} - i\frac{t^2}{\sin k_2}.
  \end{equation}
  This corresponds to a weak pairing gapless phase (independent of the considered minima). We observe that, since $g(\mathbf{k})$ splits into $g_1(k_1) + g_2(k_2)$ we obtain, as expected in the case $z=0$, that $\hat{g}(\mathbf{x})=\hat{g}_1(x_1)\delta_{x_2,0} + \hat{g}_2(x_2)\delta_{x_1,0}$; in particular $\hat{g}(\mathbf{x})\propto -it^2(-1)^{x_1}\delta_{x_2,0} - t^2(-1)^{x_2}\delta_{x_1,0}$ for $\mathbf{x}\neq(0,0)$.  Therefore the Cooper pairs, which correspond to the mesons in the gauge theory, do not decay with the distance between particle and antiparticle but are constrained to spread only in the horizontal or vertical direction.
  The Chern number at these points is trivial.

\end{enumerate}

The phase diagram is summarized in Fig. \ref{figphase}.

\section{Local Gauge Symmetry} \label{sec:local}

So far, we have introduced purely fermionic states $\left|\psi \left(T\right)\right\rangle$ which, on top of the spacetime symmetries corresponding to translation and rotation invariance, are invariant under \emph{global} $U(1)$ transformations. Now we wish to lift the global symmetry to be \emph{local}, i.e. instead of the global transformation rule (\ref{globtrans}), with a global transformation parameter (phase) $\phi$, we would like to have a local transformation rule, with vertex-dependent parameters (phases) $\phi_{\mathbf{x}}$,
\begin{equation}
\psi^{\dagger}_{\mathbf{x}} \rightarrow e^{i s_{\mathbf{x}} \phi_\mathbf{x}}\psi^{\dagger}_{\mathbf{x}}.
\label{loctrans}
\end{equation}
where $s_\mathbf{x}=(-1)^{x_1+x_2}$ accounts for the staggered charge of the fermions.
In general, the states $\left|\psi \left(T\right)\right\rangle$ defined in the previous section will not be invariant under such a local gauge transformation and need to be modified. This modification is necessary because the effect of the transformation of the phase of a single vertex \eqref{loctrans} must be compensated by the action on some additional degrees of freedom which cannot depend on other vertices if the symmetry is local. Geometry and locality considerations lead us to introduce, as customary in lattice gauge theories, new degrees of freedom on the links connecting the vertices, which participate in the local gauge transformations in a way that conserves the local charge (which will be the eigenstate of a \emph{physical} Gauss law).

\begin{figure}
  \centering
  \includegraphics[scale=0.1]{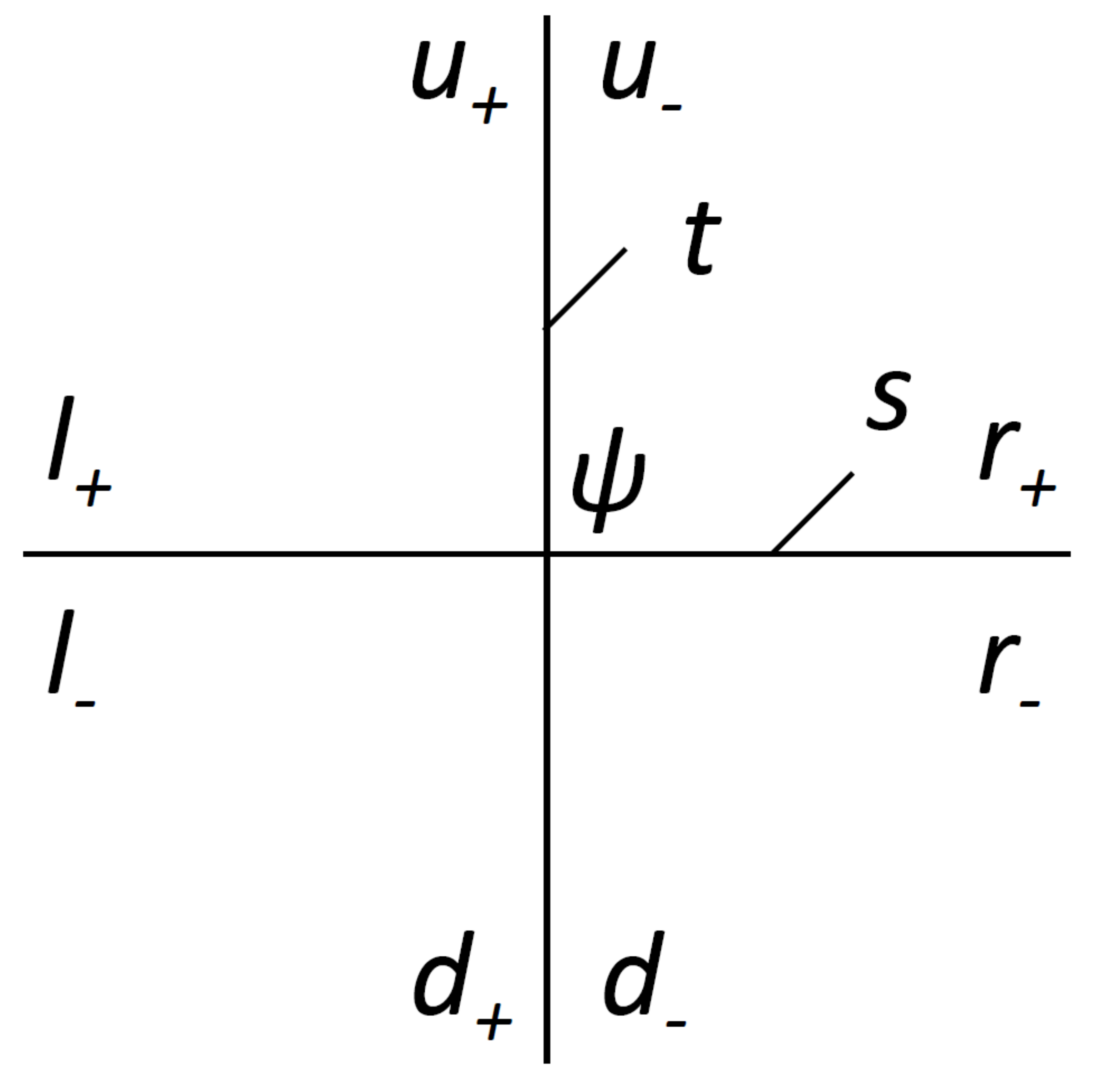}
	\caption{The Hilbert space on a vertex (composing the fiducial state), in the case of a locally gauge invariant state: a single physical fermion $\psi$ and two physical bosonic states, $s$ and $t$, with eight virtual fermions surrounding it, two on each edge intersecting at the vertex.}
  \label{fig11}
\end{figure}

Motivated by compact QED (cQED) \cite{KogutLattice}, we introduce a bosonic Hilbert space on each link, and follow a procedure similar to \cite{Zohar2015a}. These Hilbert spaces shall be denoted as  $\mathcal{H}_{\mathbf{x}}^s$ and $\mathcal{H}_{\mathbf{x}}^t$, representing the links on the right (\emph{s}ide) and above (\emph{t}op) the vertex $\mathbf{x}$ (see Figure \ref{fig11}). As we wish these Hilbert spaces to be finite, we shall work with those appearing in truncated cQED \cite{Zohar2013,AngMom}: the Hilbert spaces $\mathcal{H}_{\mathbf{x}}^{t/s}$ are spanned by the $SU(2)$ states $\left\{\left|m\right\rangle\right\}_{m=-\ell}^{\ell} \equiv \left\{\left|\ell m\right\rangle\right\}_{m=-\ell}^{\ell}$ for a given, fixed integer $\ell$ (full cQED  is obtained for $\ell \rightarrow \infty$ \cite{AngMom}).

These states are eigenstates of the \emph{electric field}, $\Sigma = L_z$. We define the raising and lowering operators
\begin{equation}
\Sigma_{\pm} = \frac{L_{\pm}}{\sqrt{\ell\left(\ell+1\right)}}
\end{equation}
having the $U(1)$ limit
\begin{equation} \label{vectorpotential}
\Sigma_{\pm} \left|m\right\rangle \underset{\ell \rightarrow \infty}{\longrightarrow} e^{\pm i \theta}\left|m\right\rangle = \left|m+1\right\rangle
\end{equation}
where $\theta$ is the cQED ``vector potential'' conjugate to the electric field.

Therefore, for each lattice vertex $\mathbf{x}$, we consider the Hilbert space of the states of both the matter fermion sitting on the vertex and the two gauge bosons on its right and top links.
Within this extended space, we have to consider the symmetry requirements associated to the new, locally gauge invariant states, $\left|\psi_b\right\rangle$, which will generalize the symmetries fulfilled by the purely fermionic states $\left|\psi\right\rangle$.

\begin{enumerate}
  \item \emph{Charge Conjugation Symmetry.} This is a generalization of translational invariance due to the effect of the staggered charge of the fermions. A translation from one sublattice to the other   corresponds to a charge conjugation that has to affect the gauge bosons as well. Therefore we extend the action of the $U_T$ operators defined for the fermions (\ref{Transop}) to include the bosonic operators acting on $\mathcal{H}^s$ and $\mathcal{H}^t$ as well
    \begin{equation}
    \begin{aligned}
    & U_T\left(\mathbf{\hat{e}}_k\right) \Sigma_{\pm}^{s/t}\left( \mathbf{x}\right) U_T^{\dagger}\left(\mathbf{\hat{e}}_k\right) = \Sigma_{\mp}^{s/t}\left( \mathbf{ x + \hat e}_k\right),\\
    & U_T\left(\mathbf{\hat{e}}_k\right) \Sigma^{s/t}\left( \mathbf{x}\right) U_T^{\dagger}\left(\mathbf{\hat{e}}_k\right) = -\Sigma^{s/t}\left( \mathbf{ x + \hat e}_k\right).
    \end{aligned}
    \label{boscc}
    \end{equation}
   Such a definition of $U_T$ introduces, in fact, a charge conjugation: it both exchanges matter particles and anti-particles, and inverts the sign of the electric field. We wish it to be a symmetry, and thus we demand
    \begin{equation}
    U_T\left(\mathbf{\hat{e}}_k\right) \left|\psi_b\left(\left\{t_i\right\}\right)\right\rangle = \left|\psi_b \left(\left\{t_i\right\}\right)\right\rangle.
    \end{equation}

  \item \emph{Rotational Invariance.} The lattice rotation has to affect both the fermionic vertices and the bosonic links. We extend the fermionic rotation (\ref{physrot}) with:
  \begin{equation}
  \begin{aligned}
    &U_p\left(\varLambda\right) \Sigma_{\pm}^{s}\left( \mathbf{x}\right) U_p^{\dagger}\left(\varLambda\right) = \Sigma_{\pm}^{t}\left(\varLambda \mathbf{ x}\right) \\
    &U_p\left(\varLambda\right) \Sigma_{\pm}^{t}\left( \mathbf{x}\right) U_p^{\dagger}\left(\varLambda\right) = \Sigma_{\pm}^{\bar s}\left(\varLambda \mathbf{ x}\right) \\
    &U_p\left(\varLambda\right) \Sigma^{s/t}\left( \mathbf{x}\right) U_p^{\dagger}\left(\varLambda\right) = \Sigma^{s/t}\left(\varLambda \mathbf{ x}\right)
    \end{aligned}
    \label{bosrot}
    \end{equation}
    where
    \begin{equation}
    \begin{aligned}
    \Sigma_{\pm}^{\bar s}\left( \mathbf{x}\right) & \equiv \Sigma_{\pm}^{ s}\left( \mathbf{x-\hat e}_1\right) \\
    \Sigma_{\pm}^{\bar t}\left( \mathbf{x}\right) & \equiv \Sigma_{\pm}^{ t}\left( \mathbf{x-\hat e}_2\right)
    \end{aligned}
    \end{equation}
    and the symmetry requirement reads
    \begin{equation}
    U_p\left(\varLambda\right) \left|\psi_b\left(\left\{t_i\right\}\right)\right\rangle = \left|\psi_b\left(\left\{t_i\right\}\right)\right\rangle.
    \end{equation}

\item \emph{Local Gauge Invariance.} We finally define the local $U(1)$ gauge transformations, generated by the \emph{Gauss law operators},
\begin{equation} \label{physgauss}
\begin{aligned}
G_{\mathbf{x}} &=  \Sigma^s\left(\mathbf{x}\right) + \Sigma^t\left(\mathbf{x}\right) - \Sigma^{\bar s}\left(\mathbf{x}\right) - \Sigma^{\bar t}\left(\mathbf{x}\right) - Q_{\mathbf{x}}  \\
&= \Sigma^s\left(\mathbf{x}\right) + \Sigma^t\left(\mathbf{x}\right) - \Sigma^s\left(\mathbf{x-\hat e}_1\right) - \Sigma^t\left(\mathbf{x-\hat e}_2\right) - Q_{\mathbf{x}}.
\end{aligned}
\end{equation}
This generator extends Eq. \eqref{fiducialG} to the gauge degrees of freedom. As one can see, the virtual electric fields $E_i$ from the previous section have been lifted to physical degrees of freedom, and the \emph{virtual} local Gauss's laws which defined the fermionic fiducial states with a global physical symmetry, have turned into \emph{physical} local conservation laws.
Therefore the physical symmetry we demand for the locally gauge invariant state is
\begin{equation}
e^{i \phi_{\mathbf{x}} G_{\mathbf{x}}} \left|\psi_b \left(\left\{t_i\right\}\right)\right\rangle = \left|\psi_b \left(\left\{t_i\right\}\right)\right\rangle \quad \forall \mathbf{x}
\label{loclaw}
\end{equation}

\end{enumerate}

Hereafter we focus on the case $\ell = 1$ which constitutes the simplest truncation scheme allowing for zero, positive and negative electric fluxes on the links. In this case the electric field states are simply $\left|0\right\rangle, \left|\pm 1\right\rangle$, and
\begin{equation}
\Sigma_{+}=\left(
                \begin{array}{ccc}
                  0 & 1 & 0 \\
                  0 & 0 & 1 \\
                  0 & 0 & 0 \\
                \end{array}
              \right)
              ; \quad
\Sigma_{-}=\left(
                \begin{array}{ccc}
                  0 & 0 & 0 \\
                  1 & 0 & 0 \\
                  0 & 1 & 0 \\
                \end{array}
              \right)
\end{equation}
in the basis $\left\{\left|+1\right\rangle,\left|0\right\rangle,\left|-1\right\rangle\right\}$.

In the following, we will present a recipe to transform  the globally-gauge-invariant PEPS of section \ref{sec:global} into a locally-gauge-invariant one, and show that the symmetries are obtained using the same parametrization presented for the purely fermionic case:  one can parametrize $\left|\psi_b\right\rangle$ with the same matrix $T\left(t,y,z\right)$ introduced for the
 fermionic Gaussian state $\left|\psi\right\rangle$. However, the bosonic state will not be Gaussian any longer, reflecting the physics of the underlying interacting theory.  The reader could also refer to other PEPS constructions, as in \cite{Tagliacozzo2014} for pure-gauge theories, or \cite{Haegeman2014} for gauge theories coupled to bosonic (Higgs) fields.

\subsection{Inclusion of Gauge Bosons in the PEPS}
Following the steps of the construction of the fermionic $\left|\psi\right\rangle$, we start by imposing the local gauge invariance for the fiducial state.
On each vertex we define a fiducial state
\begin{equation}
\left|F_b\left(\mathbf{x}\right)\right\rangle = A_b\left(\mathbf{x}\right)\left|\Omega_p \left(\mathbf{x}\right)\right\rangle
\left|0_s \left(\mathbf{x}\right)\right\rangle \left|0_t \left(\mathbf{x}\right)\right\rangle \left|\Omega_v \left(\mathbf{x}\right)\right\rangle
\equiv A_b\left(\mathbf{x}\right)\left|\Omega \left(\mathbf{x}\right)\right\rangle.
\end{equation}
with the operator $A_b$ yet to be defined. This will be a state involving one physical fermion at the vertex, and two bosonic degrees of freedom corresponding to the right and top links connected with the vertex, as shown in Fig. \ref{fig11}.
The corresponding PEPS takes the form:
\begin{equation}
\left|\psi_b\right\rangle =
\left\langle \Omega_v \right|\underset{\mathbf{x}}{\prod}\omega\left(\mathbf{x}\right)\eta\left(\mathbf{x}\right)A_b\left(\mathbf{x}\right)\left|\Omega\right\rangle
\label{locPEPS}
\end{equation}
This physical state is analogous to the fermionic PEPS $\ket{\psi}$ in (\ref{globPEPS}), with the same projectors $\omega\left(\mathbf{x}\right),\eta\left(\mathbf{x}\right)$, defined by (\ref{projdef}).

\begin{sttmnt}\label{lemma:local}
The state $\left|\psi_b\right\rangle$ is locally gauge invariant, i.e. it fulfills (\ref{loclaw}), if the following conditions on the fiducial state are met:
\begin{enumerate}[I.]
\item $U_V^d U_V^l \left|F_b\right\rangle = U_t U_s U_{\psi}^{\dagger} \left|F_b\right\rangle$
\item $U_V^r \left|F_b\right\rangle = U_s \left|F_b\right\rangle$
\item $U_V^u \left|F_b\right\rangle = U_t \left|F_b\right\rangle$
\end{enumerate}
where $U_\psi=e^{i\phi s_\mathbf{x} \psi^\dag_\mathbf{x}\psi^{\phantom{\dag}}_\mathbf{x}}$ and the virtual operators $U_V^j=e^{i\phi E_j}$ are defined the same as for the fermionic theory, whereas $U_s = e^{i \Sigma^s}$ and $U_t = e^{i \Sigma^t}$ act on the physical bosons. The previous relations can be rephrased for the $A_b$ operators as:
\begin{enumerate}[i.]
\item{$U_V^d U_V^l A_b U_V^{l \dagger} U_V^{d \dagger} = U_t U_s U_{\psi}^{\dagger} A_b U_{\psi} U_s^{\dagger} U_t^{\dagger}$}
\item{$U_V^r A_b U_V^{r \dagger}= U_s A_b U_s^{\dagger}$}
\item{$U_V^u A_b U_V^{u \dagger} = U_t A_b U_t^{\dagger}$}
\end{enumerate}
\end{sttmnt}

\emph{Proof}:
Local gauge invariance, written with the operators of the statement, is
\begin{equation}
U_{s,\mathbf{x}}U_{t,\mathbf{x}}U^{\dagger}_{\bar s,\mathbf{x}}U^{\dagger}_{\bar t,\mathbf{x}}U^{\dagger}_{\psi,\mathbf{x}}\left|\psi_b\right\rangle =
  U_{s,\mathbf{x}}U_{t,\mathbf{x}}U^{\dagger}_{s,\mathbf{x-\hat e}_1}U^{\dagger}_{ t,\mathbf{x-\hat e}_2}U^{\dagger}_{\psi,\mathbf{x}}\left|\psi_b\right\rangle = \left|\psi_b\right\rangle
\end{equation}
for any $\mathbf{x}$.
Using the PEPS definition (\ref{locPEPS}), we find out that if the properties of the statement are fulfilled, then
\begin{equation}
\begin{aligned}
e^{i \phi G_{\mathbf{x}}}\left|\psi_b\right\rangle & =
 \left\langle \Omega_v \right|\underset{\mathbf{y}}{\prod}\omega\left(\mathbf{y}\right)\eta\left(\mathbf{y}\right)
U_s\left(\mathbf{x}\right) U_t\left(\mathbf{x}\right)
U_s^{\dagger}\left(\mathbf{x-\hat e}_1\right) U_t^{\dagger}\left(\mathbf{x-\hat e}_2\right) U_{\psi}^{\dagger}\left(\mathbf{x}\right) \left|F_b\left(\mathbf{y}\right)\right\rangle \\
 & =\left\langle \Omega_v \right|\underset{\mathbf{y}}{\prod}\omega\left(\mathbf{y}\right)\eta\left(\mathbf{y}\right)
U_V^l\left(\mathbf{x}\right) U_V^d\left(\mathbf{x}\right)
U_V^{r \dagger}\left(\mathbf{x-\hat e}_1\right) U_V^{u \dagger}\left(\mathbf{x-\hat e}_2\right) U_{\psi}^{\dagger}\left(\mathbf{x}\right) \left|F_b\left(\mathbf{y}\right)\right\rangle = \left|\psi_b\right\rangle
\end{aligned}
\end{equation}
where the second equality is due to to statement's conditions, and the third one due to the local invariance of the projectors under the transformation (\ref{prjinv}). $\square$

In order to guarantee the local gauge invariance of $\left|\psi_b\right\rangle$, we have to construct a fiducial state fulfilling the conditions (I-III). To this purpose, we will adopt the following construction: we consider the fermionic state $\left|\psi\left(T\right)\right\rangle$, and lift the virtual degrees of freedom to be physical in order to enforce the local symmetry defined by the physical generator $G_\mathbf{x}$ in \eqref{physgauss}, by, as we have mentioned, lifting the virtual electric fluxes of $G_0$ (as in \eqref{Gglobgen}) into physical ones. For that, to define $A_b$, we multiply the virtual fermionic operators of the right and up edges by physical bosonic operators as follows:
\begin{equation}
\left\{
  \begin{array}{l}
 r_+^{\dagger}  \rightarrow  \Sigma_{+}^{s} r_+^{\dagger} \\
 r_-^{\dagger}  \rightarrow  \Sigma_{-}^{s} r_-^{\dagger} \\
 u_+^{\dagger}  \rightarrow  \Sigma_{+}^{t} u_+^{\dagger} \\
 u_-^{\dagger}  \rightarrow  \Sigma_{-}^{t} u_-^{\dagger}
  \end{array}
\right.
\label{reps}
\end{equation}

Let us verify that the properties of statement \ref{lemma:local} are fulfilled. Properties ii and iii are easily verified: Since
both $U_V^r r^{\dagger}_{\pm} U_V^{r \dagger} = e^{\pm i \phi}r^{\dagger}_{\pm}$ and
$U_s \Sigma^{s}_{\pm} U_s^{\dagger} = e^{\pm i \phi}\Sigma_{\pm}$, one obtains that
$U_V^r r^{\dagger}_{\pm}\Sigma^{s}_\pm  U_V^{r \dagger}  = U_s r^{\dagger}_{\pm}\Sigma^{s}_\pm U_s^{\dagger}$ and Property ii is deduced. Property iii may be similarly shown.

The first property requires more care. Using the Gaussian construction of the fermionic part, according to statement \ref{th:gauss}, we have
$e^{i \phi G_0}\left|F_b\right\rangle=\left|F_b\right\rangle$, where $G_0$ is the local fermionic virtual Gauss operator defined in (\ref{fiducialG}).
Therefore $U_V^d U_V^l \left|F_b\right\rangle = U_V^r U_V^u U_\psi^{\dagger} \left|F_b\right\rangle$. Using properties ii and iii of statement \ref{lemma:local}, which have already been proven, one obtains property i.

\subsubsection{Rotational Invariance and Charge Conjugation}
The next symmetry we check is the rotational invariance.

\begin{sttmnt}
The states $\left|\psi_b\left(T\right)\right\rangle$, obtained with the same parametrization as of $\left|\psi\left(T\right)\right\rangle$ by making the replacements (\ref{reps}), are rotationally invariant.
\end{sttmnt}

\emph{Proof}: Combining the fermionic transformation rules (\ref{physrot}),(\ref{virtrot}) with the bosonic ones (\ref{bosrot}), one gets
\begin{equation}
U_{p}\left(\varLambda\right)U_{R}\left(\varLambda\right) A_b\left(\mathbf{x}\right) U^{\dagger}_{p}\left(\varLambda\right)U^{\dagger}_{R}\left(\varLambda\right) = A_b'\left(\varLambda\mathbf{x}\right)
\end{equation}
where $A_b'$ is an operator involving a bosonic Hilbert space attached to the left edge rather than to the right, i.e. with physical states $t$ and $\bar s$.

However, thanks to the horizontal projectors $\omega$, we can convert $A_b'$ into $A_b$:
\begin{equation}
\left\langle \Omega_{v\text{Row}} \right| \underset{\mathbf{x}\in\text{Row}}{\prod} \omega\left(\mathbf{x}\right)\underset{\mathbf{x}\in\text{Row}}{\prod} A_b'\left(\mathbf{x}\right)\left|\Omega_{v\text{Row}}\right\rangle
=\left\langle \Omega_{v\text{Row}} \right| \underset{\mathbf{x}\in\text{Row}}{\prod} \omega\left(\mathbf{x}\right)\underset{\mathbf{x}\in\text{Row}}{\prod} A_b\left(\mathbf{x}\right)\left|\Omega_{v\text{Row}}\right\rangle
\end{equation}
and similarly in the vertical direction. Thus we may conclude that, once $T$ is parameterized properly,
\begin{equation}
U_p\left(\varLambda\right) \left|\psi_b\left(T\right)\right\rangle = \left|\psi_b\left(T\right)\right\rangle
\end{equation}
which proves the statement. $\square$

The last remaining symmetry is charge conjugation. We have already defined this transformation as a generalization of the fermionic translation. Combining the fermionic translation
(\ref{Transop}) with the bosonic electric field inversion (\ref{boscc}) may be straightforwardly understood as charge conjugation, i.e., both exchanging particles and anti-particles (by the translation) and the sign of the electric field. The fermionic part is guaranteed using the parametrization introduced before. All these results lead us to the final statement,

\begin{sttmnt}
 The state $\left|\psi_b\left(T\right)\right\rangle$, defined by making the replacements (\ref{reps}) in the fermionic state
$\left|\psi\left(T\right)\right\rangle$ of statement \ref{th:Tmatrix}, having the same parametrization matrix $T=T\left(t,y,z\right)$ with $t>0,y,z\in\mathbb{C}$ as in Eq. (\ref{Tmatrix}), has a local $U(1)$ gauge invariance,
as well as rotational invariance and charge conjugation symmetry.
\end{sttmnt}

\subsection{The transfer matrix associated with the PEPS}

For the numerical implementation of the locally-gauge invariant PEPS we adopt a cylindrical geometry, with periodic boundary conditions in the $\hat{\mathbf{e}}_1$ direction and open boundary conditions in the $\hat{\mathbf{e}}_2$ direction, consistent with having all the bosonic states on the lower boundary set to $\ket{0}_b$. We label by $L_{1}$ and $L_{2}$ the numbers of horizontal and vertical matter sites of the lattice, respectively.
We follow the construction presented in \cite{Yang2015} for the study of chiral fermionic systems in order to study the PEPS $\left|\psi_b\right\rangle$ in \eqref{locPEPS} . In the following we present some of its main features. In particular, in our case, the presence of two virtual fermionic modes per bond implies a bond dimension 4.
As already discussed, the state $\left|\psi_b\right\rangle$ is obtained starting from the product of all the fiducial states $\ket{F_b}$ on which the projectors $\omega$ and $\eta$ are applied.
One may define, for the whole lattice, the state:
\begin{equation}
 \ket{\mathcal{F}} \equiv
 \prod_{x_2=1}^{L_2} \left[\left( \prod_{x_1=1}^{L_1} \eta(\mathbf{x})   \omega(\mathbf{x})  \right) \prod_{x_1=1}^{L_1} A_b(\mathbf{x})
 \bigotimes_{x_1=1}^{L_1} \ket{\Omega(\mathbf{x})}\right]
\end{equation}
This state involves all the virtual and physical states in the lattice.

The expectation value of a physical observable $O_y$ supported on the $y$-th row can be expressed as:
\begin{equation} \label{expectation}
 \left\langle O_y \right\rangle = {\rm tr}\left[X_f O_y X_i \ket{\mathcal{F}}\bra{\mathcal{F}} \right]
\end{equation}
where $X_i=\prod_{x_1}\ket{\Omega_d(x_1,1)}\bra{\Omega_d(x_1,1)}$ is simply a projector, acting on the first row of the virtual states associated to the modes $d_{\pm}(x_2=1)$, that fixes the boundary conditions in such a way that there is no ingoing virtual electric flux $E_d$ in the first row of the cylinder, which can also be interpreted as the additional presence of an initial row of gauge bosons in the state $\ket{0_b}$, and $X_f$ is a similar projector setting the boundary condition for the upper row.

To evaluate the previous expression, and in particular the density matrix $\ket{\mathcal{F}}\bra{\mathcal{F}}$, it is convenient to introduce the reduced density matrix $X(x_2)$ associated with the virtual $d_\pm$ modes of the row $x_2+1$. In this way, the evaluation of the expectation value $O_y$ can be performed by dividing the calculation into one step per row. The density matrix $X(x_2)$ can be defined iteratively starting from the corresponding density matrix in the previous row \cite{Yang2015},
\begin{multline}
 X(x_2)= {\rm tr}\left[\left(\prod_{x_1}\eta(x_1,x_2)\omega(x_1,x_2)A_b(x_1,x_2) \right)\left(\bigotimes_{x_1} \ket{\Omega}\bra{\Omega}(x_1,x_2)\right)\right.\cdot \\
 \left.\left( \prod_{x_1}A^\dag_b(x_1,x_2)\omega(x_1,x_2)\eta(x_1,x_2)\right)  X(x_2-1) \right]
\end{multline}
(note that the definition of trace in a fermionic Hilbert space is non-trivial, due to the nature of fermionic Hilbert spaces mentioned before, and thus has to be done with caution) where the projectors $\ket{\Omega}\bra{\Omega}(x_1,x_2)$ are on the vacuum of all its physical and virtual fermionic modes and on the $\ket{0}$ bosonic states associated to the site $(x_1,x_2)$. The trace is taken over all the physical fermionic and bosonic states in the row $x_2$, and over all the virtual fermionic modes, in such a way that $X(x_2)$ is indeed an operator acting on the virtual modes $d_{\pm}(x_1,x_2+1)$ through the projectors $\eta(x_1,x_2)$. We have exploited the fact that, as projectors, $\omega=\omega^\dag$ and $\eta=\eta^\dag$ (see \cite{Yang2015} for more details).

The previous equation describes a mapping between $X(x_2-1)$ and $X(x_2)$ that can be expressed in terms of a transfer matrix $\mathcal{T}$ defining the mapping between the virtual density matrices:
\begin{equation}
 X(x_2)_{d,d'}=\mathcal{T}^{\tilde{d},\tilde{d}'}_{d,d'}X(x_2-1)_{\tilde{d},\tilde{d}'}.
\end{equation}
In particular we have
\begin{equation} \label{transfer}
 \mathcal{T}= {\rm tr}_{p,t,s,r,l,u}\left[\left(\prod_{x_1}\eta \omega A_b \right)\left(\bigotimes_{x_1} \ket{\Omega}\bra{\Omega}\right)\left( \prod_{x_1}A^\dag_b\omega\eta\right)  \right]
\end{equation}
where the trace is taken over all the modes associated to the row $x_2$ with the exception of the virtual $d_\pm$ states, in such a way that $\mathcal{T}$ acts on both the $d$ virtual states of the rows $x_2$ and $x_2+1$.

We can now rephrase the expression for the expectation value \eqref{expectation} of a local observable $O_y$ in the following way:
\begin{equation}
 \left\langle O_y \right\rangle = \frac{{\rm tr}\left[  X_f \mathcal{T}^{L_2-y}\tilde{\mathcal{T}}_{O_y}\mathcal{T}^{y-1} X_i \right]}{{\rm tr}\left[  X_f \mathcal{T}^{L_2} X_i \right]}
\end{equation}
where we have exploited that the observable $O_y$ has support on the row $y$ only.
$\tilde{\mathcal{T}}_{O_y}$ is the specific transfer matrix associated to the row on which the
operator $O_y$ is acting. It is defined by introducing the physical observable $O_y$ within the definition of $\mathcal{T}$ in Eq.~\eqref{transfer}
\begin{equation} \label{transfer2}
 \tilde{\mathcal{T}}_{O_y}= {\rm tr}\left[O_y\left(\prod_{x_1}\eta \omega A_b \right)\left(\bigotimes_{x_1} \ket{\Omega}\bra{\Omega}\right)\left( \prod_{x_1}A^\dag_b\omega\eta\right)  \right]
\end{equation}
This way of evaluating the expectation values emphasizes the role of the transfer matrix $\mathcal{T}$ which is fully defined by the operators $A_b$ and can be numerically evaluated. The non-degeneracy of the maximal eigenvalue of $\mathcal{T}$ is related to the presence of a finite correlation length in the physical system.

It is believed that for PEPS a gap between the two highest eigenvalues of the transfer operator $\mc T$ indicates exponentially decaying correlations in real space. This has been shown rigorously for MPS~\cite{Fannes1992} and is based on the fact that the correlation between any two local operators scales as the second highest eigenvalue to the power of their distance (the highest eigenvalue has to be normalized to 1). This observation was also made in numerical studies with PEPS, though a rigorous proof is still awaited.

\subsection{The phase diagram of the locally gauge-invariant state}

The operators $A_b$ defined by the equations (\ref{eqA},\ref{Tmatrix}) with the introduction of the bosonic operators through the substitution \eqref{reps} completely define the locally gauge invariant state $\ket{\psi_b}$ as a function of the three parameters $t \geq 0$ and $y,z\in \mathbb{C}$. In particular, the parameter $t$ couples  the physical matter fermions with the virtual and bosonic modes. For $t=0$ no physical fermionic creation operators contribute to $A_b$, therefore $\ket{\psi_b}$ becomes a bosonic state of the kind $\ket{\psi_b}_{\rm links}\ket{\Omega_p}_{\rm vertices}$, non-trivial on the bosonic links in the lattice, whereas all the fermionic matter sites remain in their vacuum state $\ket{\Omega_p}$. This limit thus corresponds to the \emph{pure} $U(1)$ truncated lattice gauge theory defined in a locally gauge invariant sector without any static charges on the lattice vertices.

Due to this feature, in the case $t=0$ only bosonic operators and observables are meaningful and the theory can be restricted only to the bosonic Hilbert space.
For all the values $t > 0$, instead, the matter fermions appear in the definition of $\ket{\psi_b}$, and this sudden addition of fermions has to be related, at least intuitively, to some discontinuity or non-analyticity in the limit $t \rightarrow 0$. When we deal with the combination of dynamical fermions and gauge fields we shall address this issue.

\begin{figure*}[t]
\centering
\includegraphics[width=0.4\textwidth]{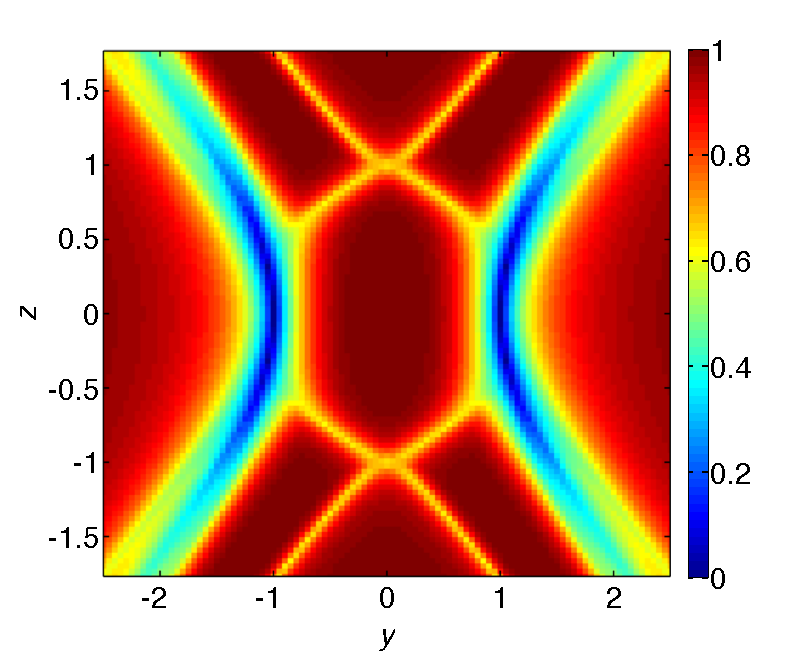} \ \includegraphics[width=0.4\textwidth]{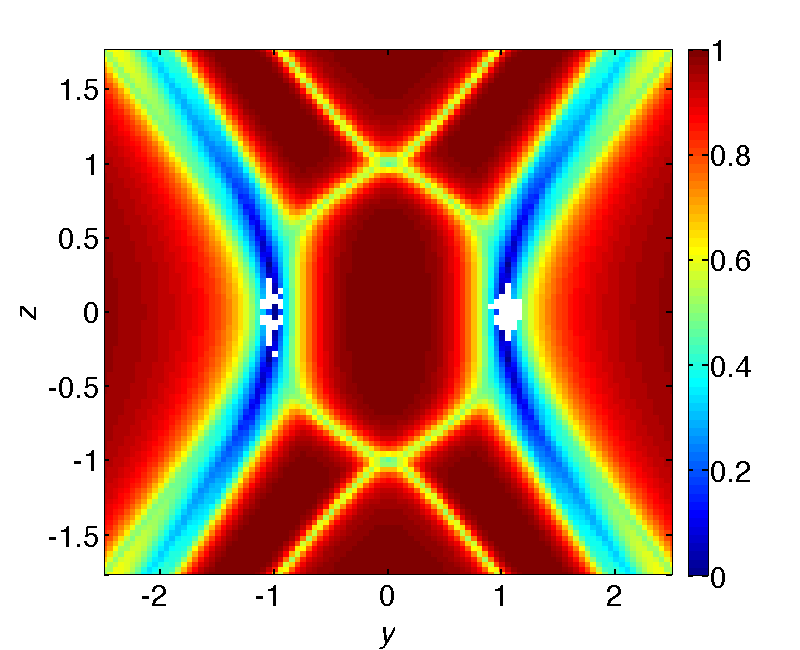} \\
\includegraphics[width=0.4\textwidth]{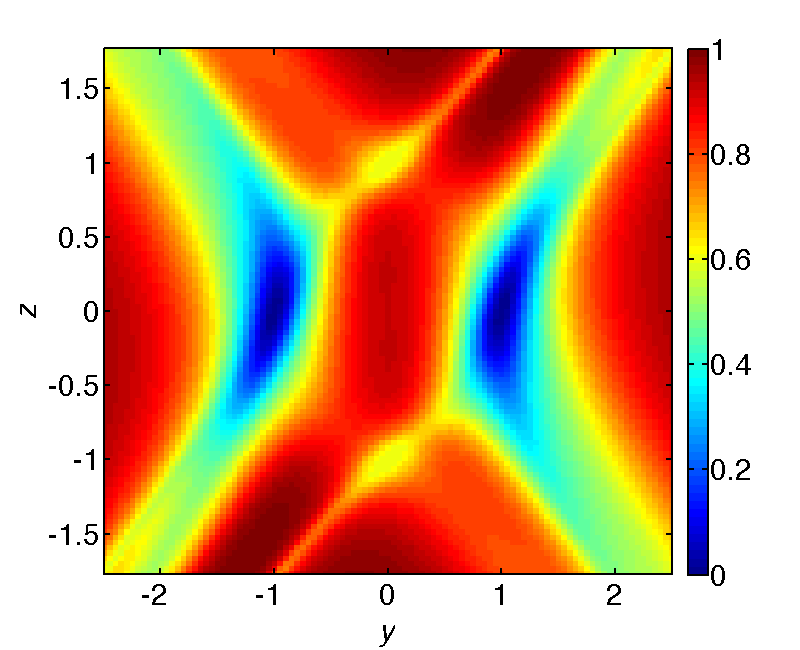} \ \includegraphics[width=0.4\textwidth]{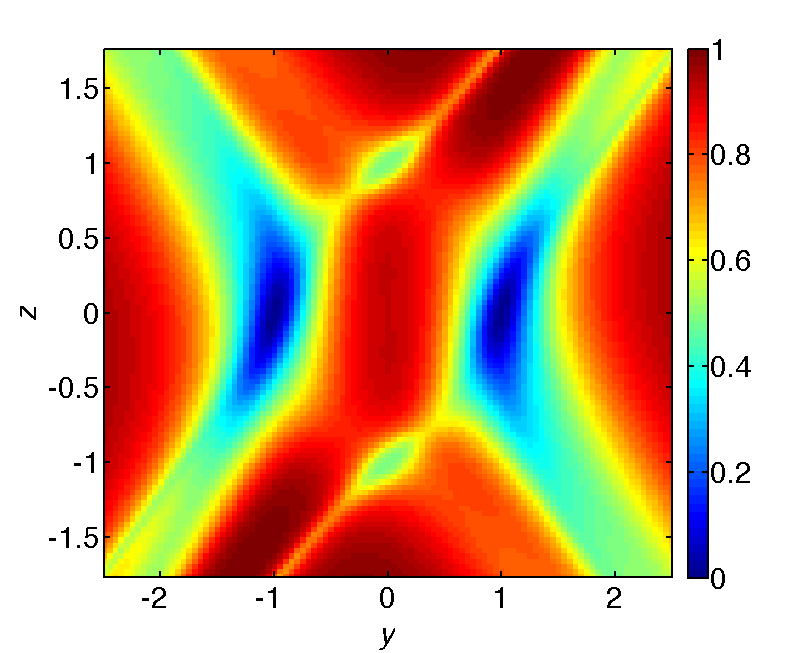}
 \caption{Left: Gap between the highest eigenvalue $\lambda_1 = 1$ and the second highest eigenvalue of the transfer matrix for $t = 0$ (top) and $t = 1$ (bottom) for a cylinder of circumference $L_1 = 6$ as a function of $y,z \in \mathbb{R}$. Right: Same plot for $L_1 = 8$. The lines with low values of the gap indeed seem to be gapless lines in the thermodynamic limit, as the corresponding values of the gap are significantly lower for $L_1 = 8$ than for $L_1 = 6$. Points were the Lanczos algorithm for calculating the eigenvalues of the transfer operator did not converge are marked in white.   } \label{fig:phases}
\end{figure*}

The calculation of the spectrum of the transfer matrix $\mathcal{T}$ presented above allows us to detect the presence of critical lines and surfaces in the phase diagram of the states $\ket{\psi_b (t,y,z)}$. In Fig.~\ref{fig:phases} we plot two sections of the phase diagram for $t=0$ and $t=1$ and real values of $y$ and $z$. The figure displays the gap $\varDelta$ between the two largest eigenvalues of the transfer matrix. In the following, the term "gap" will refer to this.

In the pure gauge theory, $t=0$, the signs of $y$ and $z$ play no role, and can be eliminated using phase transformations of the form \eqref{eqUS}, with phases $\pm \pi$. In the following we shall restrict our discussion to $y,z \in \mathbb{R}$, and thus it will be sufficient to consider $y,z \geq 0$ only. In this case,
four gapped phases can be easily spotted due to the appearance of several critical lines, defined by $\varDelta=0$ compatibly with the resolution of our numerical calculations. To verify the reliability of the existence of such critical lines in the thermodynamical limit, we have repeated the numerical calculation of the gap $\varDelta$ for different system sizes,
on cylinders of increasing width,
which seem to show that the gap $\varDelta$ is closing in the thermodynamic limit along certain lines in the phase diagram.

The analysis and characterization of the gapped phases has to be based on suitable order parameters.
In the following, we will consider separately  the pure lattice gauge theory $(t=0)$ and the full model with fermionic matter and gauge fields at $t>0$.

\subsubsection{The pure gauge theory}

In the pure bosonic model at $t=0$, the gauge invariant physical observables which can be exploited to characterize the states are of two different kinds: bosonic operators based on the local electric field $\Sigma$, which commutes with the Gauss law, and closed string operators built based on the Wilson lines $\Sigma_\pm$.
The second class commutes with the local Gauss law only if suitable products of these operators enter in such strings based on the convention about the orientation of the links on the lattice, which is defined by the signs in $G_{\mathbf{x}}$ in Eq. \eqref{physgauss}.

\begin{figure*}[t]
\centering
 \includegraphics[width=0.8\textwidth]{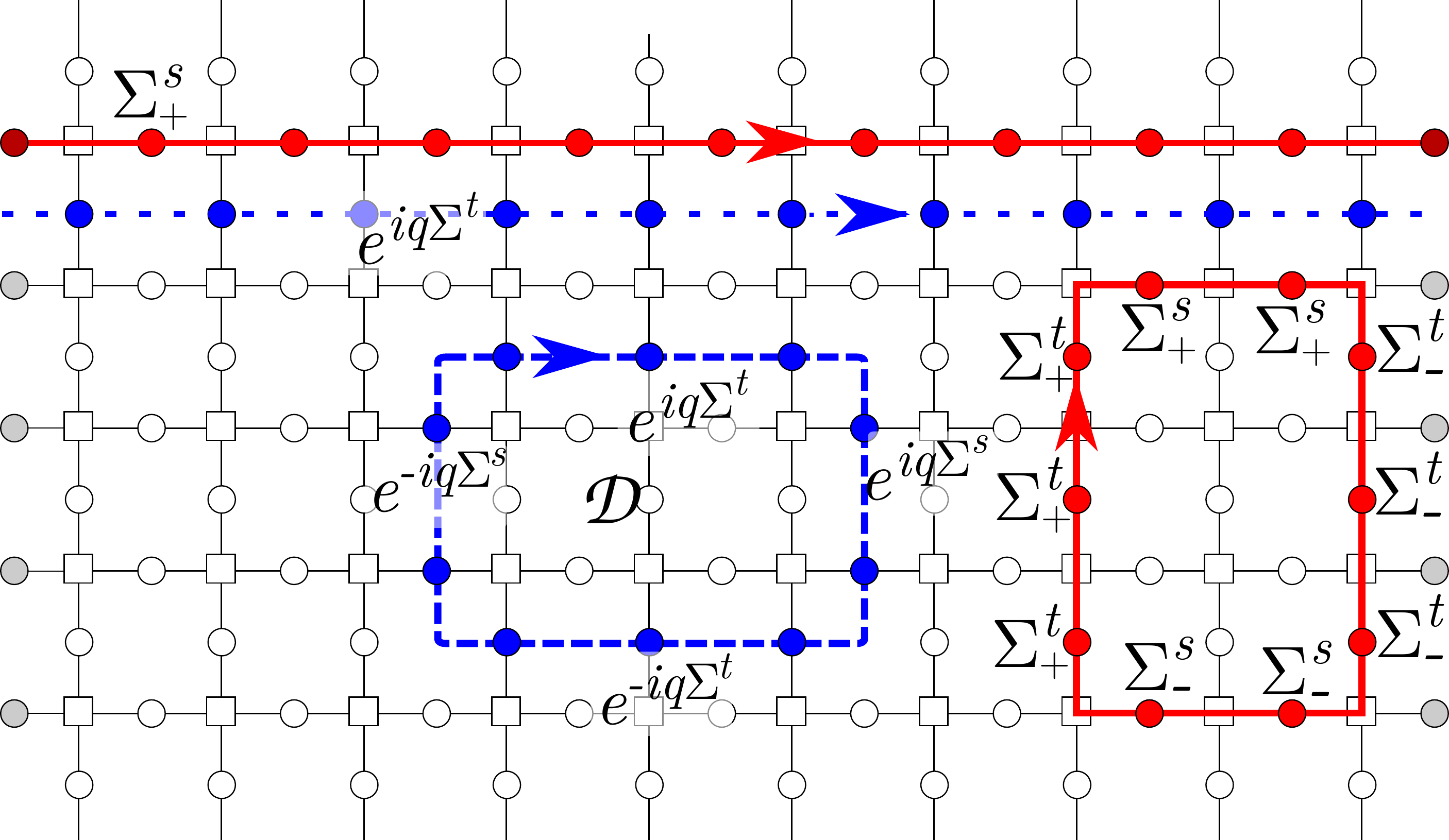}
 \caption{Wilson and 't Hooft loops are illustrated in a cylindric system of width $L_1=10$. Circles and squares correspond to physical sites associated to the gauge bosons and the matter fermions respectively. The grey circles depict periodic boundary conditions in the horizontal direction. Wilson loops are illustrated as red lines on the lattice: both a non-contractible loop and a $2\times 3$ closed loop are shown. 't Hooft loops on the dual lattice are represented by dashed blue lines:  both a non-contractible and a $3\times 2$ closed loop encircling the area $\mathcal{D}$ are shown. For each loop we specified the operators acting on the sites along each edge consistently with counterclockwise loops.} \label{fig:loops}
\end{figure*}

The local expectation value of the electric field vanishes due to the imposed charge conjugation symmetry. The operators $e^{i q \Sigma}$, with $q\in \mathbb{R}$ a generic parameter related to the elementary charge of the model, allow to define \textit{'t Hooft loops} on closed paths on the dual square lattice (see Fig. \ref{fig:loops}),
\begin{equation} \label{tHooft}
 G_{\mathcal{D}}(q) \equiv  \prod_{\mathbf{b} \in \partial \mathcal{D}} e^{iqs_\mathbf{b}\Sigma(\mathbf{b})} = \prod_{\mathbf{x} \in \mathcal{D}} e^{iqQ_\mathbf{x}}
\end{equation}
where $\mathcal{D}$ is a closed region on the square lattice delimited by the physical links on the loop of the dual lattice $\partial \mathcal{D}$ (see Fig. \ref{fig:loops}). The first product represents the application of the gauge transformation $e^{iqs_\mathbf{b}\Sigma(\mathbf{b})}$ to all the bosonic sites $\mathbf{b}$ surrounding the closed region $\mathcal{D}$. The signs $s_\mathbf{b}=\pm 1$ are defined by considering the orientation of the loop on the dual lattice, as depicted in Fig. \ref{fig:loops}.

The equality between the two products in Eq. \eqref{tHooft} is guaranteed by the Gauss law \eqref{physgauss}: the 't Hooft loop is related to both the total matter charge $\sum_{\mathbf{x}\in\mathcal{D}}Q_x$ included in the region $\mathcal{D}$ and  the electric field flux $\Phi_E=\sum_{\mathbf{b} \in \partial \mathcal{D}} s_\mathbf{b}\Sigma(\mathbf{b})$ associated to its contour $\partial \mathcal{D}$. In particular, in a pure gauge theory, the expectation value of the 't Hooft loop $G_{\mathcal{D}}(q)$ is just an exponential of the static charges enclosed in the region $\mathcal{D}$. In our PEPS construction for $t=0$, though, no static charge appears; thus $\left\langle G_{\mathcal{D}}(q) \right\rangle = 1$ for all the closed regions delimited by a contractible loop $\partial \mathcal{D}$ on the dual lattice.
The situation is different when considering a non-contractible loop on the cylinder (see Fig. \ref{fig:loops}). In this case the 't Hooft loop becomes
\begin{equation}
 G^{\rm NC}_{\mathcal{D}}(q) \equiv \prod_{x_1=1}^{L_1} e^{iq\Sigma^t(x_1,x_2)}\,.
\end{equation}
By applying the Gauss law, and considering the absence of static charges, it is easy to show that $\left\langle G^{\rm NC}_{\mathcal{D}}(q)\right\rangle $ is independent of the coordinate $x_2$ and it simply represents the electrical flux flowing along the surface of the cylinder
\begin{equation}
 \varPhi_E=\sum_{x_1}\Sigma^t(x_1,x_2)\,.
\end{equation}
Such flux, independent of $x_2$, is defined by the boundary conditions adopted in the first row and it is zero in our numerical calculations (unless stated otherwise). More in general, $G^{\rm NC}_{\mathcal{D}}(q)$ is independent of local perturbations: All the loops on the dual lattice surrounding  the cylinder once are equivalent and measure the electric flux defined by the projector operator $X_i$ in Eq. \eqref{expectation}.

The second class of gauge-invariant bosonic string operators is the one of the \textit{Wilson loops} which can be thought of as the path ordering of the exponentiated vector potential integral $\exp\left( i\oint_\mathcal{C}A_\mu dx^\mu\right) $ along a loop $\mathcal{C}$ on the lattice. Following Eq. \eqref{vectorpotential}, in our truncated lattice gauge model we apply the substitution $\exp\left(i\int_{\mathbf{x}}^\mathbf{x+e_\mu}A_\mu dx^{\mu}\right)  \equiv e^{i \theta^{s/t}(\mathbf{x})} \rightarrow \Sigma^{s/t}_+(\mathbf{x})$. Therefore the (clockwise) Wilson loop reads
\begin{equation}
 W_{\mathcal{C}}=\prod_{\mathbf{b}\in \mathcal{C}} \Sigma_{\pm}(\mathbf{b})
\end{equation}
where the choice of the operators $\Sigma_{\pm}$ depends on the orientation of the loop $\mathcal{C}$: For links oriented upward or rightward one has to apply $\Sigma_+$, whereas for downward or leftward links, $\Sigma_-$ (see Fig. \ref{fig:loops}).
It is important to notice that, in our model, the Wilson loops $W$ are not unitary operators. Furthermore they do not commute in general with each other. This is due to the relation $\left[\Sigma_+,\Sigma_- \right]=\Sigma$ valid for our truncation with $\ell=1$.

We observe that a vertical Wilson line crossing  the whole cylinder from the first to the last bosonic row is a further gauge invariant operator of the system and it does not commute with the non-contractible 't Hooft loops. Despite that, though, we have verified that our state does not show any topological degeneracy. Indeed the transfer matrix $\mathcal{T}$
is block-diagonal with the blocks labeled by $\varPhi_E$ on the ket layer and $\varPhi_E$ on the bra layer. Hence, if a density matrix $X$ has  fixed fluxes on ket and bra, it preserves them individually.
 The largest eigenvalues of different sectors of $\mathcal{T}$ labelled by $\varPhi_E$  on ket and bra are, however, non-degenerate, differently from what happens in topologically ordered models such as the toric code~\cite{Kitaev2003,Verstraete2006}. The dominating eigenvalue is the one associated with the $\varPhi_E=0$ sector. A rough finite size scaling indicates that the second largest at $\varPhi_E = \pm 1$ for ket and bra has a magnitude of $7.9\%$ as compared to the former.
The absence of topological order in spite of the presence of two non-commuting gauge-invariant string operators can be ascribed to the non-unitarity of the Wilson loop operator $W$.

\begin{figure}[t]
 \centering
 \includegraphics[width=0.5\textwidth]{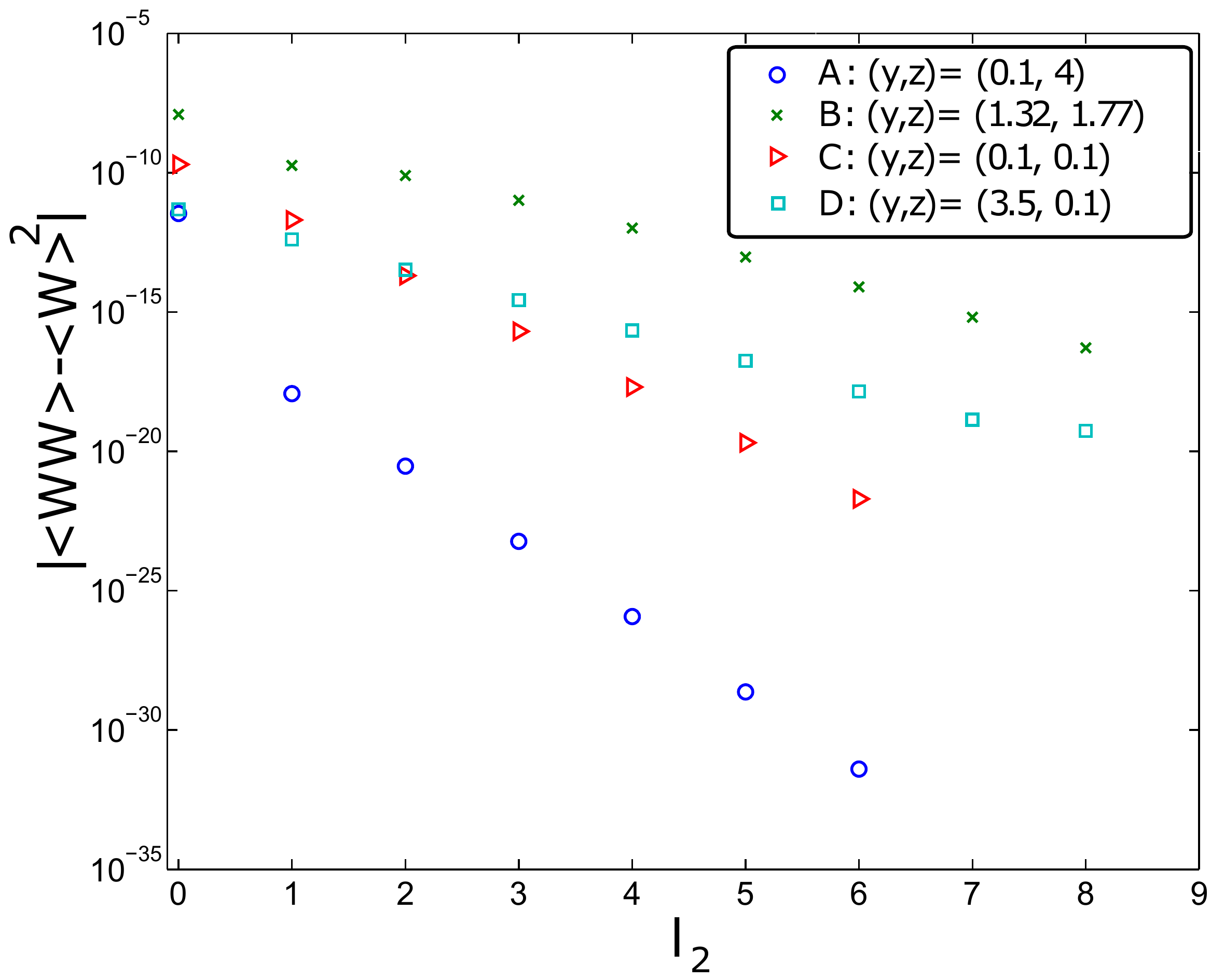}
 \caption{Semilogarithmic plot of the exponential decay of the correlation function of two non-contractible Wilson loops as a function of their distance $l_2$ [see Eq. \eqref{wilsoncorr}]. The data plotted correspond to $\left|\left\langle W^{\rm NC}(20)W^{\rm NC}(20+l_2)\right\rangle-\left\langle W^{\rm NC}(30)\right\rangle^2\right|$ in a system of size $6 \left(l_2+40\right)$. The correlation of the two Wilson loops decays exponentially in all the phases. Due to the fast exponential decay, for larger values of the distance $l_2$ the numerical errors becomes too large to obtain reliable data.}
 \label{fig:wilsoncorr}
\end{figure}

We have verified that the correlation between non-contractible Wilson loops along horizontal lines at $x_2$ and $x_2'$ decays exponentially in all the gapped phases appearing in the phase diagram at $t=0$, as expected by the non-degeneracy of the transfer matrix
\begin{equation} \label{wilsoncorr}
 \left\langle W^{\rm NC}(x_2)W^{\rm NC}(x_2')\right\rangle -\left\langle W^{\rm NC}(x_2) \right\rangle\left\langle  W^{\rm NC}(x_2')\right\rangle
 \approx C e^{-\frac{x_2'-x_2}{\lambda}}
\end{equation}
for $x_2'\gg x_2$, where we have defined the non-contractible Wilson loops as $W^{\rm NC}(x_2)=\prod_{x_1} \Sigma_+^s(x_1,x_2)$ (see Fig. \ref{fig:wilsoncorr}).

Away from the critical regions, one of the main distinctive characters of the phases in a pure lattice gauge theory is the exponential decay of the expectation value of the Wilson loops as a function of its dimension in the limit of large loops. Such exponential decay is typically dictated by either the area $\mathcal{A}$ of the loop as $e^{-\kappa_A \mathcal{A}}$ or its perimeter $\mathcal{P}$ as $e^{-\kappa_P \mathcal{P}}$. The former behavior signals a confinement of the static charges, whereas the latter would be compatible with a deconfined phase of the static charges \cite{Wilson,KogutLattice,Polyakov1987} (although in a pure compact QED, without a truncation, the latter phase cannot exist  \cite{Polyakov,DrellQuinnSvetitskyWeinstein,BanksMyersonKogut}).
To evaluate this behavior, however, the calculations need to be performed in the thermodynamic limit of the system and for large enough loops in order to properly distinguish the two behaviors and avoid finite size effects.

\begin{figure}[t]
\centering
\includegraphics[width=0.33\textwidth]{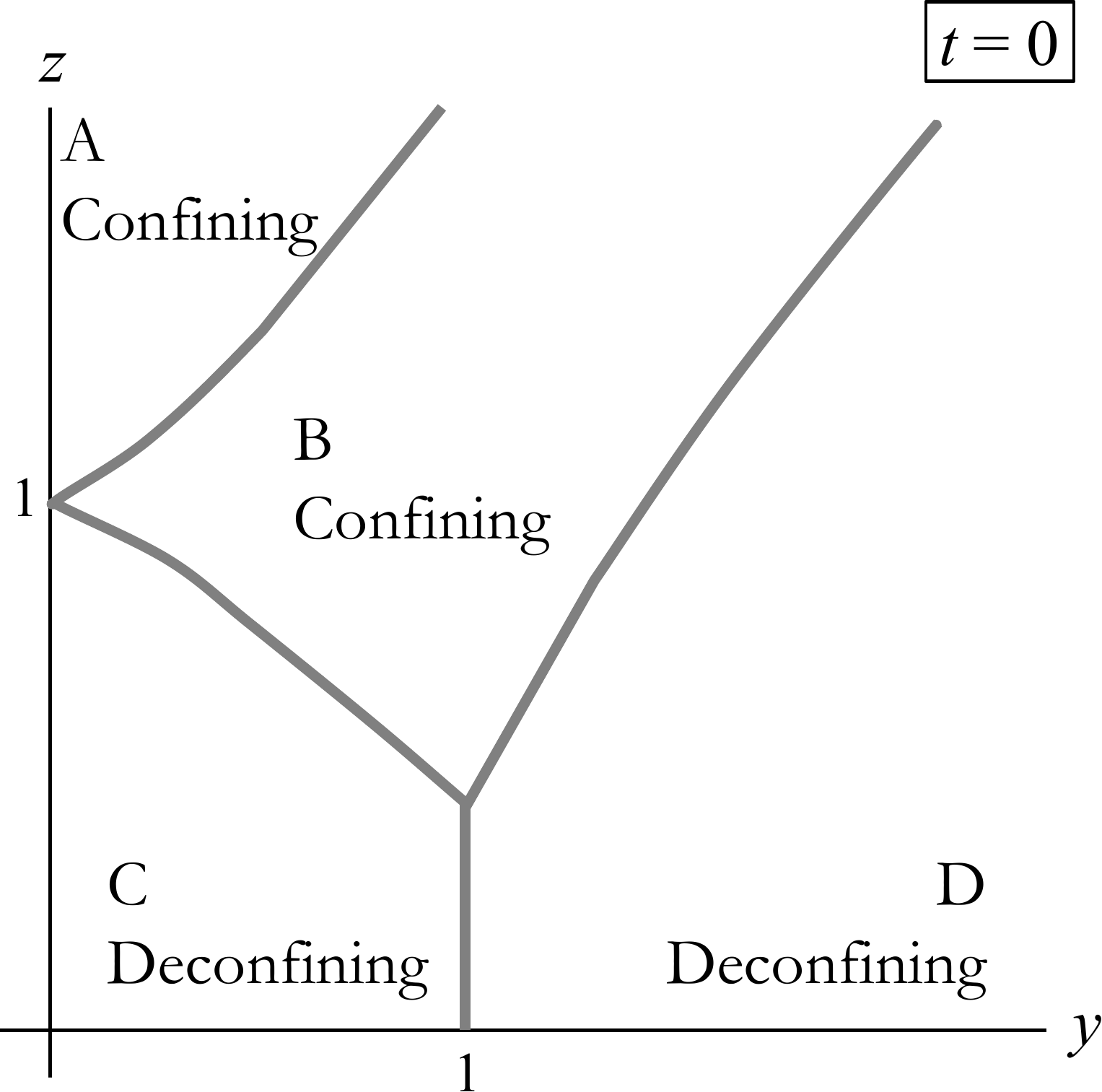}
 \caption{A schematic plot of the phase diagram for the pure gauge theory ($t=0$), with $y,z \geq 0$ (straightforwardly generalizable to any
$y,z \in \mathbb{R}$ as explained in the text). The $A,B$ phases seem to confine static charges, while the $C,D$ phases seem to be deconfined. } \label{PGPD}
\end{figure}

Our cylindrical systems are limited in the periodic dimension to a finite size of $L_1\le 9$. For the pure gauge case of $t=0$ the staggering plays no role and we can indeed exploit also odd values of the width $L_1$. When we consider rectangular Wilson loops $W(l_1,l_2)$ with width $l_1$ and length $l_2$, the maximal distance between the two vertical edges is at most $L_1/2$ due to the periodic boundary conditions, and, in our case, we were limited by $l_1<5$. This extension of the loops is sufficient to evaluate their asymptotic behavior only in the presence of a large enough gap $\varDelta$ of the transfer matrix and this implies that our calculations are reliable only far enough from the critical lines in the upper phase diagrams in Fig \ref{fig:phases}. Therefore, in the following calculations, we mainly consider points located deep inside the bulk of the gapped phases. Furthermore we label the phases $A,B,C,D$ as schematically shown in Fig. \ref{PGPD}.

Keeping this limitation in mind, let us analyze the numerical results for $L_1=8,9$ (the two sets of data are totally consistent with each other).
We have considered loops of length $l_2\le 20$ embedded in a cylinder with $L_2=l_2+40$ in such a way that the loops have a distance 20 from both the top and the bottom edges of the system.

\begin{figure*}[t]
 \includegraphics[width=1\textwidth]{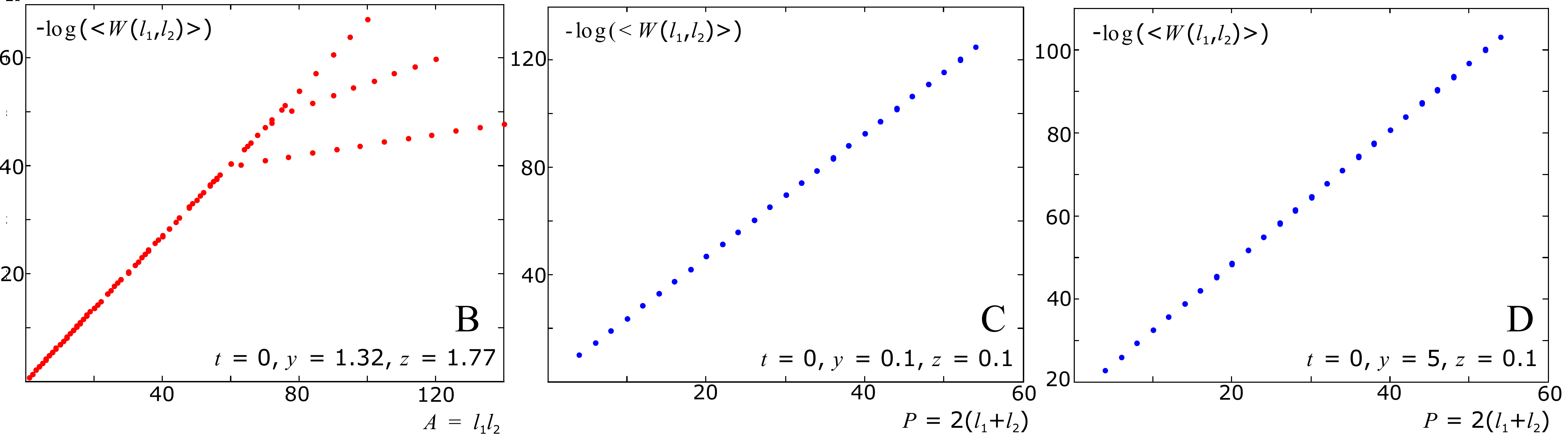}
 \caption{The area/perimeter law behavior of the Wilson loop, in the phases $B,C,D$ for the pure gauge case ($t=0$). It is clearly seen that phase B is governed by an area law, while phases $C,D$ respect a strict perimeter law. Data was computed for $L_1=8$. The two lines emanating from the main line of the area law in phase $B$ belong to the data sets of $l_1 = 6,7$ and $l_2 \geq 12,9$ respectively. These could be perimeter corrections due to finite size effects, as the widths of these loops are comparable with those of the system, $L_1$. } \label{Wloops}
\end{figure*}

The phase $D$ clearly presents a perimeter law decay of the Wilson loop for the values of the $y,z$ that we probed (see Fig. \ref{Wloops}D).
This is consistent with the PEPS structure obtained in the extreme limit $y\to \infty,\, z= 0$: an expansion of the fiducial state of a single lattice vertex around $z=0,\, {1}/{y}=0$ results in a superposition of the bosonic vacuum (in zeroth order), and $O\left(y^{-1}\right)$ terms involving either horizontal or vertical flux loops crossing the vertex (since $y$ is the $T$ parameter responsible for horizontal and vertical coupling between links). Since no flux "corners" ($z$ terms) are there, for the first nontrivial order, the most significant contributions to the physical state would be arbitrarily long (as the system size) flux lines, as expected within a deconfining phase, showing a perimeter law behavior for the Wilson loops.

The appearance of the perimeter decay of the Wilson loop can be explained from the properties of the PEPS in the limit $z\to 0$. In fact, the operator $A_b$ at $t=0$ and $z \to 0$ becomes the product of two separate operators acting on the horizontal and vertical links respectively.
We have indeed
\begin{equation}
 A_b(t=0,z=0,y)= \exp\left[y\left(l_+^\dag r_+^\dag \Sigma_+^s - r_-^\dag l_-^\dag \Sigma_-^s \right)\right] \otimes
 \exp\left[y\left(u_-^\dag d_-^\dag \Sigma_-^t - d_+^\dag u_+^\dag \Sigma_+^t \right)\right]\equiv A_h A_v
\end{equation}
This implies that the PEPS is decomposed a product state of one-dimensional systems
\begin{multline}
 \ket{\psi_b(t=0,z=0)}  =\\
 \bigotimes_{x_1} \bra{\Omega_v(x_1)}\prod_{x_2} \eta(x_1,x_2) \prod_{x_2}A_v(x_1,x_2) \ket{\Omega(x_1)}
 \bigotimes_{x_2} \bra{\Omega_v(x_2)}\prod_{x_1} \omega(x_1,x_2) \prod_{x_1}A_h(x_1,x_2) \ket{\Omega(x_2)} \\
  \equiv \ket{\rm ver}^{\otimes L_1} \ket{\rm hor}^{\otimes L_2}
\end{multline}
where we splitted all the vacuum states in the product of states for the degrees of freedom aligned along rows and colums. $\ket{\psi_b(t=0,z=0)}$ is therefore a product state of $L_1$ identical vertical 1D states, labelled by $\ket{\rm ver}$, and $L_2$ identical horizontal 1D states, labelled by $\ket{\rm hor}$. When we consider a rectangular Wilson loop $W_{\mathcal{C}}$, only four of these states are involved, one for each edge. In particular we decompose the Wilson loop into Wilson lines acting on the four 1D systems: $W_{\mathcal{C}}=W_{\rm bot}W_{\rm left}W_{\rm top}W_{\rm right}$ where these 1D operators have a structure such that $W_{\rm top}=W_{\rm bot}^\dag$ and $W_{\rm left}=W_{\rm right}^\dag$. We obtain
\begin{equation}
\left\langle W_{\mathcal{C}(l_1,l_2)} \right\rangle = \left|\bracket{{\rm hor}}{W_{\rm bot} |{\rm hor}}\right|^2 \left|\bracket{{\rm ver}}{W_{\rm left} |{\rm ver}}\right|^2.
\end{equation}
Such an expectation value vanishes exactly at $z=0$, because the finite Wilson lines violate the local Gauss law in each 1D system. In particular, in the decoupled $z=0$ limit, due to the periodic boundary conditions, each 1D horizontal state is a cat state of the form
\begin{equation}
 \ket{\rm hor} \propto \left(1+y^{2L_1} \right) \ket{0,0,0,\ldots}
 +y^{L_1} \ket{1,1,1,\ldots} + y^{L_1} \ket{-1,-1,-1,\ldots}.
\end{equation}
In the thermodynamic limit $L_1 \to \infty$, the first term prevails for every value of $y \neq \pm 1$. At $y=\pm 1$, instead, the three terms have a comparable amplitude such that the critical line originating at $y=1$ and $z=0$ is characterized by a symmetry breaking with three possible degenerate states of the corresponding parent Hamiltonian, whereas for the gapped phases only the state with no electric field survives. In the vertical direction the states $\ket{\rm ver}$ have a similar behavior but are constrained by the boundary conditions.

For small values of $z$, all these decoupled 1D states are perturbed by the introduction of domain walls with an amplitude proportional to $z$. Such domain walls describe indeed corners of the electric flux in the overall 2D state, which becomes a set of weakly coupled 1D systems. The Wilson loop $W_{\mathcal{C}}$ is an operator which can be decomposed into four of these corners, each one with amplitude $z$, in such a way that the 2D Gauss law is not violated. Therefore its expectation value decays as $Cz^4e^{-\lambda'(2l_1+2l_2)}$ in the lowest order in $z$, where the appearance of the perimeter law is due to the fact that each 1D state has a gapped transfer matrix away from the critical points, thus each 1D expectation value decays with the length of the related Wilson line. Hence, for $z\ll 1$, the decomposition of the PEPS into one-dimensional states implies that no area contribution to the decay can be present at low orders in $z$. This explains the appearance of a perimeter law for the Wilson loop in both the phases $D$ and $C$.

Our numerical results confirm that the behavior of the gapped phase $C$ is also characterized by a perimeter decay, despite being clearly separated by the phase $D$. In this central region of the phase diagram, though, the finite size effects due to the limited width of the loops seem to be more relevant. This implies that only the data at $L_1=8,9$ with $l_1$ having a maximal value 4 are reliable, whereas data taken at $L_1=6$ present large deviation with respect to the perimeter law. Furthermore, also for $L_1=8,9$ we see minor deviations from the perimeter decay due to the different widths of the loops. The perimeter law behavior may be seen in Fig. \ref{Wloops}C.

Such behavior of the phases $C$ and $D$ is fully compatible with a deconfinement of the static charges: in fact, no signature of an area law decay appears in these phases.

Let us analyze the Wilson loops in all the phases more quantitatively: To limit the effect of the perimeter contribution and evaluate the eventual decay as a function of the area (dictated by $\kappa_A$) in all the four phases, we evaluate the following parameter, introduced by Creutz \cite{Creutz1980}
\begin{equation} \label{chi}
 \chi\left(l_1,l_2 \right)\equiv -\ln\left[\frac{\left\langle W(l_1,l_2)\right\rangle\left\langle W(l_1-1,l_2-1)\right\rangle }{\left\langle W(l_1-1,l_2)\right\rangle\left\langle W(l_1,l_2-1)\right\rangle} \right].
\end{equation}
If we assume an asymptotic mixed behavior of the Wilson loop of the kind $\left\langle W(l_1,l_2)\right\rangle \propto \exp[-\kappa_A l_1 l_2-\kappa_P2(l_1+l_2)]$,  $\chi$ converges to $\kappa_A$ for large values of $l_2$ and $l_1$. In our case, however, this asymptotic decay is meaningful only for $l_1\le L_1/2$ due to the periodic boundary conditions. In the ratio defining $\chi$, the perimeter contribution disappears because it is the same in the numerator and denominator, therefore the parameter $\chi$ must go to zero in a deconfined phase (the ratio inside the logarithm tends to 1 and $\kappa_A$ results 0), whereas it must be positive in the confined ones, corresponding to $\kappa_A$ for large values of both $l_1$ and $l_2$.

As expected, the parameter $\chi$ in the phases $C$ and $D$ converges fast to zero, thus showing the absence of an area contribution to the decay.

For the phase $D$ we considered, as an example, the point $y=5$, $z=0.1$. In this point $\chi(l_1,l_2) < 5 \times 10^{-12}$ for $l_1=4$ and $l_2>4$ in a system size with $L_1=8$. The data show also a reduction by a factor of about 30 going from $l_1=2$ to $l_1=3$ and a similar reduction from 3 to 4. This is in precise agreement with the perimeter law shown in Fig. \ref{Wloops} and analogous data we obtained at smaller system sizes.

The phase $C$ presents a remarkable decay of the parameter $\chi$ as well. For $y=z=1$, the value of $\chi$ drops to $\chi \approx 5\times 10^{-7}$ for $l_1=4$ and $l_2>4$ for both $L_1=8$ and $L_1=9$. Similarly to the previous case, $\chi$ decays consistently when increasing the loop width. This behavior is therefore consistent with a perimeter decay and $C$ behaves as a deconfined phase.

Such a deconfined regime characterizing the phases $C$ and $D$ would be impossible in the compact QED due to the well-known results
 of \cite{Polyakov,DrellQuinnSvetitskyWeinstein,BanksMyersonKogut}, and therefore we interpret the presence of these deconfined phases at $t=0$ as the effect of the lattice structure of the PEPS which allows, for $z \to 0$, the decoupling of the 2D system into a collection of weakly coupled 1D chains.

The gapped phase $B$ shows, instead, a totally different behavior. Here the Wilson loop is characterized by a faster decay dominated by an area law (see Fig. \ref{Wloops}B). In a system with width $L_1=8$, though, this area law is affected by strong finite size corrections for the loops with $l_1=6,7$ and $l_2>9$, whose width is comparable with the width of the system. These finite size effects are totally absent instead for $l_1<6$. The area law is indeed confirmed by the convergence of $\chi$ to a value different from zero for $L_1=8$ and $l_1<6$: Taking as an example the point at $y=1.32$ and $z=1.77$, $\chi$ rapidly converges to the value $\chi=0.67$ for all the values of $l_1<6$ in \eqref{chi}. We observe that in this phase the decay is much faster than in phases $C$ and $D$. This leads to large numerical errors in the estimation of $\chi$ for large loops $(l_2>10)$. For all the values with $l_1<6, 2\le l_2\le 10$, though, $\chi$ behaves like a constant, clearly indicating an area law contribution with $\kappa_A>0$ (see Fig. \ref{Wloops}B). Phase $B$ displays therefore a confinement of static charges.

The data related to phase $A$ are instead more difficult to interpret: For all the calculation at $L_2=8,9$ and $2<l_1<L_2-1$, $\chi$ presents an alternation in its sign dictated by the parity of the area; however, this  cannot be considered a good order parameter for this phase, as the Wilson loops there obtain significantly low expectation values, which might lead to an accumulation of numerical errors in the calculation of $\chi$. Due to this reason, we checked also the value of the loops for smaller system sizes which reduce the decay rate: In this phase the results obtained for $L_2=6$ are very different from the ones of larger sizes and they present a  positive limit of $\chi$ without even/odd effect. If we consider, as an example, the point at $y=0.32$ and $z=2.42$ we obtain $\chi(l_1=2)\approx 1.5$ and $\chi(l_1=3)\approx 0.08$ for $l_2 \ge 10$.

From the numerical data  it is therefore difficult to understand the nature of the phase $A$. We believe, however, this phase to be confining: if one expands the fiducial state of a single vertex around $y=0, \, 1/z=0$, an opposite behavior to that of the $D$ phase is obtained: On top of the zeroth order vacuum, the leading $O\left(z^{-1}\right)$ terms involve corners of flux loops meeting on a vertex. Thus the most significant nontrivial contributions in this state are excitations of single plaquettes (``glueballs'' - the shortest possible excitations of the gauge field) which might imply that the $A$ phase is, indeed, confining.

A schematic plot of the phase diagram discussed above is given in Fig. \ref{PGPD}. As a final remark on this topic, note that since we are dealing with the ground state with no static charges, all the possible states involve closed flux loops or infinite flux lines.

\subsubsection{The matter-gauge theory}

\begin{figure}[t]
\centering
\includegraphics[width=0.5\textwidth]{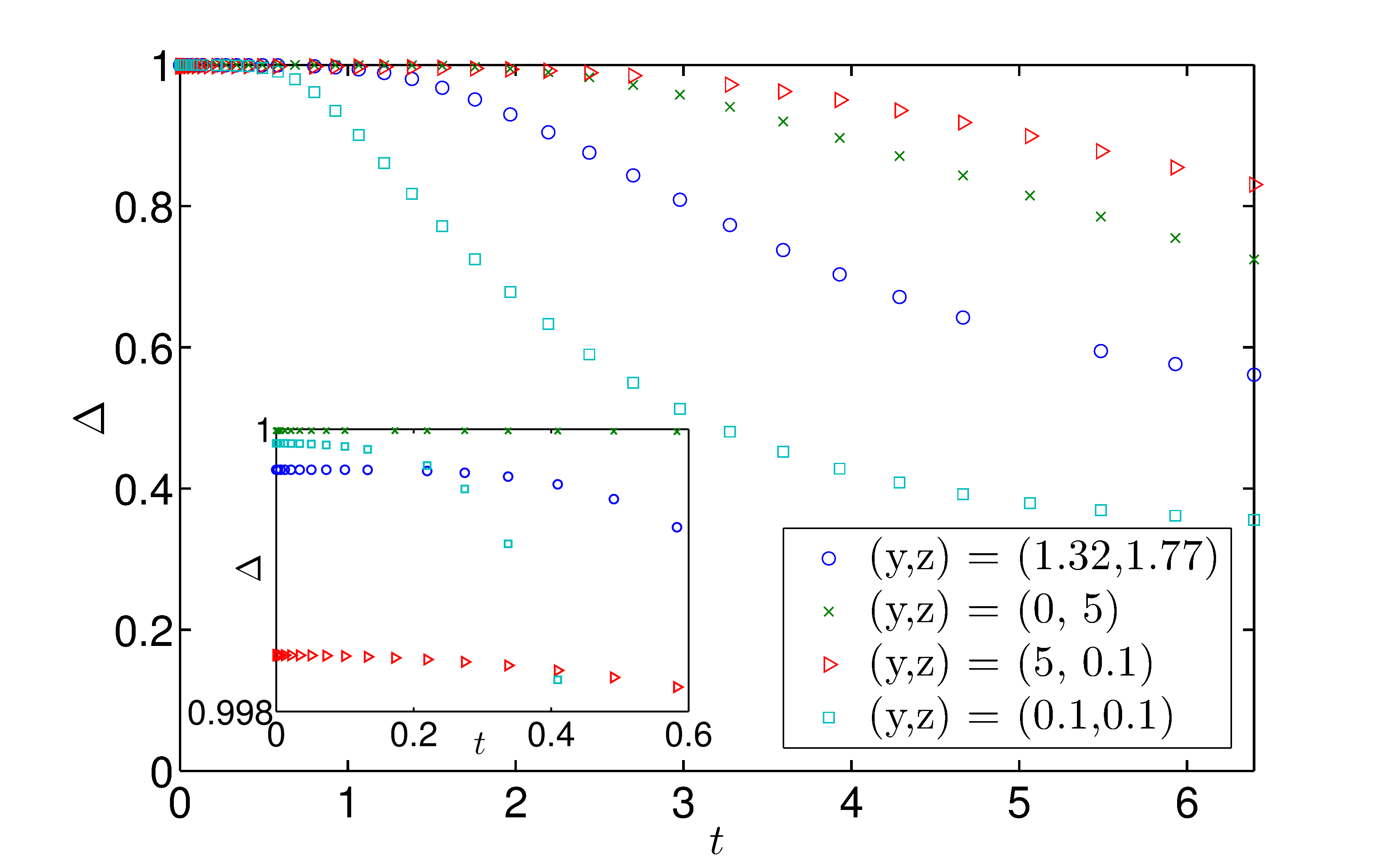}
 \caption{Gap $\varDelta$ as a function of $t$ for the given values of $y$ and $z$ and a cylinder circumference of $L_1 = 8$. The gap does not close as fermions are introduced in the system, and according to the figure closes, presumably, closes for $t \rightarrow \infty$. Inset: zoom of the data for small t} \label{fig:increase_t}
\end{figure}

The introduction of a parameter $t \neq 0$ introduces the staggered matter fermions into the system as well. Most of our numerical results do not show any particular discontinuity in the gapped phases from the pure gauge case to the complete one (see, as an example, Fig. \ref{fig:increase_t}). However, it is still reasonable to expect a discontinuity in the $t \rightarrow 0$ limit, as this point is where the fermions are coupled, or decoupled, from the gauge field.

Our numerical data imply that the phases $B$ and $C$, which used to be disconnected in the pure gauge model, seem to become adiabatically connected. The evaluation of the transfer matrix gap $\varDelta$ still displays a minimum in correspondence of the critical line at $t=0$, but, for $t>0$, there is no evidence of a closing of this minimum for increasing system sizes. Therefore, the data seem to suggest that the critical line separating these phases in the upper panels of Fig. \ref{fig:phases} actually disappears as shown in the lower panels and it constitutes an isolated line in the three-parameter phase diagram, which may be an evidence for a discontinuity at $t \rightarrow 0$.

Thanks to the introduction of matter, the set of non-trivial gauge invariant observables becomes richer and open bosonic strings delimited by fermionic operators provide further tools to examine the state $\ket{\psi_b}$. In particular, due to the staggering of the matter fermions, we can distinguish two kinds of open string operators acting on the bulk of the system:
\begin{equation}
\begin{aligned}
 M^\dag_{\mathcal{P}(\mathbf{e}_i,\mathbf{o}_f)} = \psi^\dag_{\mathbf{o}_f} \prod_{\mathbf{b}\in\mathcal{P}} \Sigma_{\pm}(\mathbf{b}) \psi^\dag_{\mathbf{e}_i}\,,\\
 J_{\mathcal{P}(\mathbf{x}_i,\mathbf{x}_f)} = \psi^\dag_{\mathbf{x}_f} \prod_{\mathbf{b}\in\mathcal{P}} \Sigma_{\pm}(\mathbf{b}) \psi_{\mathbf{x}_i}\,.
\end{aligned}
\end{equation}
$M^\dag$ represents the creation operator of a meson composed of a pair particle/antiparticle pair located on the even and odd lattice sites $\mathbf{e}_i$ and $\mathbf{o}_f$ and linked by a Wilson line defined on the path $\mathcal{P}$. As in the previous definition of the Wilson loops, the signs $+/-$ must be chosen in order to fulfill the local gauge invariance: The Wilson line is oriented from the initial site $\mathbf{e}_i$ to the final one $\mathbf{o}_f$ in such a way that upward and rightward links are associated with $\Sigma_+$, whereas downward and leftward links with $\Sigma_-$.
$J$ is instead a tunneling operator of a particle or antiparticle along the path $\mathcal{P}$, assisted by a Wilson line. $\mathbf{x}_i$ and $\mathbf{x}_f$ are both even for matter particles or both odd for antimatter particles. The signs $+/-$ along the Wilson line must be chosen consistently with gauge invariance as well.

\begin{figure}[tb]
\centering
 \includegraphics[width=0.44\textwidth]{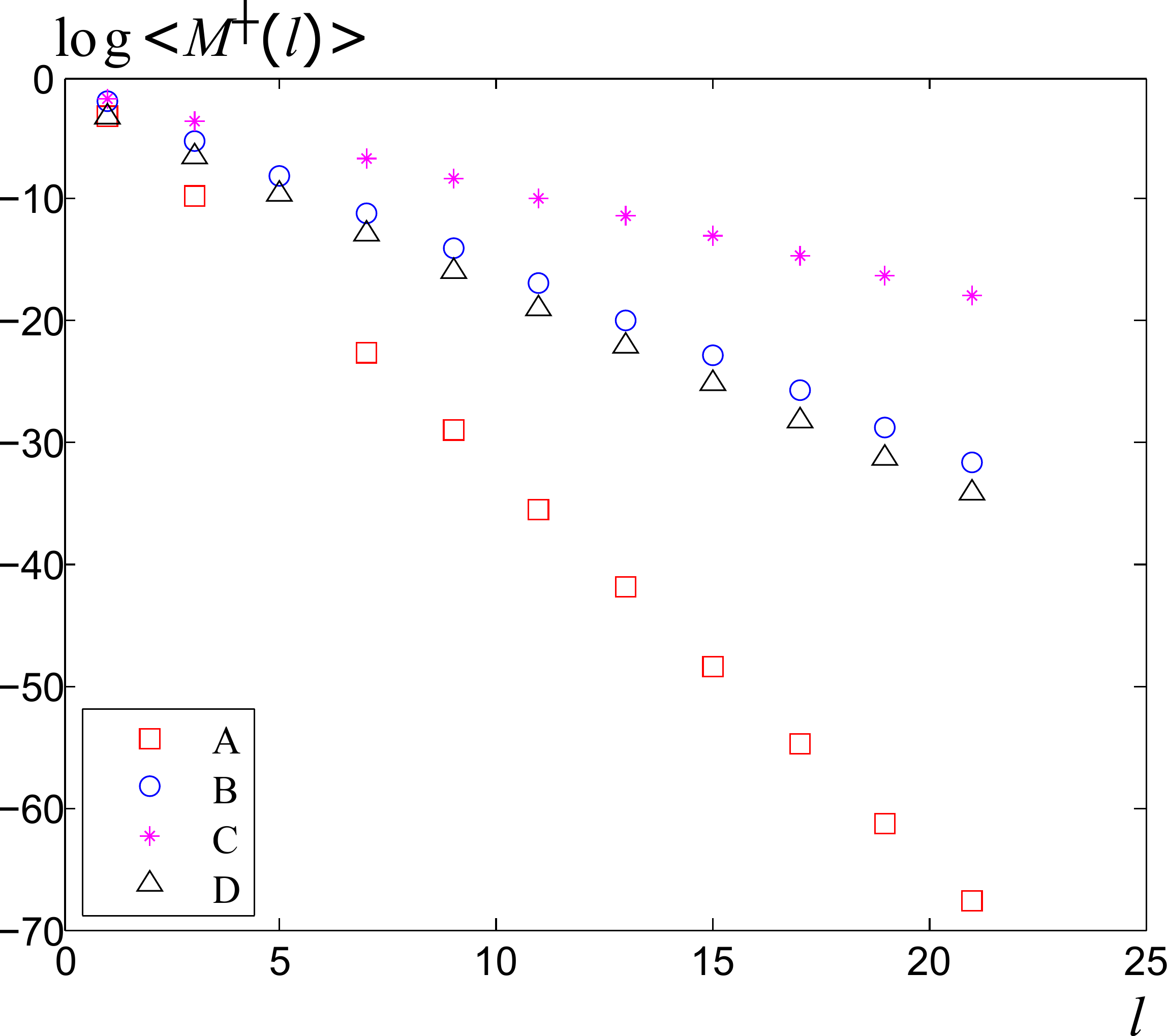}
 \caption{Expectation values of the meson operator $M^\dagger\left(l\right)$ for various lengths $l$, with $t=1$. Exponential decay is apparent in all the four phases. Data was sampled in the points $A\left(y=0,z=5\right)$, $B\left(y=1.32,z=1.77\right)$, $C\left(y=0.1,z=0.1\right)$, $D\left(y=5,z=0.1\right)$.  }
 \label{fig:mes}
\end{figure}

We have numerically evaluated the expectation value of the meson creation operator $M^\dag\left(l\right)$ for a vertical meson of variable length $l$, with $t=1$. The analysis was performed on a cylinder geometry with $L_1=8$ and $L_2=40+l$ with a starting point $\mathbf{e_i}$ located at $x_2=21$. In all the gapped phases these expectation values decay exponentially with the length of the meson (see Fig. \ref{fig:mes}). This is consistent with a picture in which the transfer matrix $\tilde{\mathcal{T}}_{\Sigma_+}$, obtained by Eq. \eqref{transfer2} with the substitution $O=\Sigma_+$, has a non-vanishing gap between its two largest eigenvalues.

\begin{figure}[tb]
\centering
 \includegraphics[width=0.50\textwidth]{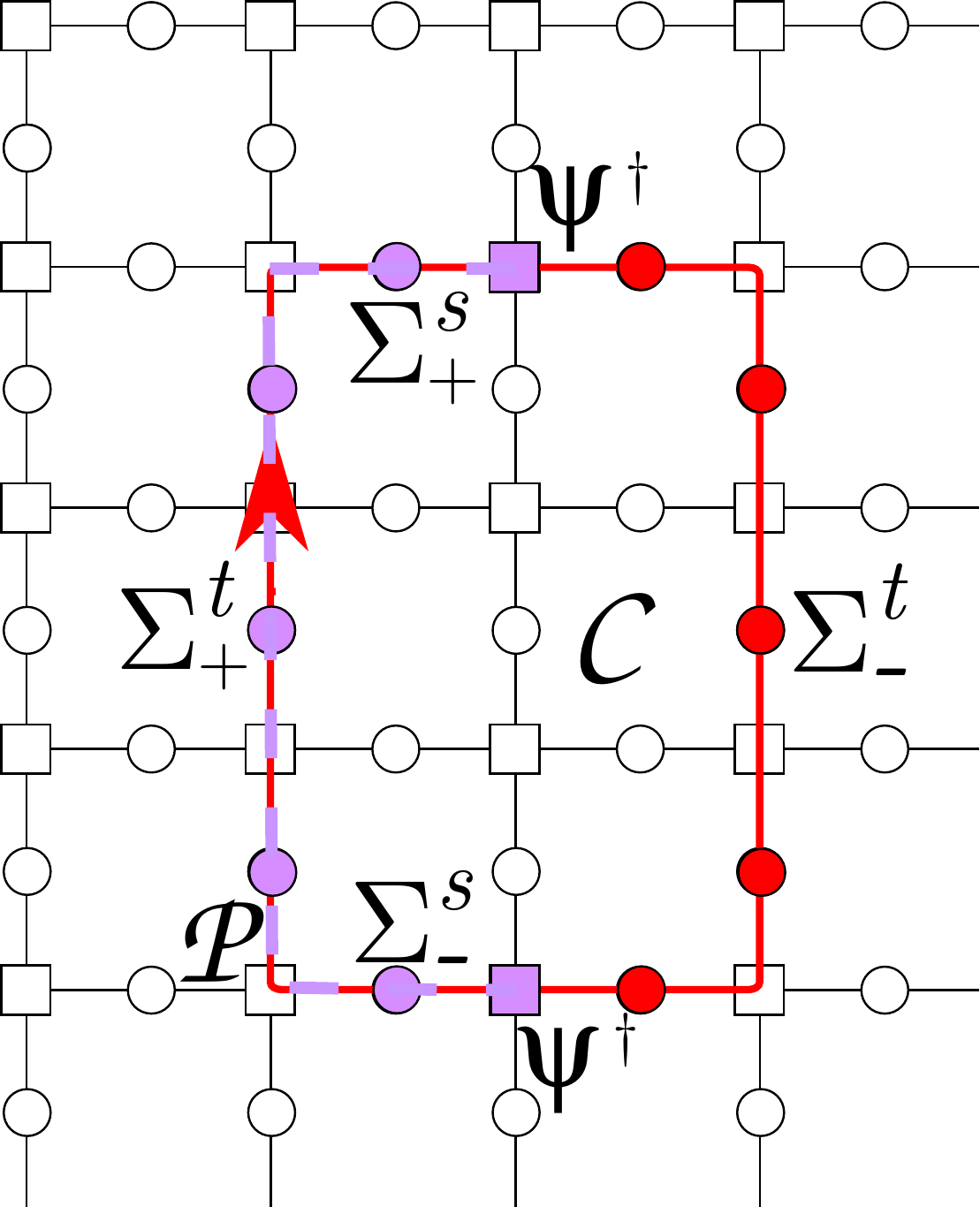}
 \caption{Schematic representation of the operators entering the horseshoe order parameter. The red rectangle $\mathcal{C}$, with dimensions $4\times 3$, represents the Wilson loop in the denominator of Eq. \eqref{FM}. The dashed violet line depicts instead the meson creation operator in its numerator, associated with half of the rectangle labelled by $\mathcal{P}$. We specify the bosonic operators involved along each edge of the loop, and the two fermionic operators (colored squares) entering the definition of $M^\dag$.}
 \label{fig:freden}
\end{figure}

Therefore the expectation value of $M^\dag$ alone is not enough to characterize the gapped phases, but it can be compared with the closed Wilson loops to provide information about the screening of the charges in our state. In particular, one can define the so-called horseshoe order parameter \cite{Gregor2011} (closely related to the Fredenhagen-Marcu order parameter \cite{Fredenhagen1986,Fredenhagen1988}) associated with a rectangular loop on the lattice. Such an order parameter corresponds to the ratio of the (squared) expectation value of the meson operator $M^\dag$ on half of the rectangle $\mathcal{C}$ and the Wilson loop associated to $\mathcal{C}$ (see Fig. \ref{fig:freden}). Due to the limitations of the size of our system, we adopt a rectangle of size $4 \times l$ (instead of $l \times l$ in the ideal case presented in \cite{Gregor2011,Fredenhagen1986,Greensite03}).
We define (the square root of) the horseshoe order parameter as
\begin{equation} \label{FM}
 \rho\left(l\right) = \frac{\left|\left\langle M^\dag_{\mathcal{P}=\mathcal{C}/2}(2,l) \right\rangle \right| }{\sqrt{\left\langle W(4,l)\right\rangle }}
\end{equation}
where the Wilson loop $W(4,l)$ is associated to a rectangle $\mathcal{C}$ of dimension $4 \times l$ with an odd $l$ (in our staggered case), the matter particle and antiparticle associated to $M^\dag$ are created in the middle of the two horizontal edges of $\mathcal{C}$ and the path $\mathcal{P}$ covers half of the rectangle $\mathcal{C}$ (see Fig. \ref{fig:freden}).
Differently from the standard evaluation of the horseshoe order parameter, we are constrained to take one dimension of the rectangle fixed to $l_1=4$ and thus we can only consider the thermodynamic limit in the vertical direction.

$\rho\left(l\right)$ is an order parameter for screening, assuming the Wilson loop has a perimeter law behavior, as expected for conventional gauge theories with a fundamental charge \cite{Gregor2011,Fredenhagen1986,Greensite03}.
Originally, it was formulated \cite{Fredenhagen1986,Greensite03} for the case in which the open edges of the horseshoe are in the temporal direction, and thus it is related to string breaking. For increasing dimensions of the considered loop, we can summarize the behavior of $\rho$ in the following way: In a phase which is neither confining nor screening for the dynamical matter, $\rho$ tends to zero; for confined or screened phases, instead, $\rho$ tends to a finite limit.
Indeed, in a deconfined phase free matter particles are strongly suppressed whereas the decay of the Wilson loop follows a slow perimeter law, therefore $\rho$ falls exponentially to zero with increasing loop sizes $l$. If, instead, the Wilson loop decay is dictated by a charge screening mechanism, the numerator and the denominator in \eqref{FM} have a comparable decay and in the limit of large loops a finite value $\rho \neq 0$ will be reached \cite{Fredenhagen1988,Greensite03}.

$\rho$  may be associated to a line tension \cite{Gregor2011} between matter charges or even to the conductance properties of the matter \cite{vanHeck2014}. Following \cite{Gregor2011}, within a theory with a bosonic ``frozen Higgs'' matter field in the limit
$l \rightarrow \infty$, $\rho \rightarrow 0$ implies deconfinement while $\rho \neq 0$ is a manifestation of a confining phase. In general, however, and in particular for a theory with fermionic matter, such as in our case, $\rho \rightarrow 0$ provides only a tool to distinguish screening from non-screening regimes, while confinement itself cannot be deduced as its consequence.

\begin{figure}[ht]
\centering
 \includegraphics[width=0.44\textwidth]{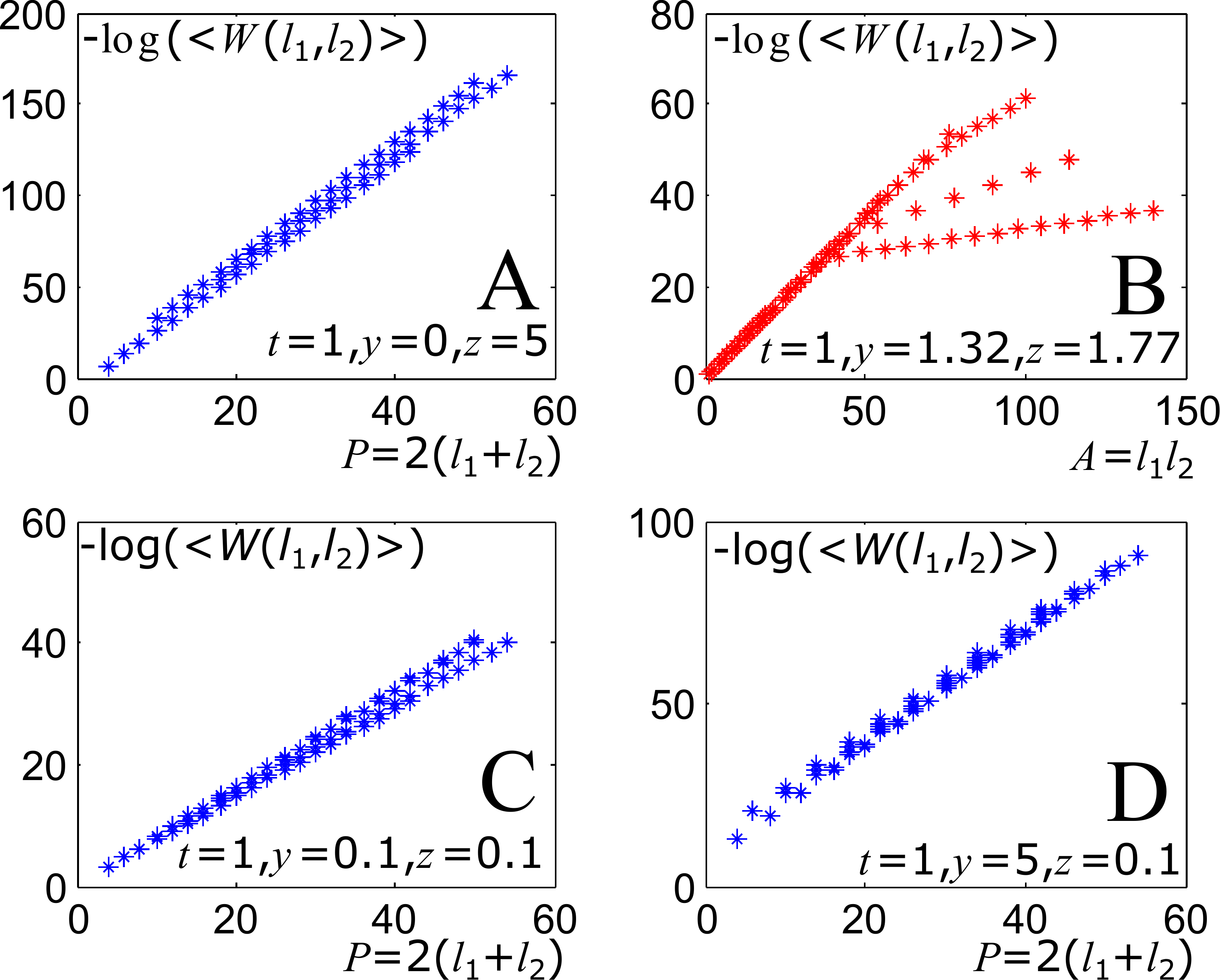}
 \caption{Area/perimeter law behavior of the Wilson loop, for $t=1$, at four representative points of the phase diagram.}
 \label{fig:Wlt1}
 \end{figure}

\begin{figure}[t]
\centering
 \includegraphics[width=0.44\textwidth]{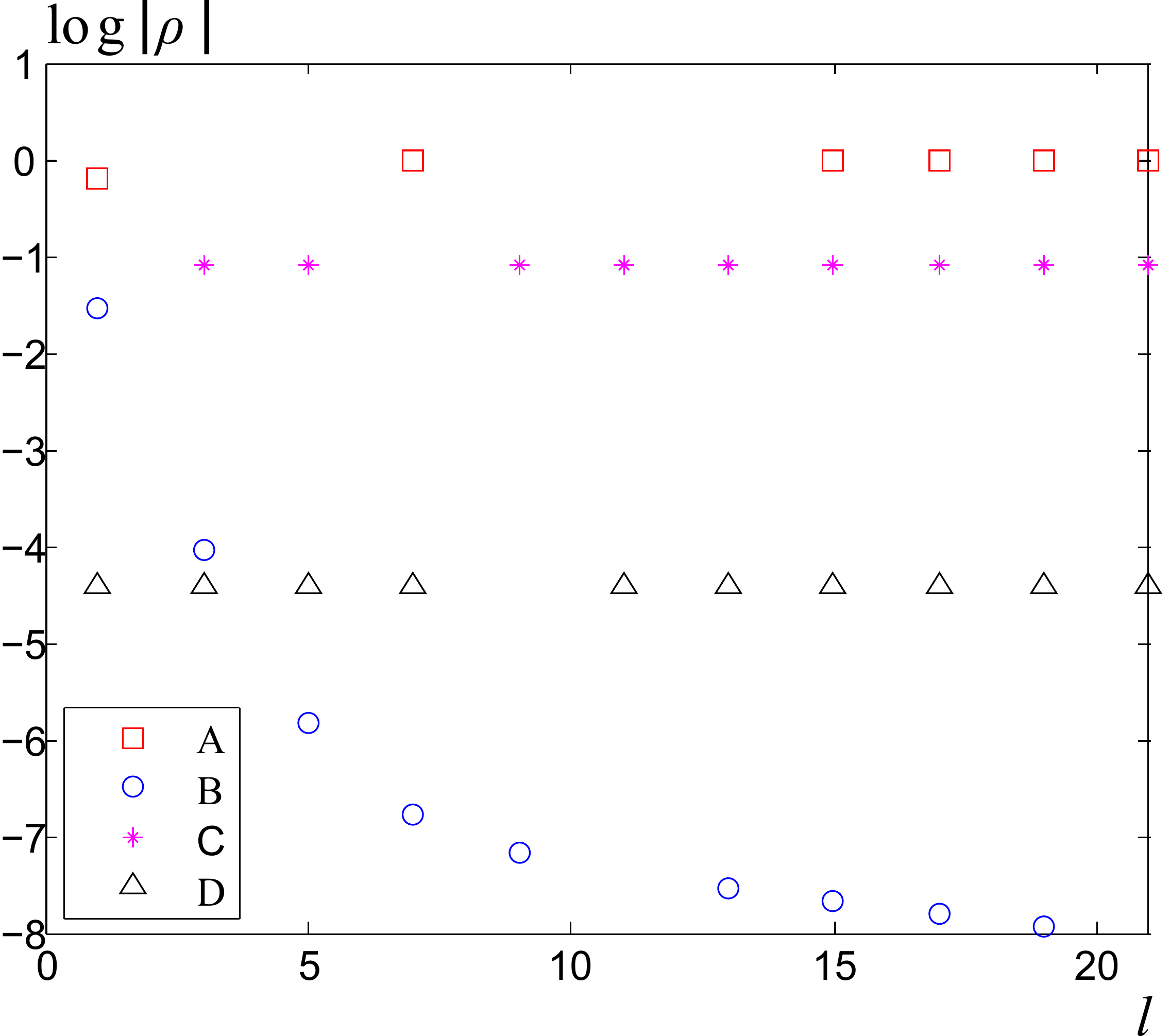} \hspace{0.05\textwidth}
\includegraphics[width=0.4\textwidth]{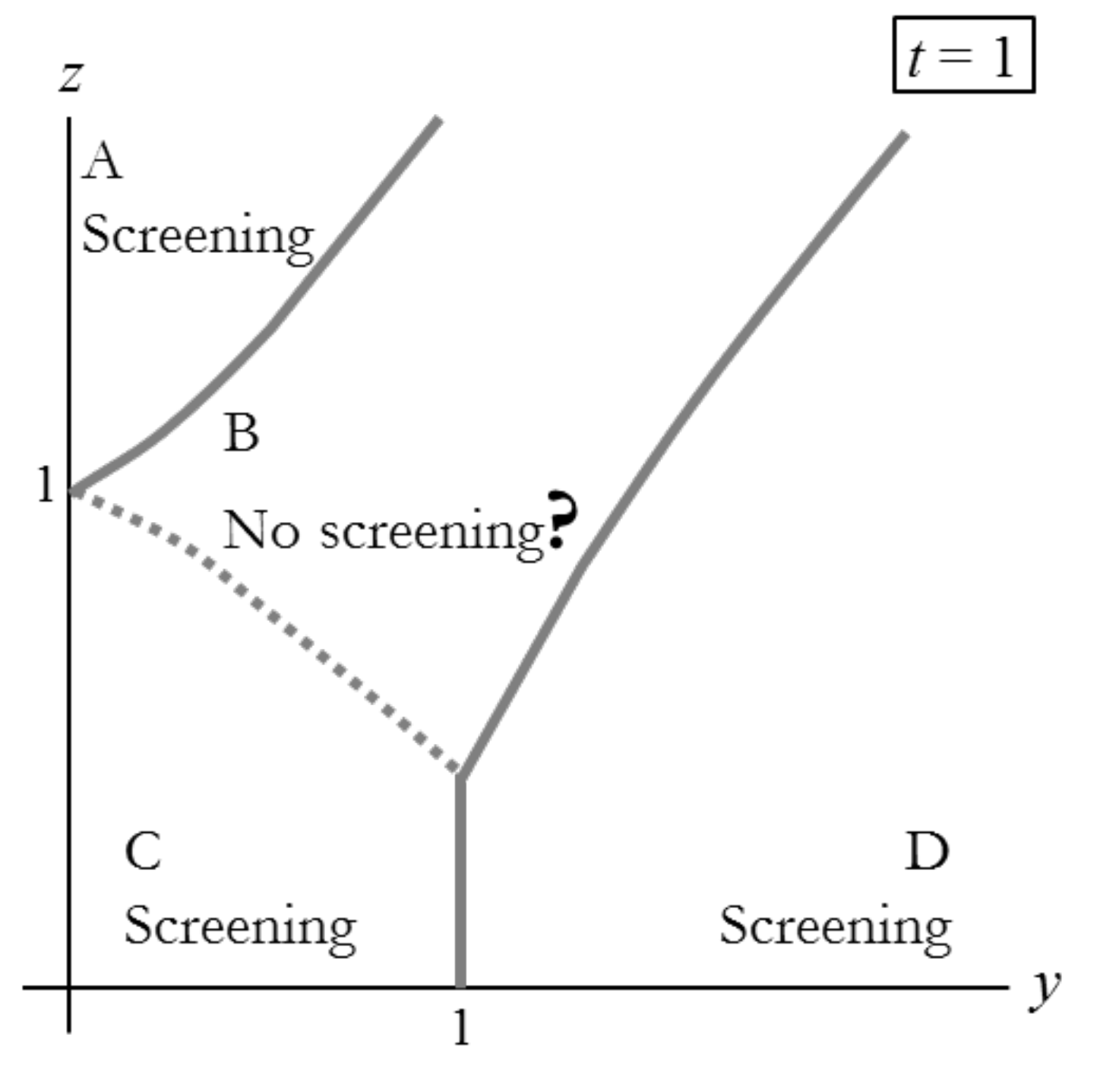}
 \caption{Left panel: The horseshoe order parameter $\rho$ for various lengths $l$, with $t=1$, for the four phases $A,B,C,D$ described in the text.  Data was sampled in the points $A\left(y=0,z=5\right)$, $B\left(y=1.32,z=1.77\right)$, $C\left(y=0.1,z=0.1\right)$, $D\left(y=5,z=0.1\right)$. Right panel: A schematic plot of the phase diagram for the gauge theory with dynamical fermionic matter ($t=1$), with $y,z \geq 0$. The $A,C,D$ phases seem to have a charge screening mechanism, while $B$'s behavior is unclear, but it might not screen.
 } \label{fig:FM}
\end{figure}

We have evaluated, for representative points at each region of the phase diagram for $t=1$, both $\rho \left(l\right)$ for various lengths ($L_1=8$, $L_2=40+l$) and the expectation values of Wilson loops with different width and length. The results may be seen in Figures \ref{fig:FM} and \ref{fig:Wlt1}. It is indeed important to consider both order parameters: as explained, the horeshoe/Fredenhagen-Marcu arguments are valid when the Wilson loops follow a perimeter decay, which is expected in the presence of a fundamental dynamical charge in a conventional (e.g. Kogut-Susskind) theory. However, in our case, due to the truncation and the PEPS construction, one should not necessarily expect only a perimeter law behavior, and, indeed, our numerical calculations show a more complicated scenario.
According to the numerical results, it seems that regions $A,C,D$ obey a perimeter law for the Wilson loop, while the region $B$ follows an area law (see Fig. \ref{fig:Wlt1}).

After having presented the Wilson loop results, we can go on to the results of the horseshoe parameter $\rho$.
It may be clearly seen from the numerical data, that $\rho \rightarrow c \neq 0$ for increasing values of $l$, in the three regions $A,C$ and $D$, whereas $\rho \rightarrow 0$ for the phase $B$. A perimeter law behavior of the Wilson loop is apparent in $A,C,D$, therefore one could most likely interpret these regions as phases which exhibit screening of the dynamical charge.
This might suggest that the limit $t \rightarrow 0$ is not analytical indeed, as the inclusion of dynamical fermions may turn the pure-gauge deconfined phases $C,D$ into screened ones in the dynamical matter case.

Region $B$  displays a less straightforward behavior. If it had a perimeter law for the Wilson loop, $\rho \rightarrow 0$ would mean that the charges in this region are not screened. However, since it manifests an area law behavior we cannot claim that with certainty. It might be still a confining/screening phase, due to the area law behaviour of the Wilson loop, although one must keep in mind that the area law is known as a valid confinement criterion only for static charges. Wilson loops with an area law in a conventional (Kogut-Susskind) theory with fundamental charges do not exist, and thus this result cannot be clearly interpreted.

Another interesting feature of this phase diagram, as mentioned before, is the lack of a phase boundary between the $B$ and $C$ phases, although they manifest different physical properties as described above. This may be understood, perhaps, as part of the discontinuity in the $t \rightarrow 0$ limit, where the phase boundary of $t=0$ disappears. Such a crossover between the two regimes is also reminiscent of the non-separated phases in the Abelian models studied by Fradkin and Shenker \cite{Fradkin1978}, although those models are very different from ours, since they involve bosonic Higgs field, rather than the fermionic matter discussed here.

%

\begin{figure}[ht]
\centering
 \includegraphics[width=0.5\columnwidth]{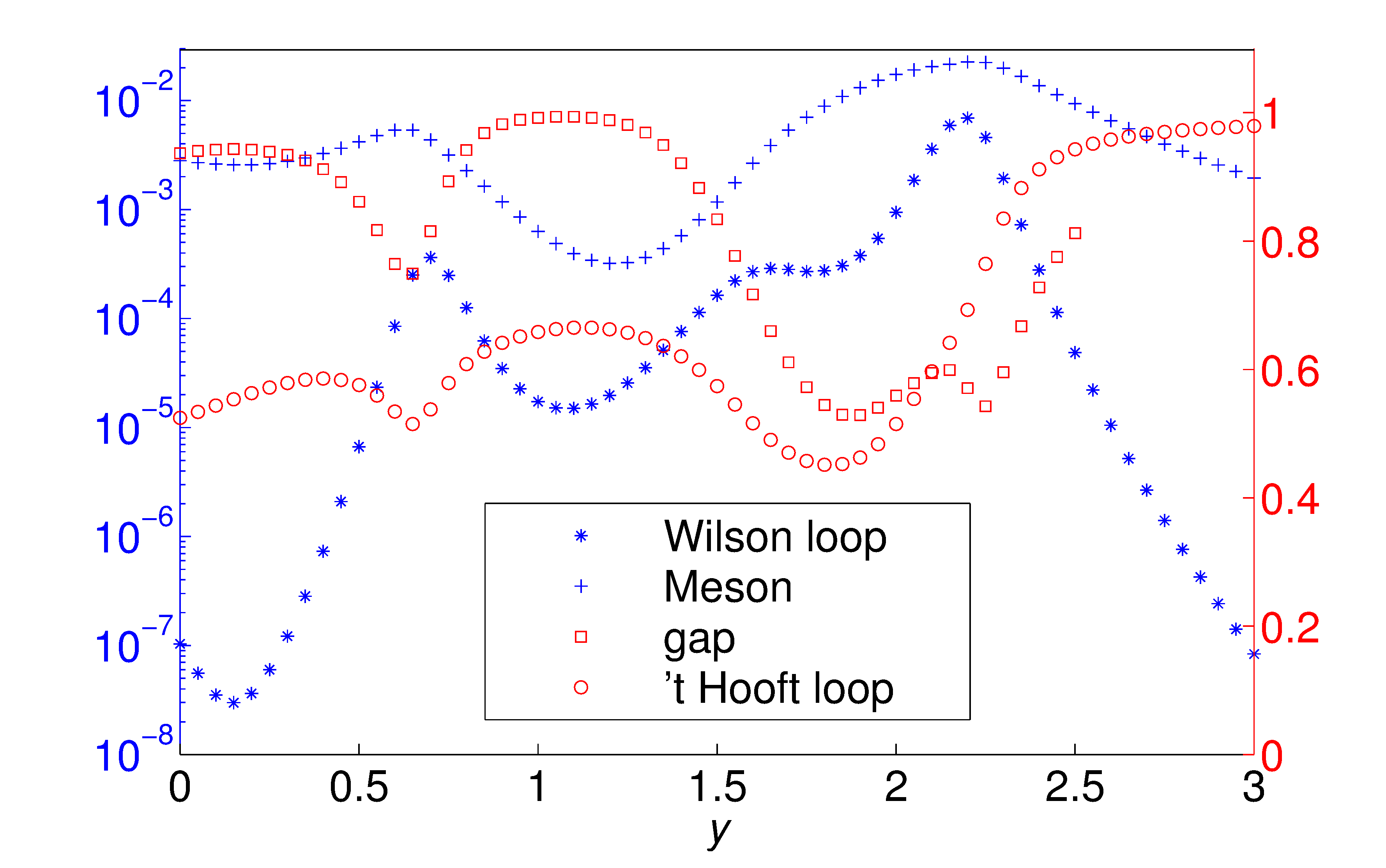}
 \caption{Magnitude of the expectation values of the Wilson loop (size $3 \times 5$), 't Hooft loop (size $3 \times 5$), meson string of length 5 and the gap $\varDelta$ for $L_1=6$, $t=1$, $z=1.5$ as a function of $y$. The expectation values seem to behave non-analytically at around $y \approx 0.65$ and in the region $1.85 \leq y \leq 2.25$ indicating transitions between the phases $A$ and $B$ and $B$ and $D$, respectively. The Wilson loop and meson expectation values are plotted in a semilogarithmic scale (left axis), whereas the gap $\varDelta$ and the 't Hooft loop refer on the normal scale (right axis). } \label{fig:phase_trans}
\end{figure}

\begin{figure}[ht]
\centering
\includegraphics[width=0.4\textwidth]{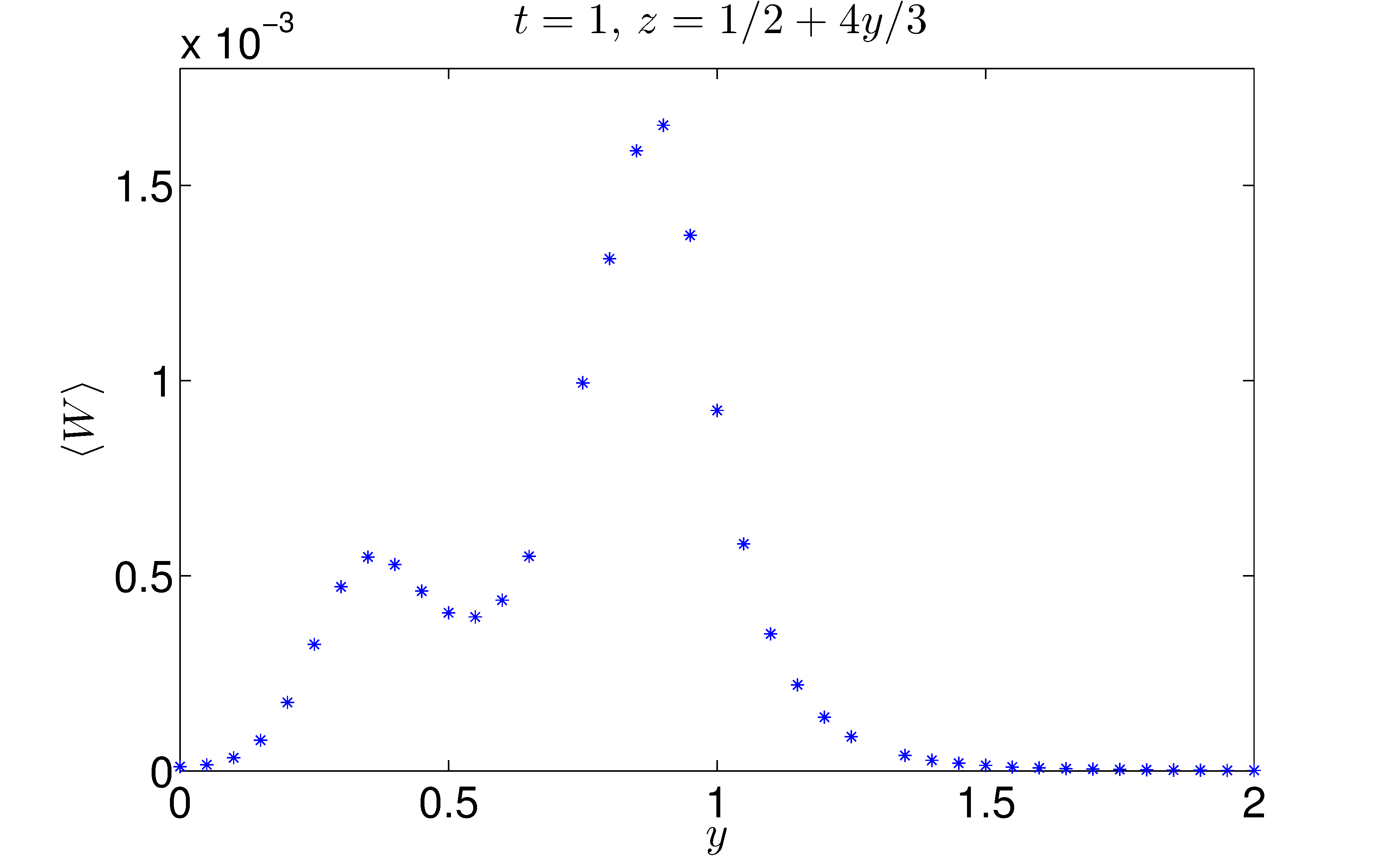}
 \caption{The expectation value of a Wilson Loop along a cut through the regions $B,C$ (the line $z=1/2 + 4/3 y$). A "transition regime" may be seen between the two peaks, however, they seem too smooth for a phase transition, agreeing with other numerical findings which refute the existence of a phase boundary between the regions. } \label{fig1243}
\end{figure}

A schematic plot of the phase diagram is given in Fig. \ref{fig:FM}.
We further corroborated the structure of the phase diagram by calculating various expectation values (Wilson loop, 't Hooft loop and meson string) for $z = 1.5$ as a function of $y \in [0,3]$. The results are shown in Fig.~\ref{fig:phase_trans}. The expectation values display pronounced peaks and cusps at the points where the gap gets small, consistently with a possible non-analytical behavior in the thermodynamic limit. This indicates quantum phase transitions at the intersections of the line $z = 1.5$ with the phase boundaries shown in Fig.~\ref{fig:FM}. On the other hand, a similar plot, of the expectation value of a $4 \times 3$ Wilson loop along the line $z=1/2 + 4/3 y$ can be seen in figure \ref{fig1243}. This line goes, as $y,z$ increase, from the $C$ region to the $B$ region, however, no non-analytical behavior related to a phase transition is seen, but rather two small, seemingly smooth peaks, presumably marking the edges of some transition region between the $B,C$ regions.

\section{Summary}
In this work, we have presented a method to extend the class of symmetric PEPS to local gauge symmetries. In particular, we have focused, as a first demonstration, on a $2+1$ dimensional truncated compact QED with fermionic matter, but the methods presented throughout this paper are generalizable to other gauge groups \cite{future}, and, theoretically speaking, also to higher dimensions.

We have shown how to construct, using the Gaussian formalism, globally invariant states for staggered fermions on a square lattice, satisfying fundamental physical symmetries; we have studied such states analytically, focusing on their parent BdG Hamiltonians and phase diagram. Then, we  introduced suitable gauge degrees of freedom on the lattice links to gauge the symmetry and made it local. In this way we  obtained a set of locally $U(1)$ invariant states of both fermionic matter and gauge fields, with similar physical symmetries (and parametrization) as those of the purely fermionic, globally invariant states.

Important properties of the PEPS have been discussed and demonstrated as tools for the study of the states we constructed. For example, we have used the spectrum of the transfer matrix of the PEPS, which is easily obtained numerically, to define the phase diagram of the system, by distinguishing gapped and gapless regions. We have also demonstrated the possibility of measuring several key physical observables for the study of the PEPS, such as Wilson loops or horseshoe order parameters. Through these results we have been able to sketch and characterize a phase diagram for truncated compact QED in $2+1$ dimensions, which presumably includes both confined and deconfined phases in the pure gauge case and possibly some screening phases in the presence of dynamical matter.
 However, the main goal of this paper is to provide a demonstration and a proof of principle of the power of PEPS for lattice gauge theory calculations, and the ability of PEPS to encode local gauge symmetries and describe the states of lattice gauge theories.

The methods presented here can be further generalized and exploited for the study of other lattice gauge theories \cite{future}. With the  application of efficient numerical techniques and methods, they may be used for a massive, systematic study of lattice gauge theories, and help shed light on their physics - e.g., such states, presumably with a larger bond dimension, may be used as variational ans\"atze for the Kogut-Susskind Hamiltonian \cite{KogutSusskind}. The road along this direction is still long, but with the use of current ideas and techniques, as given in this work and other ones recently published (see for example \cite{Tagliacozzo2014,Haegeman2014}), one could add tensor networks, and PEPS in particular, to the stack of available computational methods in high energy physics, and benefit a lot from the rich variety of options they suggest.

\subsection*{Acknowledgments}
The authors would like to thank Marcello Dalmonte, Jutho Haegeman, Tao Shi, Luca Tagliacozzo and Frank Verstraete for helpful discussions. The authors acknowledge support from the EU Integrated Project SIQS. EZ acknowledges the support of the Alexander-von-Humboldt foundation, through its fellowship for postdoctoral researchers.
\appendix

\setcounter{sttmnt}{0}
\renewcommand{\thesttmnt}{A\arabic{sttmnt}}

\section{The Covariance Matrix Parametrization} \label{app:covariance}

In this Appendix we investigate the properties of the Gaussian fiducial states $\ket{F}$ fulfilling the global U(1) gauge symmetry under the point of view of their covariance matrix. This analysis provides a deeper insight of statement \ref{th:gauss} and poses the basis for the explicit calculation of the parent Hamiltonian and correlations of the state $\ket{\psi(T)}$.

A covariance matrix for (Dirac) fermionic
operators $\alpha_j, \alpha_j^\dag$ may be decomposed into the following block structure \cite{Kraus2009b}:
\begin{equation}
\varGamma=\left(\begin{array}{cc}
\overline{\mathcal{R}} & \overline{\mathcal{Q}}\\
\mathcal{Q} & \mathcal{R}
\end{array}\right)
\end{equation}
where
\begin{equation}
\begin{aligned}
&\mathcal{Q}_{kl}=\frac{i}{2}\left\langle \left[\alpha_{k},\alpha_{l}\right]\right\rangle ,\\
&\mathcal{R}_{kl}=\frac{i}{2}\left\langle \left[\alpha_{k},\alpha_{l}^{\dagger}\right]\right\rangle.
\end{aligned}
\end{equation}
From these definitions, it follows that $\mathcal{R}$ is anti-Hermitian, $\mathcal{R}=-\mathcal{R}^{\dagger}$,
and that $\mathcal{Q}$ is anti-symmetric, $\mathcal{Q}=-\mathcal{Q}^{T}$. A pure state satisfies
$\varGamma\varGamma^{\dagger}=\frac{1}{4}\Id$.

Our PEPS construction for the globally invariant state relies on the fiducial Gaussian states $\ket{F}$ defined by the Equations (\ref{fiducial1},\ref{eqA}).
Let us concentrate on an even vertex  (the case of an odd vertex will be obtained by exchanging positive with negative modes). There, if we order the operators such that the negative modes come before the positive ones, we obtain the following block structure:
\begin{equation}
\varGamma=\left(\begin{array}{cccc}
\overline{\mathcal{R}}_{aa} & \overline{\mathcal{R}}_{ab} & \overline{\mathcal{Q}}_{aa} & \overline{\mathcal{Q}}_{ab}\\
\overline{\mathcal{R}}_{ba} & \overline{\mathcal{R}}_{bb} & \overline{\mathcal{Q}}_{ba} & \overline{\mathcal{Q}}_{bb}\\
\mathcal{Q}_{aa} & \mathcal{Q}_{ab} & \mathcal{R}_{aa} & \mathcal{R}_{ab}\\
\mathcal{Q}_{ba} & \mathcal{Q}_{bb} & \mathcal{R}_{ba} & \mathcal{R}_{bb}
\end{array}\right).
\label{covblock}
\end{equation}

Let us consider $\widetilde{\varGamma}$, the covariance matrix of
the state obtained from the fiducial state by a \emph{gauge tranformation}:
\begin{equation}
\ket{\widetilde{F}} =e^{iG\phi}\left|F\right\rangle.
\end{equation}
From the definition of $G$ (\ref{fiducialG}) we obtain,
\begin{equation}
\widetilde\varGamma=\left(\begin{array}{cccc}
\overline{\mathcal{R}}_{aa} & e^{2i\phi}\overline{\mathcal{R}}_{ab} & e^{2i\phi}\overline{\mathcal{Q}}_{aa} & \overline{\mathcal{Q}}_{ab}\\
e^{-2i\phi}\overline{\mathcal{R}}_{ba} & \overline{\mathcal{R}}_{bb} & \overline{\mathcal{Q}}_{ba} & e^{-2i\phi}\overline{\mathcal{Q}}_{bb}\\
e^{-2i\phi}\mathcal{Q}_{aa} & \mathcal{Q}_{ab} & \mathcal{R}_{aa} & e^{-2i\phi}\mathcal{R}_{ab}\\
\mathcal{Q}_{ba} & e^{2i\phi}\mathcal{Q}_{bb} & e^{2i\phi}\mathcal{R}_{ba} & \mathcal{R}_{bb}
\end{array}\right)
\end{equation}
or simply
\begin{equation}
\widetilde{\varGamma}=V\left(\phi\right)\varGamma V^{\dagger}\left(\phi\right)
\end{equation}
where
\begin{equation}
V\left(\phi\right)=e^{i\phi V_0}
\end{equation}
and
\begin{equation}
V_0=\left(\begin{array}{cccc}
\Id & 0 & 0 & 0\\
0 & -\Id & 0 & 0\\
0 & 0 & -\Id & 0\\
0 & 0 & 0 & \Id
\end{array}\right)
\end{equation}
is the generator of the covariance matrix transformation correponding
to the gauge transformation.

\begin{sttmnt} $\widetilde{\varGamma}=\varGamma$, if and only if
the fiducial state is gauge invariant, i.e. $G\left|F\right\rangle =q\left|F\right\rangle $
for some integer $q$.
\end{sttmnt}
\emph{Proof}:
\begin{enumerate}
\item $G\left|F\right\rangle =q\left|F\right\rangle \Longrightarrow\widetilde{\varGamma}=\varGamma$:
Since $\left|F\right\rangle $ is an eigenstate of the gauge transformation,
$\ket{\widetilde{F}} =e^{i\phi q}\left|F\right\rangle $
- the two states only differ by a global phase,
thus the corresponding expectation values of all operators
are identical, in particular the ones from which the covariance matrices
are constructed.
\item $\widetilde{\varGamma}=\varGamma\Longrightarrow G\left|F\right\rangle =q\left|F\right\rangle $:
If the covariance matrices of both states are equal, and the related states are Gaussian,
they correspond to the same state and thus may be described by equivalent
density matrices (as Gaussian states are completely classified by
their covariance matrices). Moreover, the states are pure, and thus
may differ only by a global phase. Thus we deduce that $\ket{\widetilde{\psi}} =e^{i\theta}\left|\psi\right\rangle $,
or that $e^{iG\phi}\left|F\right\rangle =e^{i\theta}\left|F\right\rangle $.
Since the spectrum of $G$ contains only integer eigenvalues, we deduce
that there exists an integer $q$ such that $e^{iG\phi}\left|F\right\rangle =e^{iq\phi}\left|F\right\rangle $,
and that completes the proof. $\square$
\end{enumerate}

What, then, does the equality of the covariance matrices imply on the block
structure?

\begin{sttmnt} $\widetilde{\varGamma}=\varGamma$, if and only if
$\mathcal{R}_{ab}=0,\mathcal{R}_{ba}=0,\mathcal{Q}_{aa}=0,\mathcal{Q}_{bb}=0$.
\end{sttmnt}

\emph{Proof}: $\widetilde{\varGamma}=\varGamma$ if and only if $\left[V_0,\varGamma\right]=0$
(considering an infinitesimal transformation). We denote
\begin{equation}
V_0=\left(\begin{array}{cc}
\tilde V_0 & 0\\
0 & -\tilde V_0
\end{array}\right),\quad \tilde V_0=\left(\begin{array}{cc}
\Id_{aa} & 0\\
0 & -\Id_{bb}
\end{array}\right),
\end{equation}
where $\Id$ labels suitable identity matrices, and we calculate the commutator:
\begin{equation}
\left[V_0,\varGamma\right]=-\left(\begin{array}{cc}
\left[\bar{\mathcal{R}},\tilde V_0\right] & -\left\{ \overline{\mathcal{Q}},\tilde V_0\right\} \\
\left\{ \mathcal{Q},\tilde V_0\right\}  & \left[\tilde V_0,\mathcal{R}\right]
\end{array}\right)
\end{equation}
this will vanish as long as $\mathcal{R}$ commutes with $\tilde V_0$ and $\mathcal{Q}$ anti-commutes
with it, which ensures the block structure. $\square$

Note that if $\left|F\right\rangle $ is gauge invariant, the expectation
values of all the non gauge invariant operators with respect to it
must vanish, implying the block structure: this implies one direction
in the combination of the two above statements.

\begin{sttmnt} \label{lemma:A3} If $G\left|F\right\rangle =q\left|F\right\rangle $,
then $\mathrm{Tr}\left(\mathcal{R}_{bb}\right)-\mathrm{Tr}\left(\mathcal{R}_{aa}\right)=-iq+\frac{i}{2}\left(N_{p}-N_{n}\right)$.
\end{sttmnt}
\emph{Proof:} $G\left|F\right\rangle =q\left|F\right\rangle $ means that
\begin{equation}
\left\langle F\right|G\left|F\right\rangle =q\label{eq:G0-1}
\end{equation}
Recall that
\begin{equation}
\left\langle F\right|G\left|F\right\rangle =\underset{k}{\sum}\left\langle F\right|b_{k}^{\dagger}b_{k}\left|F\right\rangle
-\underset{k}{\sum}\left\langle F\right|a_{k}^{\dagger}a_{k}\left|F\right\rangle
\end{equation}
and
\begin{equation}
\left\langle F\right|\alpha_{k}^{\dagger}\alpha_{k}\left|F\right\rangle =-\frac{1}{2}\left\langle F\right|\left[\alpha_{k},\alpha_{k}^{\dagger}\right]-1\left|F\right\rangle =i\mathcal{R}_{kk}+\frac{1}{2}\label{eq:n}
\end{equation}
therefore,
\begin{equation}
\left\langle F\right|G\left|F\right\rangle =i\left(\mathrm{Tr}\left(\,\mathcal{R}_{bb}\right)-\mathrm{Tr}\left(\mathcal{R}_{aa}\right)\right)+\frac{1}{2}\left(N_{p}-N_{n}\right)
\end{equation}
which completes the proof. $\square$

As a corollary, we may state:

\begin{sttmnt} $G\left|F\right\rangle =q\left|F\right\rangle $
if and only if the corresponding covariance matrix (\ref{covblock}) satisfies the following
conditions:
\begin{enumerate}
\item $\mathcal{R}_{ab}=0,\mathcal{R}_{ba}=0,\mathcal{Q}_{aa}=0,\mathcal{Q}_{bb}=0$
\item $\mathrm{Tr}\left(\mathcal{R}_{bb}\right)-\mathrm{Tr}\left(\mathcal{R}_{aa}\right)=-iq+\frac{i}{2}\left(N_{p}-N_{n}\right)$
\end{enumerate}
\end{sttmnt}
Note that $\varGamma\varGamma^{\dagger}=\frac{1}{4}\Id$ as well, but
this is a result of the purity of the state, independent of gauge invariance.
The statement is almost equivalent to statement \ref{th:gauss}; the only missing part is $q=0$, which may be also proven using the covariance matrix approach
(see below).

The block structure is left invariant under canonical (unitary) transformations
of the type $\mathcal{R}\longrightarrow \mathcal{URU}^{\dagger}$, $\mathcal{Q}\longrightarrow \mathcal{UQU}^{T}$, as long as $\mathcal{U}$ is
decomposed into
\begin{equation}
\mathcal{U}=\left(\begin{array}{cc}
\mathcal{U}_{n} & 0\\
0 & \mathcal{U}_{p}
\end{array}\right)\label{eq:CanTrans}
\end{equation}
i.e., $\mathcal{U}\in U\left(N_{n}\right)\times U\left(N_{p}\right)$ and $\left[\mathcal{U},V_0 \right]=0 $. These
transformations are passive, and furthermore do not mix the positive
and negative modes, which guarantees that the symmetry, and thus the
structure of the covariance matrix, is conserved; it is trivial for
$\mathcal{R}$, and straightforward for $\mathcal{Q}$, as $\mathcal{Q}\longrightarrow \mathcal{UQU}^{T}$
results in
\begin{equation}
\mathcal{Q}\longrightarrow\left(\begin{array}{cc}
0 & \mathcal{U}_{n}\mathcal{Q}_{ab}\mathcal{U}_{p}^{T}\\
-\mathcal{U}_{p}\mathcal{Q}_{ab}^{T}\mathcal{U}_{n}^{T} & 0
\end{array}\right).
\end{equation}

Since $\left|F\right\rangle$ is a pure fermionic Gaussian state, it may be written in a BCS form \cite{Kraus2010},
\begin{equation}
\left|F\right\rangle =\underset{k}{\prod}\left(u_{k}+v_{k}\widetilde{a}_{k}^{\dagger}\widetilde{b}_{k}^{\dagger}\right)\left|\Omega\right\rangle
\end{equation}
where $u_{k}^{2}+v_{k}^{2}=1$ (we make the assumption they are real, as shall later be clear from the singular value decomposition of statement \ref{lemma:A4}),
and the covariance matrices take the form (in the alternating ordering $a,b,a,b,..$)
\begin{equation}
\mathcal{Q}_{0}=\left(\underset{k}{\oplus}v_{k}u_{k}\sigma_{y}\right)\oplus 0_{1} , \label{eq:Q0def}
\end{equation}
\begin{equation}
\mathcal{R}_{0}=\frac{i}{2}\left(\left(\underset{k}{\oplus}\left(1-2v_{k}^{2}\right)\Id\right)\oplus \Id_{1}\right), \label{eq:R0def}
\end{equation}
where the uncoupled mode is present since we have an odd number of modes,
$0_1$ is the $1 \times 1$ null matrix, and $\Id_1$ the the $1 \times 1$ identity matrix.

The question then, is, whether this canonical BCS form is the result
of a transformation $\mathcal{U}\in U\left(N_{n}\right)\times U\left(N_{p}\right)$.
We shall see that this is, indeed, the case: intuitively, this must
hold for symmetry reasons. We now show this explicitly, using statement \ref{th:gauss}.

\begin{sttmnt} \label{lemma:A4} The Bogoliubov transformation into the BCS form preserves the gauge symmetry.
\end{sttmnt}

\emph{Proof:} First, perform a singular value decomposition of $T$ in Eq. \eqref{eqA}, to obtain
\begin{equation}
T=W_{n}\Lambda W_{p}^{\dagger}
\end{equation}
where $W_{n}$ is a unitary $N_{n}\times N_{n}$ matrix, $W_{p}$
is a unitary $N_{p}\times N_{p}$ matrix, and $\Lambda$ is a $N_{n}\times N_{p}$
diagonal matrix, whose diagonal (square) part includes the eigenvalues
$\left\{ \lambda_{k}\right\} _{k=1}^{\min\left(N_{n},N_{p}\right)}$,
which are all real and non-negative.

The canonical transformation $\mathcal{U}$ of the kind \eqref{eq:CanTrans},
such that
\begin{equation}
\begin{aligned}
a_i \longrightarrow \,&\tilde a_i = \left(\mathcal{U}_n\right)_{ij}a_j, \\
b_i \longrightarrow \,&\tilde b_i= \left(\mathcal{U}_p\right)_{ij}b_j
\end{aligned}
\end{equation}
with
\begin{equation}
\mathcal{U}_{n}=W_{n}^{\dagger};\quad \mathcal{U}_{p}=W_{p}^{\intercal}
\end{equation}
brings the state into the (normalized) form
\begin{equation}
\left|F\right\rangle =\mathcal{N}^{-1/2}\exp\left(\underset{k}{\sum}\lambda_{k}\widetilde{a}_{k}^{\dagger}\widetilde{b}_{k}^{\dagger}\right)\left|\Omega\right\rangle
\end{equation}
which is the desired BCS state. We identify
\begin{equation}
\mathcal{N}^{-1/2}=\underset{k}{\prod}u_{k},\quad \lambda_{k}u_{k}=v_{k}.
\end{equation}
Using $u_{k}^{2}+v_{k}^{2}=1$ we finally obtain
\begin{equation}
u_{k}=\frac{1}{\sqrt{1+\lambda_{k}^{2}}}, \quad
v_{k}=\frac{\lambda_{k}}{\sqrt{1+\lambda_{k}^{2}}}
\end{equation}
and
\begin{equation}
\mathcal{N}=\underset{k}{\prod}\left(1+\lambda_{k}^{2}\right).
\end{equation}

From this, we can evaluate the BCS covariance matrices, $\mathcal{R}_{0}$ and
$\mathcal{Q}_{0}$. Let us now write them in our usual ordering of the modes
(unlike in equations (\ref{eq:Q0def})-(\ref{eq:R0def})). In the
explicit form below, we assume that $N_{n}=N_{p}+1$ (even vertices), but the following
arguments will also hold for $N_{n}=N_{p}-1$ (odd vertices).
\begin{equation}
\mathcal{R}_{0}=\left(\begin{array}{ccccccc}
\frac{i}{2} & 0 & 0 & 0 & 0 & 0 & 0\\
0 & \frac{i}{2}\frac{1-\lambda_{1}^{2}}{1+\lambda_{1}^{2}} & 0 & 0 & 0 & 0 & 0\\
0 & 0 & \ddots & 0 & 0 & 0 & 0\\
0 & 0 & 0 & \frac{i}{2}\frac{1-\lambda_{N_p}^{2}}{1+\lambda_{N_p}^2} & 0 & 0 & 0\\
0 & 0 & 0 & 0 & \frac{i}{2}\frac{1-\lambda_{1}^{2}}{1+\lambda_{1}^{2}} & 0 & 0\\
0 & 0 & 0 & 0 & 0 & \ddots & 0\\
0 & 0 & 0 & 0 & 0 & 0 & \frac{i}{2}\frac{1-\lambda_{N_p}^{2}}{1+\lambda_{N_p}^{2}}
\end{array}\right)
\end{equation}
\begin{equation}
\mathcal{Q}_{0}=\left(\begin{array}{ccccccc}
0 & 0 & 0 & 0 & 0 & 0 & 0 \\
0 & 0 & 0 & 0 & -\frac{i\lambda_{1}}{\left(1+\lambda_{1}^{2}\right)} & 0 & 0\\
0 & 0 & 0 & 0 & 0 & \ddots & 0\\
0 & 0 & 0 & 0 & 0 & 0 & -\frac{i\lambda_{N_p}}{\left(1+\lambda_{N_p}^{2}\right)}\\
0 & \frac{i\lambda_{1}}{\left(1+\lambda_{1}^{2}\right)} & 0 & 0 & 0 & 0 & 0\\
0 & 0 & \ddots & 0 & 0 & 0 & 0\\
0 & 0 & 0 & \frac{i\lambda_{N_p}}{\left(1+\lambda_{N_p}^{2}\right)} & 0 & 0 & 0
\end{array}\right)
\end{equation}
According to statement \ref{lemma:A3}, $\mathrm{Tr}\left(\mathcal{R}_{bb}\right)-\mathrm{Tr}\left(\mathcal{R}_{aa}\right)=-iq+\frac{i}{2}\left(N_{p}-N_{n}\right)$.
From this, assuming that $\left|N_{p}-N_{n}\right|=1$, we immediately get that $q=0$:
the only states suitable for this representations are the ones with zero static charge, which agrees with statement \ref{th:gauss}.
 $\square$

If we denote the covariance matrix of the BCS modes by $\varGamma_{0}$,
the covariance matrix $\varGamma$ for the original modes contains
the blocks
\begin{equation}
\mathcal{R}=\mathcal{U}^{\dagger}\mathcal{R}_{0}\mathcal{U}
\end{equation}
and
\begin{equation}
\mathcal{Q}=\mathcal{U}^{\dagger}\mathcal{Q}_{0}\overline{\mathcal{U}}
\end{equation}
and can be obtained as
\begin{equation}
\varGamma=\left(\begin{array}{cc}
\mathcal{U}^{T} & 0\\
0 & \mathcal{U}^{\dagger}
\end{array}\right)\varGamma_{0}\left(\begin{array}{cc}
\overline{\mathcal{U}} & 0\\
0 & \mathcal{U}
\end{array}\right).
\end{equation}

As a corollary, we deduce that the most general gauge invariant fermionic
Gaussian state with no static charges can be parametrized by either the set of real numbers $\left\{ \lambda_{k}\right\} _{k=1}^{\min\left(N_{n},N_{p}\right)}$
and the canonical transformation $\mathcal{U}\in U\left(N_{n}\right)\times U\left(N_{p}\right)$, or the $5 \times 4$ complex matrix $T$. We shall adopt the second parametrization,
and see how one may reduce the number of parameters when the desired symmetries are demanded. For that we shall now turn to the construction of the non-local state.

The covariance matrix of the fiducial states, both even and odd, is the one previously investigated,
up to the difference between even and odd sites (staggering). For
even sites $N_{n}=5,N_{p}=4$, and the other way around for odd ones.
We may deduce that both the even and odd sites possess the same covariance
matrix, with a different ordering of the modes: in even sites, the
negative modes will come first, thus having the ordering $\left\{\psi,l_{+},r_{-},u_{-},d_{+},l_{-},r_{+},u_{+},d_{-}\right\}$,
while on the odd sites, the positive modes come first, i.e. $\left\{\psi,l_{-},r_{+},u_{+},d_{-},l_{+},r_{-},u_{-},d_{+}\right\}$.

\section{The Gaussian Mapping} \label{app:gaussian}
Since we have a translationally invariant state, the PEPS may be calculated from the fiducial states and the bond states using a Gaussian mapping \cite{Bravyi05}.

This is done in terms of Majorana fermions $\left\{c_k\right\}$: for any fermionic mode $\alpha_k$, either positive or negative, one defines the Majorana operators
\begin{equation}
c_{2k-1} = \alpha_k + \alpha_k^{\dagger} \,,\quad c_{2k} = i\left(\alpha_k - \alpha_k^{\dagger}\right).
\end{equation}
The covariance matrix of a state is proportional to the commutator of the corresponding Majorana operators:
\begin{equation}
\Gamma_{lm} = \frac{i}{2}\left\langle\left[c_l,c_m\right]\right\rangle
\end{equation}

We denote the Majorana covariance matrix of the fiducial state of a single vertex, $\left|F\right\rangle$, by $M$. It is an $18 \times 18$ real, anti-symmetric matrix, using the following convention for the mode ordering: $\{\psi, l_+, r_-, l_-, r_+, u_-, d_+, u_+, d_- \}$ for even vertices and $\{\psi, l_-, r_+, l_+, r_-, u_+, d_-, u_-, d_+\}$ for odd ones. Note the positive-negative correspondence, in accordance with the staggering and the translational invariance. That means that the $M$ matrices will take the same form
on both the even and odd sublattices, \emph{but in two different bases}. $M$ will be used as the Gaussian channel, which, following \cite{Bravyi05}, may be decomposed   into
\begin{equation}
M=\left(
  \begin{array}{cc}
    A & B \\
    -B^\top & D \\
  \end{array}
\right) \label{gaussm}
\end{equation}
with $A$ being the sub-block for the physical Majorana modes (a $2 \times 2$ matrix), $D$ for the virtual ones ($16 \times 16$) and $B$ is the sub-block of physical with
virtual modes ($2 \times 16$).

The bond states $\left|H\right\rangle ,\left|V\right\rangle $ have the same covariance matrices:
\begin{equation}
\Gamma_{0}= \left(\begin{array}{cccc}
0 &\sigma_{x} & 0 & 0\\
-\sigma_{x} & 0 & 0 & 0\\
0 & 0 & 0 & \sigma_{x}\\
0 & 0 & -\sigma_{x} & 0
\end{array}\right)
\end{equation}
where the ordering of the blocks is according to the ordering of the
corresponding blocks in the covariance matrices $M$ of the fiducial states, defined above. Due
to the positive-negative correspondence, this means, for example, that
the variance matrix of the first virtual mode with the second one
is $\sigma_{x}$; the identity of these ``first'' and ``second''
modes changes depending on the parity of the bond, whether it connects an even vertex to an odd one, or an odd to an even. However,
the mathematical form is identical.

The complete covariance matrix for the bonds will take the form (for
an $L\times L$ lattice):
\begin{equation}\small
\Gamma_{\text{in}}=\bigoplus_{i=2,1}\bigoplus_{x_i=1}^L\left[\mathrm{Perm}\left(L,1,2,...,L-1\right)\otimes\left(\begin{array}{cccc}
0 & \sigma_{x} & 0 & 0\\
0 & 0 & 0 & 0\\
0 & 0 & 0 & \sigma_{x}\\
0 & 0 & 0 & 0
\end{array}\right)+
 \mathrm{Perm}\left(2,3,...,L,1\right)\otimes\left(\begin{array}{cccc}
0 & 0 & 0 & 0\\
-\sigma_{x} & 0 & 0 & 0\\
0 & 0 & 0 & 0\\
0 & 0 & -\sigma_{x} & 0
\end{array}\right)\right]_{x_i}
\label{Gin}
\end{equation}
where $x_i$ labels either the rows or the columns
and $\mathrm{Perm}\left(j_1,...,j_L\right)$ is a permutation matrix whose nonzero elements are the entries $\left\{\left(j_n,n\right)\right\}_{n=1}^L$
The Fourier transform of $\Gamma_{\text{in}}$ is simply:
\begin{equation} \small
G_{\text{in}}\left(k_{1},k_{2}\right)=\left(\begin{array}{cccc}
0 & \sigma_{x}e^{i k_{1}} & 0 & 0\\
-\sigma_{x}e^{-ik_1} & 0 & 0 & 0\\
0 & 0 & 0 & \sigma_{x}e^{ik_1}\\
0 & 0 & -\sigma_{x}e^{-ik_1} & 0
\end{array}\right)\oplus  \left(\begin{array}{cccc}
0 & \sigma_{x}e^{-ik_2} & 0 & 0\\
-\sigma_{x}e^{ik_2} & 0 & 0 & 0\\
0 & 0 & 0 &\sigma_{x}e^{-ik_2}\\
0 & 0 & -\sigma_{x}e^{ik_2} & 0
\end{array}\right).
\end{equation}

We are now ready to apply $M$ as a Gaussian channel on $\Gamma_{\text{in}}$.
All the modes are aligned in a proper way, and since everything is
translationally invariant we can work in the momentum space and obtain the momentum space covariance matrix of the state $\left|\psi\left(T\right)\right\rangle$ \cite{Kraus2010}:
\begin{equation}
G_{\text{out}}\left(\mathbf{k}\right)=A+B\left(D-G_{\text{in}}\left(\mathbf{k}\right)\right)^{-1}B^{\top} \label{eq:Gaussian_map}
\end{equation}
with $A,B,D$ being the blocks in \eqref{gaussm}. Following \cite{Kraus2010}, one obtains
\begin{equation}
\begin{aligned}
G_{\text{out}}\left(\mathbf{k}\right)&=\left(\begin{array}{cc}
iP\left(\mathbf{k}\right) & Q\left(\mathbf{k}\right) \\
-\overline{Q}\left(\mathbf{k}\right) & -iP\left(\mathbf{k}\right)\end{array}\right)
= \frac{1}{\mathcal{D}\left(\mathbf{k}\right)}\left(\begin{array}{cc}
iP_0\left(\mathbf{k}\right) & Q_0\left(\mathbf{k}\right)\\
-\overline{Q}_0\left(\mathbf{k}\right) & -iP_0\left(\mathbf{k}\right)
\end{array}\right),
\label{Gout2}
\end{aligned}
\end{equation}
where
\begin{equation}
\mathcal{D}\left(\mathbf{k}\right)=\det\left(D-G_{\text{in}}\left(\mathbf{k}\right)\right).
\label{ddef}
\end{equation}
By writing the inverse in Eq.~\eqref{eq:Gaussian_map} in terms of the adjugate matrix, it is easy to discern that the functions $P_0(\mb k)$ and $Q_0(\mb k)$ are trigonometric polynomials with a maximum order of 4 in $k_1$ and in $k_2$ individually (the same is true for the determinant $\mathcal{D}(\mb k)$). However, since in the fiducial state $|F\rangle$ positive and negative modes are not mutually entangled due to the gauge symmetry condition, these polynomials might turn out to have lower maximum orders in practice.

Furthermore, due to $G_\mr{out}^\dg(\mb k) = -G_\mr{out}(\mb k)$, $P\left(\mathbf{k}\right)$ is a real function. $Q\left(\mathbf{k}\right)$ may be decomposed into its real and imaginary parts,
\begin{equation}
Q\left(\mathbf{k}\right)=R\left(\mathbf{k}\right)+i I\left(\mathbf{k}\right)
\end{equation}

Using $G_{\text{in}}\left(-\mathbf{k}\right) = \overline{G_{\text{in}}}\left(\mathbf{k}\right)$, and similarly for $G_{\text{out}}$, as $A,B,D$ are real, one obtains from (\ref{Gout2}) that
the following relations apply:
\begin{equation}
P\left(-\mathbf{k}\right) = -P\left(\mathbf{k}\right),R\left(-\mathbf{k}\right) = R\left(\mathbf{k}\right),I\left(-\mathbf{k}\right) = -I\left(\mathbf{k}\right)
\end{equation}
and from the purity of the final state
\begin{equation}
\mathcal{D}\left(\mathbf{k}\right) = \sqrt{R_0^2\left(\mathbf{k}\right)+I_0^2\left(\mathbf{k}\right)+P_0^2\left(\mathbf{k}\right)}
\end{equation}
Furthermore, since $d\left(\mathbf{k}\right)$ is real,
\begin{equation}
\mathcal{D}\left(-\mathbf{k}\right)=\overline{\mathcal{D}}\left(\mathbf{k}\right)=\mathcal{D}\left(\mathbf{k}\right)
\end{equation}
and thus
\begin{equation}
P_0\left(-\mathbf{k}\right) = -P_0\left(\mathbf{k}\right),R_0\left(-\mathbf{k}\right) = R_0\left(\mathbf{k}\right),I_0\left(-\mathbf{k}\right) = -I_0\left(\mathbf{k}\right)
\end{equation}
hold as well.

\section{Proofs of statements 5-7} \label{app:C}
Some of the analytical results for the globally invariant case, namely statements 5-7, are strongly based on the PEPS construction, and thus their proofs have been omitted from the main text. We hereby give the full proofs of these three statements. But for that, we begin, first, with a formulation of the PEPS in momentum space.

Following the convention for physical fermions (\ref{FTconv}), we define a similar Fourier transform for virtual fermions:
\begin{equation}
\begin{aligned}
&a_{\mathbf{x}}^{\dagger} = \frac{1}{\sqrt{L_1 L_2}} \underset{\mathbf{k}}{\sum} e^{i  \mathbf{k} \cdot \mathbf{x}} a_{\mathbf{k}}^{\dagger}, \\
&b_{\mathbf{x}}^{\dagger} = \frac{1}{\sqrt{L_1 L_2}} \underset{\mathbf{k}}{\sum} e^{i  \mathbf{k} \cdot \mathbf{x}} b_{\mathbf{k}}^{\dagger}.
\end{aligned}
\end{equation}

To manifest translational invariance better, we swap the $a$ and $b$ operators on odd sites, i.e. exchange all the $a$ operators there by $b$ operators and vice versa. This results
in a uniform expression for the operator $A$ everywhere,
\begin{equation}
A\left(\mathbf{x}\right) = \exp \left(T_{ij} a_{i,\mathbf{x}}^{\dagger} b_{j,\mathbf{x}}^{\dagger}\right)
\end{equation}
where the $j$ indices are $1,...,4$, involving only virtual fermions, while these of $i$ are $0,...,4$, with $a_0 \equiv \psi$ the physical fermion.

The state $\ket{\psi(T)}$ defined in \eqref{globPEPS} involves a product of these operators everywhere, which results in a summation, in the exponent, over all the sites; we can perform a Fourier transform,
and obtain
\begin{equation}
\underset{\mathbf{x}}{\prod} A\left(\mathbf{x}\right) = \underset{\mathbf{k}'}{\prod} A\left(\mathbf{k}'\right) A\left(0,0\right) A\left(\pi,\pi\right) A\left(\pi,0\right) A\left(0,\pi\right)
\end{equation}
where  $\mathbf{k}'$ involves only half of the $\mathbf{k}$ values in the Brillouin zone (the ``positive momenta''), except for
$\left(0,0\right) , \left(\pi,\pi\right), \left(\pi,0\right), \left(0,\pi\right)$,
and, as a result of the Fourier transform,
\begin{equation}
A\left(\mathbf{k}\right) = \exp\left(T_{ij}  a_i^{\dagger}\left(\mathbf{k}\right) b_j^{\dagger}\left( - \mathbf{k}\right)\right)
\exp\left(T_{ij}  a_i^{\dagger}\left(-\mathbf{k}\right) b_j^{\dagger}\left(  \mathbf{k}\right)\right)
\end{equation}

We continue with Fourier transforming the projectors. The product of all projectors may be denoted as
\begin{equation}
\left| \left\{HV\right\}\right\rangle \left\langle\left\{HV\right\}\right| \equiv \bigotimes \left|H\right\rangle \left\langle H\right| \otimes \left|V\right\rangle \left\langle V\right|
\end{equation}
(note that as these operators are quadratic in the fermionic operators, a tensor product is well defined)
with
\begin{equation}
\left|\left\{HV\right\}\right\rangle \equiv
\prod_{\mathbf{k}'} B^{\dagger}\left(\mathbf{k}'\right) B^{\dagger}\left(0,0\right) B^{\dagger}\left(\pi,\pi\right)
B^{\dagger}\left(\pi,0\right) B^{\dagger}\left(0,\pi\right)
\left|\Omega_v\right\rangle
\end{equation}
and
\begin{equation}
B\left(\mathbf{k}\right) = \exp \left(S_{ij}\left(\mathbf{k}\right)a_i\left(\mathbf{k}\right)a_j\left(-\mathbf{k}\right)\right) \times
\exp \left(S_{ij}\left(\mathbf{k}\right)b_i\left(\mathbf{k}\right)b_j\left(-\mathbf{k}\right)\right),
\end{equation}
where
\begin{equation}
S\left(k_x,k_y\right) = \left(
                          \begin{array}{cccc}
                            0 & -e^{-i  k_1} & 0 & 0 \\
                            e^{i  k_1} & 0 & 0 & 0 \\
                            0 & 0 & 0 & -e^{-i k_2} \\
                            0 & 0 & e^{i k_2} & 0 \\
                          \end{array}
                        \right).
\end{equation}
Since $\left\langle \Omega_v | \left\{HV\right\}\right\rangle$ is just an irrelevant constant (equal to 1), the physical state is simply
\begin{equation}
\left|\psi\left(T\right)\right\rangle = \left\langle \Omega_v\right| \prod B \prod A \left|\Omega_v\right\rangle \left|\Omega_p\right\rangle
\end{equation}
and we can decompose it into a tensor product of momentum states,
\begin{equation}
\left|\psi\left(T\right)\right\rangle =
\underset{\mathbf{k}'}{\bigotimes} \left|\psi\left(\mathbf{k}'\right)\right\rangle \otimes \left|\psi\left(0,0\right)\right\rangle
\left|\psi\left(\pi,\pi\right)\right\rangle \left|\psi\left(\pi,0\right)\right\rangle \left|\psi\left(0,\pi\right)\right\rangle
\end{equation}
where we separated paired and unpaired modes. The paired ones have the form:
\begin{equation}
\left|\psi\left(\mathbf{k'}\right)\right\rangle = \left\langle\Omega_v\right|B\left(\mathbf{k'}\right)A\left(\mathbf{k'}\right) \left|\Omega\right\rangle
\end{equation}
with $\left|\Omega_v\right\rangle, \left|\Omega\right\rangle $ being the vacua of the relevant momentum subspace (virtual and general respectively),
and a tensor product structure is again well defined as all the states in the product have an even fermionic parity.

For $\mathbf{k} = \left(0,0\right) , \left(\pi,\pi\right), \left(\pi,0\right), \left(0,\pi\right)$, one defines instead $\tilde{S}\left(\mathbf{k}\right) = S\left(\mathbf{k}\right)/2$: the operators $B$ are defined using the $\tilde{S}$ matrices instead of the $S$ ones, and
$A\left(\mathbf{k}\right) = \exp \left(T_{ij} a_i^{\dagger}\left(\mathbf{k}\right) b_j^{\dagger}\left(\mathbf{k}\right)\right)$.

\emph{Proof of statement \ref{th:dispersion}}:
Denote the blocks of the covariance matrix of the fiducial state by $\tilde A,\tilde B,\tilde D$, the initial bond state by $\left|\psi_{\text{in}}\right\rangle$ and the output state by
$\left|\psi_{\text{out}}\right\rangle$. Then (assuming the number of virtual fermionic modes is even, which is, indeed, our case), the norm of the output state is given by
 \cite{Bravyi05}
\begin{equation}
\left\langle \psi_{\text{out}} | \psi_{\text{out}} \right\rangle  =
 \text{Pf} \left(\tilde D - \tilde G_{\text{in}}\right)  \left\langle \psi_{\text{in}} | \psi_{\text{in}} \right\rangle
\end{equation}
where $\text{Pf}$ denotes the Pfaffian.

In our case, the fiducial state was not normalized, and thus the fact that $\left|\psi_{\text{out}}\right\rangle$ is not normalized is not a mere outcome of the Gaussian mapping.
Therefore we modify this formula by multiplying by a further constant $\gamma$. That shall be done carefully, considering the paired and unpaired momenta separately.
For the paired momenta, $\tilde D = D \oplus D$ and
$\tilde G_{\text{in}} = G_{\text{in}}\left(\mathbf{k}\right) \oplus G_{\text{in}}\left(-\mathbf{k}\right)$, while for the unpaired ones,
$\tilde D = D$ and $\tilde G_{\text{in}} = G_{\text{in}}\left(\mathbf{k}\right)$. Next, since the fiducial states lack some normalization factors, we deduce that such factors should appear twice for
paired momenta, and once for unpaired momenta, i.e.:
\begin{equation}
\left\langle \psi\left(\mathbf{k}\right) | \psi\left(\mathbf{k}\right) \right\rangle = \\
\gamma
 \text{Pf} \left[ D \oplus D - G_{\text{in}}\left(\mathbf{k}\right) \oplus G_{\text{in}}\left(-\mathbf{k}\right)\right]
 \left\langle \psi_{\text{in}}\left(\mathbf{k}\right) | \psi_{\text{in}}\left(\mathbf{k}\right) \right\rangle
 \end{equation}
for paired momenta, and
\begin{equation}
\left\langle \psi\left(\mathbf{k}\right) | \psi\left(\mathbf{k}\right) \right\rangle = \sqrt{\gamma}
 \text{Pf} \left[ D  - G_{\text{in}}\left(\mathbf{k}\right) \right]  \left\langle \psi_{\text{in}}\left(\mathbf{k}\right) | \psi_{\text{in}}\left(\mathbf{k}\right) \right\rangle
\end{equation}
for unpaired momenta.

For the paired momenta, since $\mathcal{D}\left(\mathbf{k}\right) = \mathcal{D}\left(-\mathbf{k}\right) = \det\left(D-G_{\mr{in}}\left(\mathbf{k}\right)\right)$,
one simply obtains:
\begin{equation}
\mathcal{D}\left(\mathbf{k}\right) =
\frac{\left\langle \psi\left(\mathbf{k}\right) | \psi\left(\mathbf{k}\right) \right\rangle}{\gamma   \left\langle \psi_{\text{in}}\left(\mathbf{k}\right) | \psi_{\text{in}}\left(\mathbf{k}\right) \right\rangle}
=\frac{\left|\alpha\left(\mathbf{k}\right)\right|^2+\left|\beta\left(\mathbf{k}\right)\right|^2}{256 \gamma },
\end{equation}
whereas for the unpaired ones:
\begin{equation}
\mathcal{D}\left(\mathbf{k}\right) =
\frac{\left\langle \psi\left(\mathbf{k}\right) | \psi\left(\mathbf{k}\right) \right\rangle ^2}
{\gamma  \left\langle \psi_{\text{in}}\left(\mathbf{k}\right) | \psi_{\text{in}}\left(\mathbf{k}\right) \right\rangle ^2}
=\frac{\left|\tilde \alpha\left(\mathbf{k}\right)\right|^4}{256 \gamma}
\end{equation}
thus we obtain continuity if $\tilde \alpha \left(\mathbf{k}\right) = \sqrt{\alpha \left(\mathbf{k}\right)}$ -
which, as we shall show, holds. Given this assumption which will shortly be proven, we define
\begin{equation}
E\left(\mathbf{k}\right) = \left|\alpha\left(\mathbf{k}\right)\right|^2+\left|\beta\left(\mathbf{k}\right)\right|^2
\end{equation}
as the dispersion relation we work with. This proves statement \ref{th:dispersion} $\square$

We shall now turn to the calculation of the functions $\alpha\left(\mathbf{k}\right),\beta\left(\mathbf{k}\right)$ - the proofs of statements \ref{th:BCScoeff} and \ref{th:unpaired}. For that, we first prove a more general statement.

\begin{sttmnt} \label{lemma:expect}
The expectation value
\begin{equation}
F\left(A,B,C,D\right) = \left\langle \Omega \right|
e^{A_{ij}\left(\mathbf{k}\right) a_i\left(\mathbf{k}\right) a_j\left(-\mathbf{k}\right)}
e^{B_{kl}\left(\mathbf{k}\right) b_k\left(\mathbf{k}\right) b_l\left(-\mathbf{k}\right)}
e^{C_{\alpha \beta}\left(\mathbf{k}\right) a^{\dagger}_{\alpha}\left(\mathbf{k}\right) b^{\dagger}_{\beta}\left(-\mathbf{k}\right)}
e^{D_{\gamma \delta}\left(\mathbf{k}\right) a^{\dagger}_{\gamma}\left(-\mathbf{k}\right) b^{\dagger}_{\delta}\left(\mathbf{k}\right)}
\left|\Omega\right\rangle
\label{Fdef}
\end{equation}
is given by:
\begin{equation}
F\left(A,B,C,D\right) = \det\left(ADBC^\intercal+\Id\right)
\end{equation}
\end{sttmnt}
\emph{Proof}:
Let us calculate $F$ explicitly. The exponentials have to be expanded, but since the creation and annihilation operators must be balanced,
they  should all be expanded to the same order, and thus one obtains
\begin{equation}
\begin{aligned}
 &F = \underset{N}{\sum}\left(\frac{1}{N!}\right)^4 A_{i_1 j_1} ... A_{i_N j_N} B_{k_1 l_1} ... B_{k_N l_N} \times \\
 & C_{\alpha_1 \beta_1} ... C_{\alpha_N \beta_N}      D_{\gamma_1 \delta_1} ... D_{\gamma_N \delta_N}
 Z^{i_1 ... i_N j_1 ... j_N k_1 ... k_N l_1 ... l_N}_{\alpha_1 ... \alpha_N \beta_1 ... \beta_N \gamma_1 ... \gamma_N \delta_1 ... \delta_N}
\end{aligned}
\end{equation}
where $Z$ is the vacuum expectation value of the product of creation and annihilation operators, which we shall now calculate:
\begin{equation}
\begin{aligned}
 & Z^{i_1 ... i_N j_1 ... j_N k_1 ... k_N l_1 ... l_N}_{\alpha_1 ... \alpha_N \beta_1 ... \beta_N \gamma_1 ... \gamma_N \delta_1 ... \delta_N} =
 \left\langle \Omega \right|
   a_{i_1}\left(\mathbf{k}\right)a_{j_1}\left(-\mathbf{k}\right)...a_{i_N}\left(\mathbf{k}\right)a_{j_N}\left(-\mathbf{k}\right) \times \\
  & b_{k_1}\left(\mathbf{k}\right)b_{l_1}\left(-\mathbf{k}\right)...b_{k_N}\left(\mathbf{k}\right)b_{l_N}\left(-\mathbf{k}\right) \times
 a^{\dagger}_{\alpha_1}\left(\mathbf{k}\right)b^{\dagger}_{\beta_1}\left(-\mathbf{k}\right)...a^{\dagger}_{\alpha_N}\left(\mathbf{k}\right)b^{\dagger}_{\beta_N}\left(-\mathbf{k}\right) \times \\
 & a^{\dagger}_{\gamma_1}\left(-\mathbf{k}\right)b^{\dagger}_{\delta_1}\left(\mathbf{k}\right)...a^{\dagger}_{\gamma_N}\left(-\mathbf{k}\right)b^{\dagger}_{\delta_N}\left(\mathbf{k}\right)
\left| \Omega \right\rangle.
\end{aligned}
\end{equation}

After changing the operators' positions, in a way which leaves the sign invariant, we can decompose this term into a product of four vacuum expectation values,
\begin{equation}
\begin{aligned}
& Z^{i_1 ... i_N j_1 ... j_N k_1 ... k_N l_1 ... l_N}_{\alpha_1 ... \alpha_N \beta_1 ... \beta_N \gamma_1 ... \gamma_N \delta_1 ... \delta_N} =
 \left\langle a_{i_1}\left(\mathbf{k}\right)...a_{i_N}\left(\mathbf{k}\right) a^{\dagger}_{\alpha_1}\left(\mathbf{k}\right) ... a^{\dagger}_{\alpha_N}\left(\mathbf{k}\right) \right\rangle \times \\
  & \left\langle a_{j_1}\left(-\mathbf{k}\right)...a_{j_N}\left(-\mathbf{k}\right) a^{\dagger}_{\gamma_1}\left(-\mathbf{k}\right) ... a^{\dagger}_{\gamma_N}\left(-\mathbf{k}\right) \right\rangle \times
 \left\langle b_{k_1}\left(\mathbf{k}\right)...b_{k_N}\left(\mathbf{k}\right) b^{\dagger}_{\delta_1}\left(\mathbf{k}\right) ... b^{\dagger}_{\delta_N}\left(\mathbf{k}\right) \right\rangle \times \\
  & \left\langle b_{l_1}\left(-\mathbf{k}\right)...b_{l_N}\left(-\mathbf{k}\right) b^{\dagger}_{\beta_1}\left(-\mathbf{k}\right) ... b^{\dagger}_{\beta_N}\left(-\mathbf{k}\right) \right\rangle =
 \delta^{i_1 ... i_N}_{\alpha_N ... \alpha_1}
\delta^{j_1 ... j_N}_{\gamma_N ... \gamma_1}
\delta^{k_1 ... k_N}_{\delta_N ... \delta_1}
\delta^{l_1 ... l_N}_{\beta_N ... \beta_1} = \\
& \delta^{i_1 ... i_N}_{\alpha_1 ... \alpha_N}
\delta^{j_1 ... j_N}_{\gamma_1 ... \gamma_N}
\delta^{k_1 ... k_N}_{\delta_1 ... \delta_N}
\delta^{l_1 ... l_N}_{\beta_1 ... \beta_N}
\end{aligned}
\end{equation}
where the Kronecker symbol is defined by $\delta^{i_1 ... i_N}_{\alpha_1 ... \alpha_N} = \pm 1$ if $i_1,...,i_N$ are a cyclic/anti-cyclic permutation of
$\alpha_1 ... \alpha_N$, and $0$ otherwise \cite{Synge1969,Bowen2008},
and the last equality results from exchanging the order of the lower indices in all the Kronecker symbols, which leads to the same sign in all four of them and thus
leaves the total sign invariant.

Note that $\delta^{i_1 ... i_N}_{\alpha_1 ... \alpha_N} A_{i_1 j_1} ... A_{i_N j_N}$ is antisymmetric under an exchange of any two $j$ indices: let us demonstrate that by exchanging $j_1,j_2$.
Then,
\begin{equation}
\delta^{i_1 ... i_N}_{\alpha_1 ... \alpha_N} A_{i_1 j_2}  A_{i_2 j_1} ... A_{i_N j_N} =
\delta^{i_2 i_1 ... i_N}_{\alpha_1 ... \alpha_N} A_{i_2 j_2}  A_{i_1 j_1} ... A_{i_N j_N} = -\delta^{i_1 ... i_N}_{\alpha_1 ... \alpha_N} A_{i_1 j_1} ... A_{i_N j_N}.
\end{equation}
An important property of the Kronecker symbol is that if $M_{i_1...i_N}$ is antisymmetric \cite{Synge1969} then:
\begin{equation}
\frac{1}{N!}\delta^{i_1 ... i_N}_{j_1 ... j_N} M_{i_1...i_N} = M_{j_1...j_N}
\end{equation}
and thus, using the antisymmetric behavior of $\delta^{i_1 ... i_N}_{\alpha_1 ... \alpha_N} A_{i_1 j_1} ... A_{i_N j_N}$ and similar symbols, we can significantly simplify $F$
(\ref{Fdef}):
\begin{equation}
\begin{aligned}
 &F = \underset{N}{\sum}\left(\frac{1}{N!}\right)^4
 \delta^{i_1 ... i_N}_{\alpha_1 ... \alpha_N}
\delta^{j_1 ... j_N}_{\gamma_1 ... \gamma_N}
\delta^{k_1 ... k_N}_{\delta_1 ... \delta_N}
\delta^{l_1 ... l_N}_{\beta_1 ... \beta_N}
A_{i_1 j_1} ... A_{i_N j_N} B_{k_1 l_1} ... B_{k_N l_N} \times \\
 & C_{\alpha_1 \beta_1} ... C_{\alpha_N \beta_N}      D_{\gamma_1 \delta_1} ... D_{\gamma_N \delta_N}
 = \underset{N}{\sum}\left(\frac{1}{N!}\right)^3
 \delta^{i_1 ... i_N}_{\alpha_1 ... \alpha_N}
\delta^{k_1 ... k_N}_{\delta_1 ... \delta_N}
\delta^{l_1 ... l_N}_{\beta_1 ... \beta_N} \times \\
& \left(AD\right)_{i_1 \delta_1} ... \left(AD\right)_{i_N \delta_N}
B_{k_1 l_1} ... B_{k_N l_N} C_{\alpha_1 \beta_1} ... C_{\alpha_N \beta_N} = \\
&\underset{N}{\sum}\left(\frac{1}{N!}\right)^2
 \delta^{i_1 ... i_N}_{\alpha_1 ... \alpha_N}
\delta^{l_1 ... l_N}_{\beta_1 ... \beta_N}
 \left(ADB\right)_{i_1 l_1} ... \left(ADB\right)_{i_N l_N}
C_{\alpha_1 \beta_1} ... C_{\alpha_N \beta_N} = \\
&\underset{N}{\sum}\left(\frac{1}{N!}\right)
 \delta^{i_1 ... i_N}_{\alpha_1 ... \alpha_N}
 \left(ADBC^\intercal\right)_{i_1 \alpha_1} ... \left(ADBC^\intercal\right)_{i_N \alpha_N}.
\end{aligned}
\end{equation}

One may also use the Kronecker symbol for the calculation of the characteristic polynomial of a matrix $M$ \cite{Bowen2008},
\begin{equation}
p_M\left(x\right) = \det\left(M-x\Id\right)= \underset{N}{\sum}\left(\frac{1}{N!}\right) \left(-x\right)^N
 \delta^{i_1 ... i_N}_{\alpha_1 ... \alpha_N}
 M_{i_1 \alpha_1} ... M_{i_N \alpha_N}
\end{equation}
and thus we finally obtain that
\begin{multline}
F\left(A,B,C,D\right) = p_{ADBC^\intercal}\left(-1\right) = \det\left(ADBC^\intercal+\Id\right) = \\
= \det\left(DBC^\intercal A+\Id\right)  =  \det\left(BC^\intercal AD+\Id\right)  =  \det\left(C^\intercal ADB+\Id\right)
\end{multline}
where the many possible forms are possible thanks to the symmetry properties of $F$. $\square$

Now we can finally turn to the \emph{proof of statement \ref{th:BCScoeff}}:
We start with the calculation of $\alpha$. Note that
\begin{equation}
\alpha\left(\mathbf{k}\right) = \left\langle \Omega_p | \psi \right\rangle \equiv
\left\langle \Omega \right| B\left(\mathbf{k}\right) A\left(\mathbf{k}\right) \left|\Omega\right\rangle
\end{equation}
where $\left|\Omega\right\rangle$ is the total, both physical and virtual, vacuum of the relevant momentum subspace.

Using the definition of the state $\left|\psi\left(\mathbf{k}\right)\right\rangle$, one can simply write $\alpha$ as
\begin{equation}
\alpha\left(\mathbf{k}\right) = \left\langle \Omega \right|
e^{S_{ij} a_i\left(\mathbf{k}\right) a_j\left(-\mathbf{k}\right)}
e^{S_{kl} b_k\left(\mathbf{k}\right) b_l\left(-\mathbf{k}\right)}
e^{T_{\alpha \beta} a^{\dagger}_{\alpha}\left(\mathbf{k}\right) b^{\dagger}_{\beta}\left(-\mathbf{k}\right)}
e^{T_{\gamma \delta} a^{\dagger}_{\gamma}\left(-\mathbf{k}\right) b^{\dagger}_{\delta}\left(\mathbf{k}\right)}
\left|\Omega\right\rangle
\end{equation}
Since there are no physical annihilation operators, the physical part of $A$ does not contribute, and one may replace the vacuum by the virtual vacuum, and $T$ by $\tau$.
In this form, $\alpha\left(\mathbf{k}\right) = F\left(S,S,\tau,\tau\right)$. Then one simply obtains
\begin{equation}
\alpha\left(\mathbf{k}\right) = \det\left(S \tau^\intercal S \tau + \Id\right)
\end{equation}
whose straightforward calculation leads to Eq. (\ref{alpha}).

Similarly,
\begin{equation}
\begin{aligned}
 & \beta\left(\mathbf{k}\right) = \left\langle \Omega \right|
\psi\left(-\mathbf{k}\right) \psi\left(\mathbf{k}\right)
e^{S_{ij} a_i\left(\mathbf{k}\right) a_j\left(-\mathbf{k}\right)}
e^{S_{kl} b_k\left(\mathbf{k}\right) b_l\left(-\mathbf{k}\right)}
e^{T_{\alpha \beta} a^{\dagger}_{\alpha}\left(\mathbf{k}\right) b^{\dagger}_{\beta}\left(-\mathbf{k}\right)}
e^{T_{\gamma \delta} a^{\dagger}_{\gamma}\left(-\mathbf{k}\right) b^{\dagger}_{\delta}\left(\mathbf{k}\right)}
\left|\Omega\right\rangle = \\
 & \underset{X \rightarrow \infty}{\lim} \frac{1}{X} \left\langle \Omega \right|
e^{S_{Xij} a_i\left(\mathbf{k}\right) a_j\left(-\mathbf{k}\right)}
e^{S_{kl} b_k\left(\mathbf{k}\right) b_l\left(-\mathbf{k}\right)}
e^{T_{\alpha \beta} a^{\dagger}_{\alpha}\left(\mathbf{k}\right) b^{\dagger}_{\beta}\left(-\mathbf{k}\right)}
e^{T_{\gamma \delta} a^{\dagger}_{\gamma}\left(-\mathbf{k}\right) b^{\dagger}_{\delta}\left(\mathbf{k}\right)}
\left|\Omega\right\rangle = \\
& \underset{X \rightarrow \infty}{\lim} \frac{1}{X} F\left(S_X,S,T,T\right) = \underset{X \rightarrow \infty}{\lim} \frac{1}{X} \det\left(ST^\intercal S_XT + \Id\right)
\end{aligned}
\end{equation}
where $S_X = -X \Id_1 \oplus S$ ($\Id_1$ is the $1 \times 1$ identity matrix). This results in (\ref{beta}) and completes the proof of statement \ref{th:BCScoeff}. $\square$

The only remaining task, which shall result in the proof of statement \ref{th:unpaired}, is the calculation of  $\tilde \alpha$ for the unpaired modes, $\mathbf{k}=\left(0,0\right) , \left(\pi,\pi\right), \left(\pi,0\right), \left(0,\pi\right)$.
\begin{equation}
\tilde \alpha\left(0,0\right)=\left\langle \Omega \right|e^{\tilde{S}_{ij}\left(0,0\right)a_i a_j}e^{\tilde{S}_{kl}\left(0,0\right)b_k b_l}e^{\tau_{\alpha \beta}\left(0,0\right)
a^{\dagger}_{\alpha} b^{\dagger}_{\beta}} \left| \Omega \right\rangle
\end{equation}
with all the operators at $\mathbf{k}=0$, and similarly for $\tilde \alpha\left(\pi,\pi\right)$. We expand the exponentials such that all the creation and annihilation operators are balanced,
and obtain
\begin{equation}
\begin{aligned}
& \tilde \alpha\left(0,0\right)=\underset{N}{\sum}\left(\frac{1}{N}\right)^2 \frac{1}{\left(2N\right)!}\tilde{S}_{i_1 j_1}...\tilde{S}_{i_N j_N}\tilde{S}_{k_1 l_1}...\tilde{S}_{k_N l_N}
\tau_{\alpha_1 \beta_1}...\tau_{\alpha_{2N} \beta_{2N}} \times
\\ & \left\langle a_{i_1}a_{j_1}...a_{i_N}a_{j_N} b_{k_1}b_{l_1}...b_{k_N}b_{l_N}
a^{\dagger}_{\alpha_1}b^{\dagger}_{\beta_1}...a^{\dagger}_{\alpha_{2N}}b^{\dagger}_{\beta_{2N}}\right\rangle
\end{aligned}
\end{equation}
Note that $\tilde{S}\left(\pi,\pi\right) = -\tilde{S}\left(0,0\right)$, but as the $\tilde{S}$ matrices always appear an even number of times in the expansion,
$\tilde \alpha\left(\pi,\pi\right)=\tilde \alpha\left(0,0\right)$.

Here, only $N=0,1,2$ contribute - higher orders vanish due to the fermionic statistics. The zeroth order is trivially 1. The first one is
\begin{equation}
\begin{aligned}
&\tilde \alpha_1\left(0,0\right) = \frac{1}{2}\tilde{S}_{ij}\tilde{S}_{kl}
\left(\delta_{j \alpha_1}\delta_{i \alpha_2}-\delta_{i \alpha_1}\delta_{j \alpha_2}\right)
\left(\delta_{l \beta_1}\delta_{k \beta_2}-\delta_{k \beta_1}\delta_{l \beta_2}\right) \tau_{\alpha_1 \beta_1} \tau_{\alpha_2 \beta_2} = \\
& -2 \text{Tr}\left(\tilde{S}^\intercal \tau \tilde{S} \tau^\intercal\right) = -2\left(y^2+z^2\right).
\end{aligned}
\end{equation}
And the second, last order is
\begin{equation}
\begin{aligned}
&\tilde \alpha_2\left(0,0\right) = \frac{1}{4} \frac{1}{4!} \delta^{i_1 j_1 i_2 j_2}_{\alpha_1 \alpha_2 \alpha_3 \alpha_4} \delta^{k_1 l_1 k_2 l_2}_{\beta_1 \beta_2 \beta_3 \beta_4}
\tilde{S}_{i_1 j_1}\tilde{S}_{i_2 j_2}\tilde{S}_{k_1 l_1}\tilde{S}_{k_2 l_2}\tau_{\alpha_1 \beta_1} \tau_{\alpha_2 \beta_2}\tau_{\alpha_3 \beta_3} \tau_{\alpha_4 \beta_4}.
\end{aligned}
\end{equation}
Since the number of possible indices is $4$, the Kronecker symbols may be decomposed as
$\delta^{i_1 j_1 i_2 j_2}_{\alpha_1 \alpha_2 \alpha_3 \alpha_4} = \epsilon^{i_1 j_1 i_2 j_2}\epsilon_{\alpha_1 \alpha_2 \alpha_3 \alpha_4}$ etc., and then
$\frac{1}{4!}\epsilon_{\alpha_1 \alpha_2 \alpha_3 \alpha_4}\epsilon_{\beta_1 \beta_2 \beta_3 \beta_4}\tau_{\alpha_1 \beta_1} \tau_{\alpha_2 \beta_2}\tau_{\alpha_3 \beta_3} \tau_{\alpha_4 \beta_4} = \det\left(\tau\right)$
as well as $\frac{1}{4}\left(\epsilon_{i_1 j_1 i_2 j_2}\tilde{S}_{i_1 j_1}\tilde{S}_{i_2 j_2}\right)^2 = 1$, hence the second order is simply $\det\left(\tau\right) = \left(y^2-z^2\right)^2$.
Altogether we obtain
\begin{equation}
\tilde \alpha\left(0,0\right)=\tilde\alpha\left(\pi,\pi\right)=1-2\left(y^2+z^2\right)+ \left(y^2-z^2\right)^2
= \left(1-\left(y+z\right)^2\right)\left(1-\left(y-z\right)^2\right).
\end{equation}
A similar calculation for the other unpaired momenta yields
\begin{equation}
\tilde \alpha\left(\pi,0\right)=\tilde\alpha\left(0,\pi\right)=1-2\left(y^2-z^2\right)+ \left(y^2-z^2\right)^2
= \left(1-\left(y^2-z^2\right)\right)^2.
\end{equation}

Note that the right limit is obtained for the unpaired momenta $\mathbf{k}$: indeed, $\alpha \rightarrow \tilde{\alpha}^2$ there,
as a straightforward consequence of Eqs. (\ref{alpha}),(\ref{beta}). This proves statement \ref{th:unpaired}. $\square$

\section*{References}

\bibliography{ref}

\end{document}